\documentclass[fleqn,usenatbib]{mnras}
\usepackage[T1]{fontenc}
\usepackage{ae,aecompl}

\usepackage[pdfpagelabels=false]{hyperref}
\usepackage{graphicx}	% Including figure files
\usepackage{amsmath}	% Advanced maths commands
\usepackage{amssymb}	% Extra maths symbols
\usepackage{subfigure}
\title[Fe\,{\sc i} and Ca\,{\sc ii} H$\&$K in KELT-20b]{Searching for Thermal Inversion Agents in the Transmission Spectrum of KELT-20b/MASCARA-2b: Detection of Neutral Iron \textbf{and Ionised Calcium H$\&$K Lines}}

\author[S.K. Nugroho et al.]{Stevanus K. Nugroho$^{1}$\thanks{E-mail: s.nugroho@qub.ac.uk},
Neale P. Gibson$^{2}$,
Ernst J. W. de Mooij$^{1}$,
Chris A. Watson$^{1}$,
\newauthor
Hajime Kawahara$^{3,4}$,
and Stephanie Merritt$^{1}$
\smallskip
\\
$^{1}$School of Mathematics and Physics, Queen's University Belfast, University Road, Belfast, BT7 1NN, United Kingdom\\
$^{2}$School of Physics, Trinity College Dublin, The University of Dublin, Dublin 2, Ireland\\
$^{3}$Department of Earth and Planetary Science, The University of Tokyo, Tokyo 113-0033, Japan\\
$^{4}$Research Center for the Early Universe, School of Science, The University of Tokyo, Tokyo 113-0033, Japan\\
}

\date{Accepted 2020 May 20. Received 2020 May 19; in original form 2019 November 20}

\pubyear{2020}

\begin{document}
\label{firstpage}
\pagerange{\pageref{firstpage}--\pageref{lastpage}}
\maketitle

%@arxiver{fig8c.pdf, fig11c.pdf, fig16.pdf}

% Abstract of the paper
\begin{abstract}
We analyse the transmission spectra of KELT-20b/MASCARA-2b to search for possible thermal inversion agents. The data consist of three transits obtained using HARPSN and one using CARMENES. We removed stellar and telluric lines before cross-correlating the residuals with spectroscopic templates produced using a 1D plane-parallel model assuming an isothermal atmosphere and chemical equilibrium at solar metallicity. Using a likelihood-mapping method, we detect Fe\,{\sc i} at $>$ 13-$\sigma$, Ca\,{\sc ii} H$\&$K at $>$ 6-$\sigma$ and confirm the previous detections of Fe\,{\sc ii}, Ca\,{\sc ii} IRT and Na\,{\sc i} D. The detected signal of Fe\,{\sc i} is shifted by -3.4$\pm$0.4 km s$^{-1}$ from the planetary rest frame, which indicates a strong day-night wind. Our likelihood-mapping technique also reveals that the absorption features of the detected species extend to different altitudes in the planet's atmosphere. Assuming that the line lists are accurate, we do not detect other potential thermal inversion agents (NaH, MgH, AlO, SH, CaO, VO, FeH and TiO) suggesting that non-chemical equilibrium mechanisms (e.g. a cold-trap) might have removed Ti- and V-bearing species from the upper atmosphere. Our results, therefore, shows that KELT-20b/MASCARA-2b cannot possess an inversion layer caused by a TiO/VO-related mechanism. The presence of an inversion layer would therefore likely be caused by metal atoms such as Fe\,{\sc i} and Fe\,{\sc ii}. Finally, we report a double-peak structure in the Fe\,{\sc i} signal in all of our data-sets that could be a signature of atmospheric dynamics. However, further investigation is needed to robustly determine the origin of the signal.

\end{abstract}

% Select between one and six entries from the list of approved keywords.
% Don't make up new ones.

%%%%%%%%%%%%%%%%%%%%%%%%%%%%%%%%%%%%%%%%%%%%%%%%%%

%%%%%%%%%%%%%%%%% BODY OF PAPER %%%%%%%%%%%%%%%%%%

\section{Introduction}
To date, we have detected more than 4000 planets orbiting stars other than the Sun. Despite this ever-growing sample of planets ripe for study, many important mysteries remain regarding their most fundamental properties. One such mystery is the physics and chemistry driving the presence of thermal inversion layers in ultra-hot Jupiters ($>$2000 K). Inversion layers were predicted by \citet{Hubeny2003} and \citet{Fortney2008} for highly-irradiated atmospheres, caused by the strongly-absorbing molecular features of TiO and VO, as seen in M-dwarfs. These species are expected to be present in the hottest exoplanet atmospheres, absorbing incoming stellar irradiation at optical wavelengths and depositing a significant amount of energy in the upper atmosphere. \citet{Spiegel2009} showed that a hot Jupiter with solar C/O requires TiO of at least solar abundance to create an observable temperature inversion.

While there have been several claims of the detection of thermal inversions \citep[e.g. ][]{Machalek2008, Knutson2008, Todorov2010, O'Donovan2010}, and tentative evidence of TiO in the atmosphere of hot Jupiters \citep[e.g. ][]{Desert2008, Evans2016}, it was not until the detection of an emission feature of H$_{2}$O on the day-side of WASP-121b by \citet{Evans2017} using WFC3/HST that the existence of a stratospheric inversion layer was directly confirmed. Meanwhile, the first detection of TiO was claimed on the atmospheric limb of WASP-19b by \citet{Sedaghati2017} using low-resolution spectroscopy with VLT/FORS2. In contrast with \citet{Sedaghati2017}, however, \citet{Espinoza2019} reported that five out of six optical transmission spectra of WASP-19b (taken using IMACS/Magellan) were featureless and hence consistent with high altitude clouds. A significant slope as a function of wavelength and a tentative detection of TiO absorption features were found in only one of the spectra, which interestingly also had the clearest signature of stellar contamination. One of the possible explanations for this results is there were stellar contamination effects they did not consider when modelling the stellar component in the analysis which potentially could mimic the TiO feature in the extracted transmission spectrum of the planet.

Unlike low-resolution spectroscopy, high-resolution spectroscopy (R $>$ 25,000) is able to resolve molecular bands into individual absorption/emission lines. The variation of Doppler shifts caused by planetary orbital motion enables absorption/emission lines in the exoplanet spectrum to be distinguished from telluric and/or stellar lines and ensures unambiguous detection of specific atomic or molecular species. This has become one of the most robust approaches in the attempt to characterize the atmospheres of exoplanets \citep[e.g.][]{Redfield2008,Snellen2008,Snellen2010,Jensen2011,Brogi2012,Brogi2013,Birkby2013,Snellen2014,Brogi2014,Lockwood2014,Schwarz2015,Hoeijmakers2015,Wyttencbach2015,Birkby2017,Hawker2018, Brogi2018,Pino2018, Wang2018, Sanchez2019,Alonso-Floriano2019,Brogi2019, Hoeijmakers2019, gibson2020, Merrit2020}. Using this technique, \citet{Nugroho2017} directly detected a high-resolution emission signature of TiO on the day-side of WASP-33b using the Subaru telescope, providing simultaneous evidence of the existence of both this molecule and a hot stratosphere in the atmosphere of the planet.
While there are a number of ultra-hot Jupiters which has the evidence of possessing a thermal inversion layer in their atmosphere, e.g. WASP-103b \citep{Kreidberg2018}, WASP-18b \citep{Sheppard2017,Arcangeli2018}, HAT-P-7b \citep{Wong2016,Mansfield2018}, WASP-33b \citep{Haynes2015,Nugroho2017}, WASP-121b \citep{Evans2017}, TiO has only been detected in WASP-33b and, possibly, WASP-19b \citep{Sedaghati2017}.

It is still possible for a hot Jupiter to exhibit a stratospheric thermal inversion without TiO/VO if H$^{-}$ present in the atmosphere and the infrared opacity is low enough/lack of infrared coolant, for example through the thermal/photo-dissociation of H$_{2}$O, or if it has high C/O so that all of the main coolant (CH4, H$_{2}$O, and HCN) are depleted \citep{Molliere2015, Parmentier2018, Lothringer2018, Arcangeli2018}. Another scenario is if there are other strong visible opacity sources present besides TiO/VO. \citet{Gandhi2019} suggested that molecular species such as AlO, CaO, NaH and MgH could provide an optical opacity comparable to TiO/VO, thus also enabling the existence of thermal inversions. The existence of thermal inversion in an ultra-hot Jupiter, however, is also influenced by the spectral type of the host star. \citet{Lothringer2018} and \citet{Lothringer2019} showed that the thermal inversion layer could also be observed even in a planet with equilibrium temperature ($T_{\mathrm{eq}}$) of 2250 K if it is orbiting an early-type host star. Even without TiO and VO, metal atoms like Fe\,{\sc i}, Fe\,{\sc ii}, C\,{\sc i} and Ti\,{\sc ii} are large contributors to the absorption of significant UV and optical wavelength stellar flux, and are also able to create an observable inversion layer \citep[see Figure 5 in ][]{Lothringer2019}. Most of the metals that could potentially create inversion layers were recently observed for the first time in KELT-9b ($T_{\mathrm{eq}}$= 4050 K) using high-resolution spectroscopy \citep{Hoeijmakers2018}. 

For a cooler planet, \citet{Casasayas-Barris2018} and \citet{Casasayas-Barris2019} detected Fe\,{\sc ii}, the Ca\,{\sc ii} IR Triplet (hereafter Ca\,{\sc ii} IRT), the Na\,{\sc i} doublet (hereafter Na\,{\sc i} D) and the Balmer series of H\,{\sc i} in the transmission spectrum of KELT-20b/MASCARA-2b (hereafter KELT-20b). There have been no constraints on its temperature structure and no detection of Fe\,{\sc i} or any molecular thermal inversion agents. However, as it orbits a bright, early-type star (A2V, V= 7.6) and $T_{\mathrm{eq}}$ of 2260 K, it is a prime target to search for potential thermal inversion agents that have been suggested in the literature. Therefore, we re-analysed the transmission spectrum of KELT-20b to search for additional species using the cross-correlation technique.

In this paper, we present the detection of Fe\,{\sc i} and Ca \,{\sc ii} H$\&$K in the transmission spectrum of KELT-20b, as well as independently confirming the previously reported presence of Fe\,{\sc ii}, Na\,{\sc i} D and Ca\,{\sc ii} IRT. In Section \ref{sec:obsdata}, we describe the observations and data reduction. We then detail our analysis to search for planetary atmosphere signals in Section \ref{sec:planetsignal}. In Section \ref{sec:resanddis} we outline and discuss our findings and, finally, we summarise the conclusions of our study in Section \ref{sec:concl}.

\section{Observations and data reduction} \label{sec:obsdata}
\begin{table}
\centering 
\caption{Physical and system parameters of KELT-20/MASCARA-2 and KELT-20b. Almost all of the parameters in the table are adopted from \citet{Talens2018}, values with (a) are taken from \citet{Lund_2017}, while values from this work are marked with (b) for the result using HARPSN N1 and (c) for the result using CARMENES data.\label{table1}}
\begin{tabular}{lc}
\hline
\hline
Parameter & Value  \\
\hline
\textbf{KELT-20} &  \\
$M_{\star}$ ($\text{M}_{\bigodot}$)    & 1.89$^{+0.07}_{-0.05}$\\ 
$R_{\star}$ ($\text{R}_{\bigodot}$)    & 1.60$\pm$0.06\\
Spectral type & A2V \\
$T_{\text{eff}}$ (K) & 8720 $^{+250}_{-260}$ $^{\mathrm{a}}$\\
$\log$ $g$ & 4.31$\pm$0.02\\
$[$Fe/H$]$ & -0.29$^{+0.22}_{-0.36}$ $^{\mathrm{a}}$\\
Age (Myr) & 200$^{+100}_{-50}$ \\
$\textit{v}_{\mathrm{rot}}$ sin $\textit{i}_{\star}$ (km s$^{-1}$)& 114.0 $\pm$3\\
$v_{\mathrm{sys}}$ (km s$^{-1}$)&-21.30 $\pm$ 0.30\\
& -23.30  $\pm$ 0.40 $^{\mathrm{a}}$\\
& -22.06  $\pm$ 0.35 $^{\mathrm{b}}$\\
& -22.02  $\pm$ 0.47 $^{\mathrm{c}}$\\
\vspace{0.1pt}\\
\hline\

\textbf{KELT-20b} \\
$T_{0}$ $(\text{BJD}_{\text{TBD}})$         & 2457909.5906 $^{+0.0003}_{-0.0002}$ \\
$P$ (days)                                  & 3.474119 $^{+0.000005}_{-0.000006}$\\
$T_{\text{14}}$ (hours)                     & 3.57552$^{+0.02184}_{-0.02112}$\\
$M_{\mathrm{P}}$ ($\text{M}_{\text{J}}$)    & $<$ 3.382 (3$\sigma$)\\
$R_{\mathrm{P}}$/$\text{R}_{\star}$      & 0.117$\pm$0.009\\
$a/\text{R}_{\star}$                     & 7.66$\pm$1.09\\
$\textit{i}$ ($^{\circ}$)                   & 86.4$^{+0.5}_{-0.4}$\\
\end{tabular}
\end{table}
The transits of KELT-20b were observed on the night of 16 August 2017 (hereafter N1, PID: CAT17A$\_$38, PI: Rebolo), 13 July 2018 and 20 July 2018 (hereafter N2 and N3, PID: CAT18A$\_$34, PI: Casasayas-Barris) using the HARPS-North spectrograph (R$\sim$115,000, ) on the 3.58 m Telescopio Nazionale Galileo (TNG) at the Observatorio del Roque de Los Muchachos, La Palma and on the night of 23 August 2017 using the CARMENES spectrograph (R$\sim$94,600 for the VIS channel, \citet{quirrenback2016}) at the Calar Alto Observatory (PID: DDT175, PI: Czesla). As has been described in Casasayas-Barris et al. (2018, 2019), the spectra were taken before, during and after the transit, resulting in 90 and 116 spectra obtained for the N1 and N2 data-sets (t$_{\text{exp}}$= 200 s), respectively; 78 spectra for the N3 data-set (t$_{\text{exp}}$= 300 s); and 74 spectra for the CARMENES data-set (t$_{\text{exp}}$= 192 s). Due to a low signal-to-noise ratio (S/N), we discarded 8 spectra from the N2 data-set (frame numbers of 29-31, 39-43).

We obtained the extracted two-dimensional spectra for the HARPSN data (the e2ds data) from the Italian centre for Astronomical Archive (IA2)\footnote{ \url{http://archives.ia2.inaf.it/tng/faces/search.xhtml?dswid=-3493}} which were reduced using the HARPS-North Data Reduction Software (DRS), version 3.7. The HARPSN spectrograph covers a wavelength range from $\sim$3800 to 6900 \AA \ divided into 69 echelle orders (with mean velocity dispersion of $\sim$0.8\,km s$^{-1}$ pixel$^{-1}$). The wavelength solution for each order and exposure is calculated using a third-degree polynomial following the DRS User Manual\footnote{\url{http://www.tng.iac.es/instruments/harps/data/HARPS-N_DRSUserManual_1.1.pdf}} resulting in air wavelength at the observer rest frame. As explained in \citet{Casasayas-Barris2019}, there was a problem with the Atmospheric Dispersion Corrector (ADC) during the observation of N2 and N3 which manifest as continuum profile variations in some of the spectra. This effect was effectively removed after we corrected the blaze function variation in the data-set. For the CARMENES data, the one-dimensional reduced spectra were obtained from the Calar Alto Archive\footnote{\url{http://caha.sdc.cab.inta-csic.es/calto/}}. The spectra were reduced using the CARMENES pipeline v2.01,CARACAL \citep[CARMENES Reduction And Calibration; ][]{Caballero2016}, covering the optical range from 5171 to 9634 \AA, \ which is divided into 60 echelle orders (with a mean velocity sampling of $\approx$1.2\,km s$^{-1}$ pixel$^{-1}$). However, due to low S/N and high telluric contamination, only the first 50 bluest orders were used. The remaining orders cover the wavelength from $\approx$5163 to 8934 \AA. The wavelength value per pixel for each order is given in vacuum at the observer rest frame and was converted into the air wavelength using the formula from \citet{Morton2000}. Hereafter, we start labelling the order from 1 from the blue.
 
\subsection{Data preparation} \label{sec:dataprep}
The data from HARPSN and CARMENES were treated similarly. First, pixels with NaN values were masked from the data, then the spectra of each order were stacked into a two-dimensional matrix with the column as the wavelength bin and the row as the frame number. To normalise the data, the spectrum with the highest S/N from both data-sets was chosen as a reference. The continuum of the reference spectra was fit using the \textit{continuum} task in {\sc IRAF}\footnote{The Image Reduction and Analysis Facility ({\sc IRAF}) is distributed by the US National Optical Astronomy Observatories, operated by the Association of Universities for Research in Astronomy, Inc., under a cooperative agreement with the National Science Foundation.} and then divided out from all the spectra. 

To make sure that all of the spectra share a similar blaze function, we performed the following procedure. We calculated the ratio between the spectrum of each exposure and the reference spectrum: if the blaze function was stable during the observation, the continuum profile of the ratio should be flat. Since the blaze function variation only affects the continuum profile of the spectrum, we removed any outliers from the ratio caused by the variation of telluric and/or stellar lines (e.g. airmass, water vapour column, unstable wavelength solution) by performing 3-$\sigma$ clipping relative to the pseudo-continuum approximated using a smoothing function\footnote{Using {\sc PyAstronomy.pyasl.smooth} \url{https://github.com/sczesla/PyAstronomy}} with a flat window function spanning 51 pixels. We then fit the residuals using a Chebyshev polynomial\footnote{Using {\sc numpy.polynomial.Chebyshev.fit}} and divided this fit from the corresponding spectrum, resulting in a spectrum with a similar blaze function to the reference spectrum. 

The low S/N regions (e.g. at the edge of the spectral order) and pixels that have values of less than 20 per cent of the continuum level (e.g. due to strong telluric absorption) are masked from the data. Then we performed 5-$\sigma$ clipping in each wavelength bin and masked any outliers, e.g. due to cosmic rays. In total, we masked 0.71 per cent, 4.98 per cent, 1.07 per cent and 3.93 per cent of the total number of pixels from the data of N1, N2, N3 and CARMENES respectively.

\begin{figure*}
    \centering
    \subfigure{\includegraphics[width=0.4\linewidth]{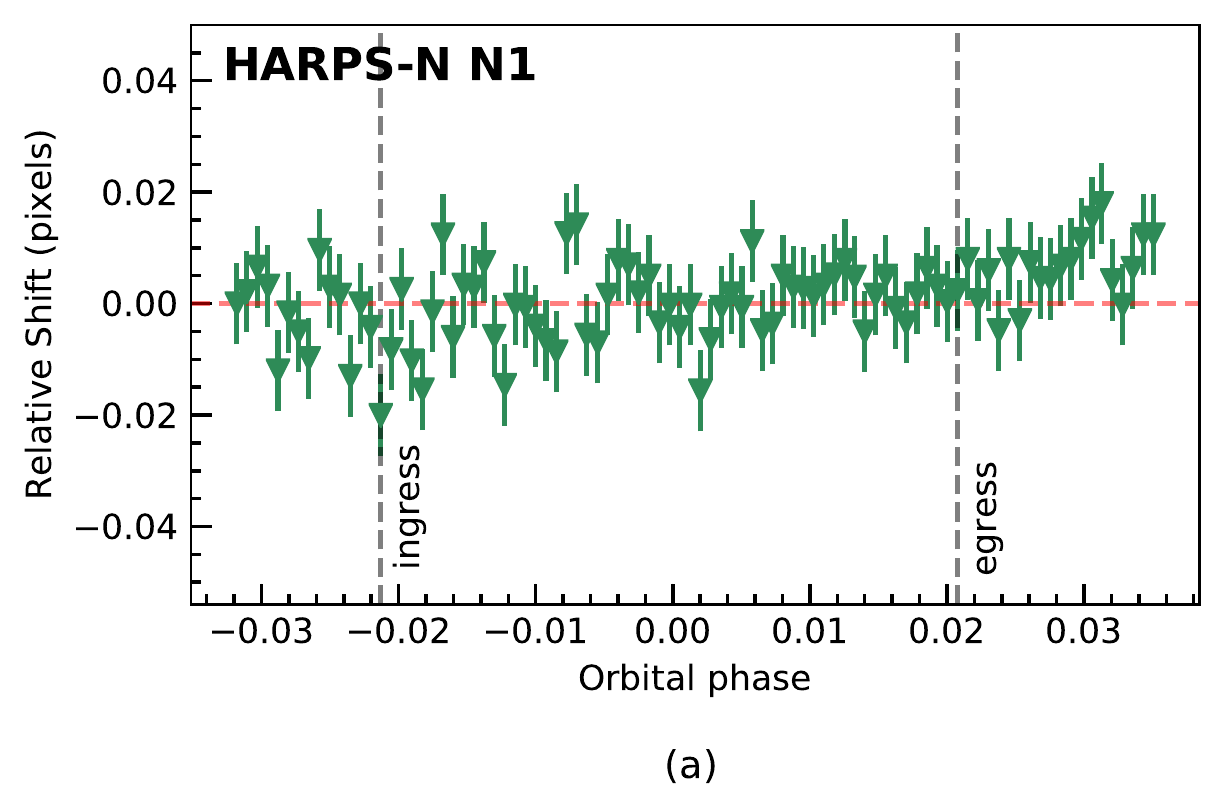}\label{fig:wavecalibN1}}
    \subfigure{\includegraphics[width=0.4\linewidth]{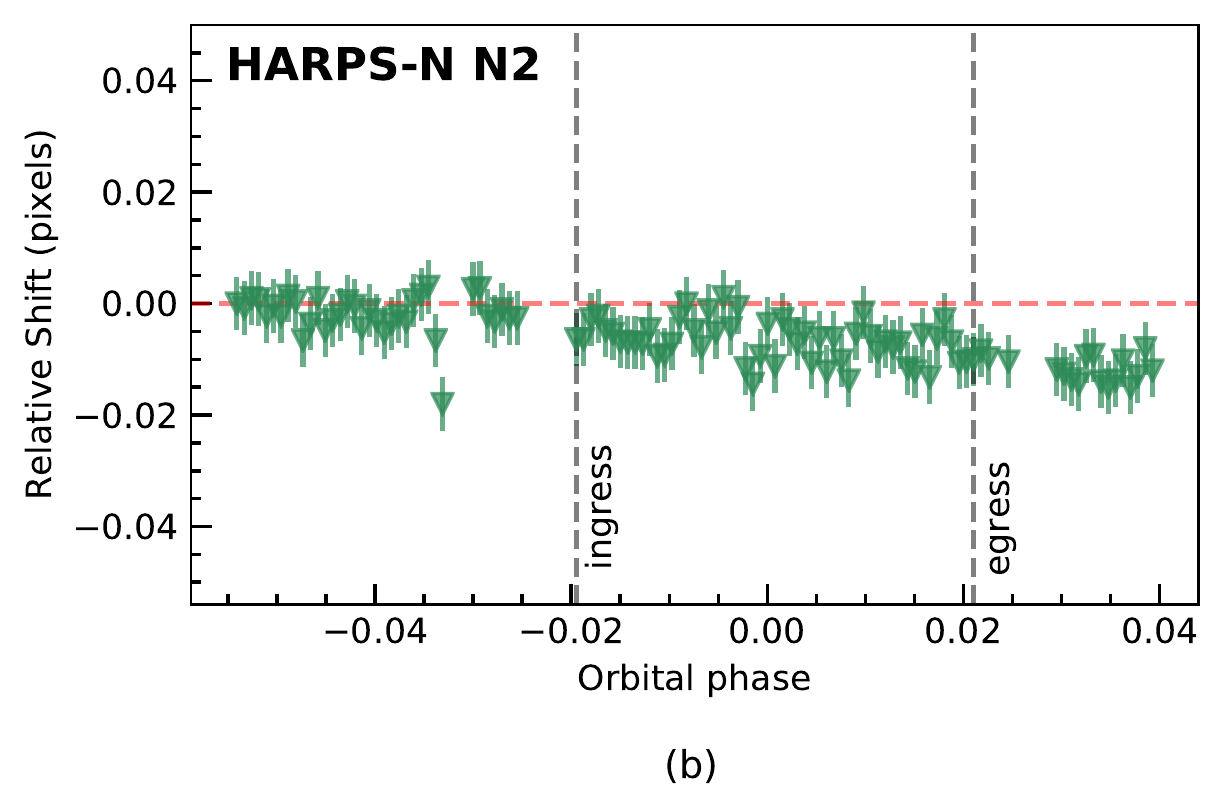}\label{fig:wavecalibN2}}
    \subfigure{\includegraphics[width=0.4\linewidth]{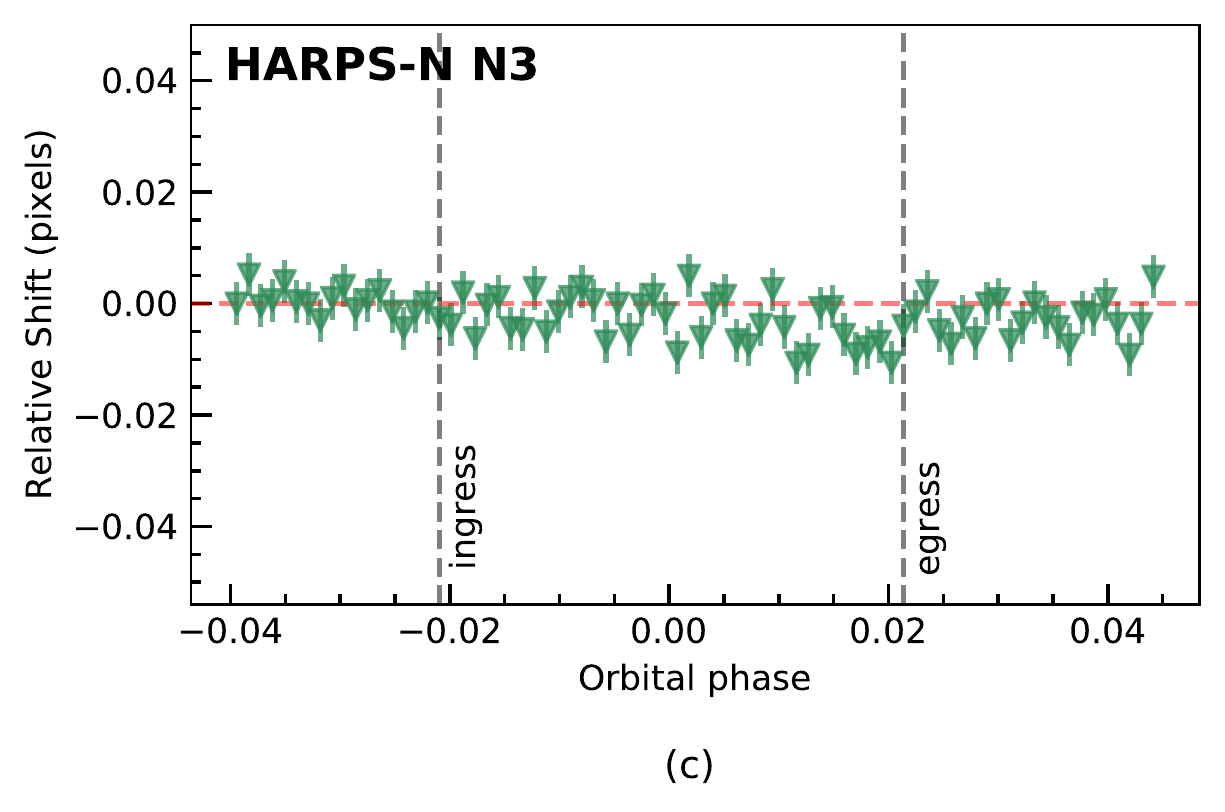}\label{fig:wavecalibN3}}
    \subfigure{\includegraphics[width=0.4\linewidth]{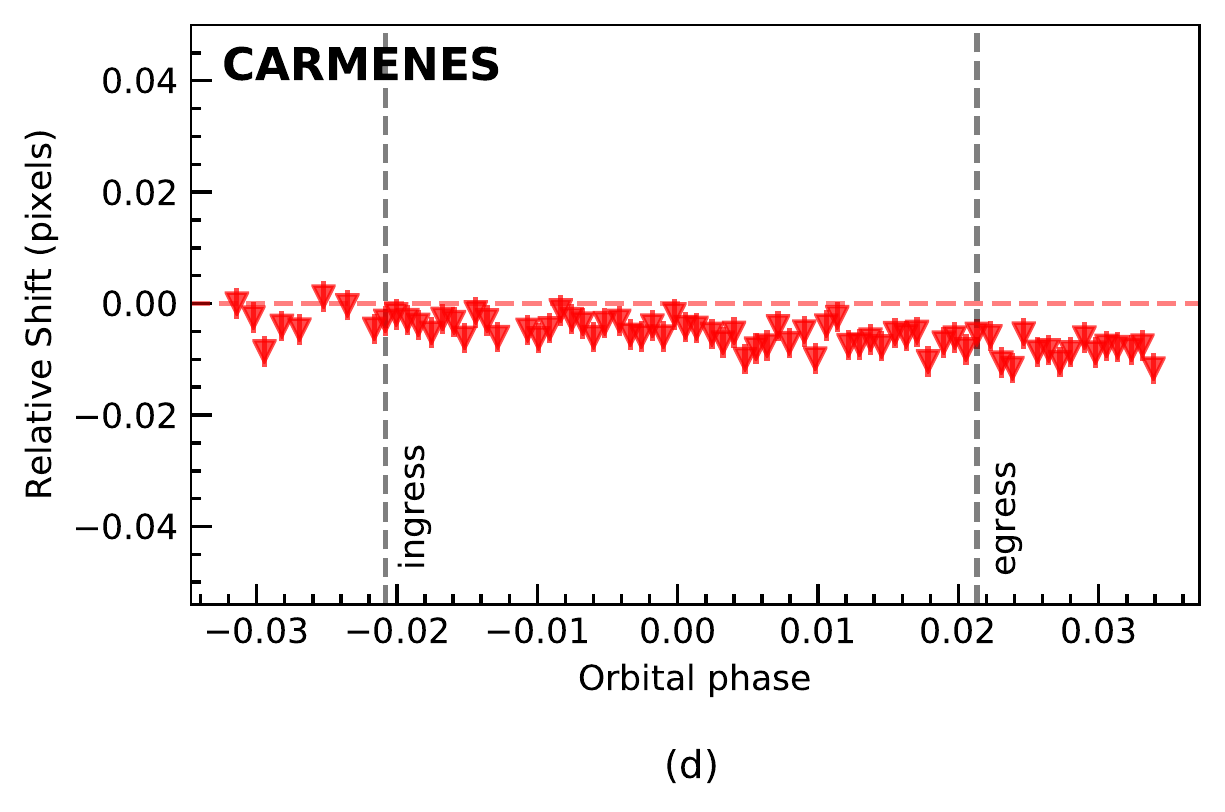}\label{fig:wavecalibCarm}}

    \caption{The relative wavelength shifts during the observations for HARPSN data (a, b and c) and CARMENES data (d). The black dashed lines mark the beginning of the ingress and the ending of the egress, while the red dashed lines mark the zero value for each data-set. The maximum relative wavelength shifts are less than 0.1 pixels and 0.01 pixels for HARPSN and CARMENES data, respectively. These results demonstrate that both spectrographs were relatively stable during the observations.
    \label{fig:wavecalib}}
\end{figure*} 

Before removing the telluric and stellar lines, we checked the stability of the wavelength calibration by cross-correlating the orders that contain telluric lines (orders 54, 55, 61, 64, and 69 for the HARPSN data, and orders 22, 25, 30-36, 39, 42-46 and 51 for the CARMENES data) with the Doppler-shifted telluric spectrum model produced by the {\sc Cerro Paranal Sky Model} \citep[]{Noll2012, Jones2013}. The correlation was done assuming lags of 0.001-pixel steps over a range spanning -1 to 1 pixel. The line depth difference of the telluric lines in the spectrum model and the observed data does not affect the result since the purpose of this cross-correlation is to measure possible wavelength shifts. We calculated the cross-correlation order-by-order and measured the shift of the spectrum during the observation relative to the spectrum of the first exposure. Then, to estimate the final relative shift of one exposure, we calculated the mean of the relative shift of all selected orders within one exposure and took the standard deviation as the error. The result is shown in Figure \ref{fig:wavecalib}. For the HARPSN data, the wavelength calibration is stable within $\pm$ 0.04 pixels and there is no apparent long term trend in the shifts except for the N2 data-set. For CARMENES, the wavelength calibration is more precise than HARPSN (probably due to the fact it covers more telluric lines than HARPSN), although a long term trend in the shift similar to that found for the N2 data-set was observed. The maximum relative wavelength shifts during the observations are 0.1 pixels and 0.01 pixels, which is insignificant compared to the precision that we require, and therefore we did not perform any further wavelength alignment.
    
\subsection{Removal of the Doppler shadow} \label{subsec:dopshadow}
Following \citet{Yan2017, Casasayas-Barris2017, Casasayas-Barris2018, Cauley2019, Yan2019, Turner2020}, we modelled and removed the Doppler shadow from the data. The stellar and planetary parameters used in the model are in Table \ref{table1}. Using {\sc Spectroscopy Made Easy} \citep[SME, ][]{Piskunov2017}, we generated stellar spectra at 21 different limb angles ($\mu= \cos{\theta}$) with the line data obtained from VALD3\footnote{http://vald.astro.uu.se \url{http://vald.astro.uu.se}} \citep{Ryabchikova2015} using the atmospheric model of ATLAS9 \citep{Heiter2002}. The stellar disk was modelled using a regular grid with a radius of 510 pixels assuming a solid body rotation while ignoring the effect of differential rotation and gravity darkening. For each pixel, we linearly interpolated the $\mu$-dependent stellar spectrum to the correct $\mu$, taking the Doppler shift due to stellar rotation into account. The Doppler shadow was then modelled by masking the pixel occulted by the planetary disk during the transit, after which the stellar spectrum was integrated over the entire disk and subsequently convolved to the instrument resolution. This model corrects both the Rossiter-McLaughlin (R-M) and the Center-to-Limb-Variation (CLV) effects. We note that the planetary disk ignores the wavelength-dependent effect of the planetary radius caused by its atmosphere. As mentioned in \citet{Turner2020}, the R-M effect is underestimated at the wavelength where the absorption from the planetary atmosphere is strong. To ensure that the model only include the R-M and CLV effects, for each observation we divided the model by the out-of-transit spectrum after continuum normalisation, and this was subsequently divided out of the data.

\subsection{Removal of stellar and telluric lines} \label{subsec:sysrem}
The telluric and the stellar lines should be removed from the data before searching for any planetary signal using the cross-correlation method. We removed these lines using {\sc SysRem} \citep{Tamuz2005}. {\sc SysRem} is a de-trending algorithm developed to remove linear systematic effects or common-mode signals from a large photometric survey data-set. In the high-resolution spectroscopy analysis, {\sc SysRem} treats the wavelength bins as a large set of light-curves. The strength of telluric lines changes during the observations (e.g. due to changes in airmass, weather, or water vapour column level) while the position of the line centre is static (or quasi-static in case of the stellar lines). This variation is recognised by {\sc SysRem} as a common-mode signal that can be fit and removed effectively by taking into account the uncertainty of each data point. In the first order, {\sc SysRem} will not remove the planetary signal since the position of planetary absorption lines changes during the observation due to its orbital motion. During the in-transit phase, the expected radial velocity (RV) of KELT-20b changes from $-33$\,km s$^{-1}$ to $+33$\,km s$^{-1}$, which corresponds to 82 pixels and 55 pixels shift from ingress to egress for HARPSN and CARMENES respectively. This, therefore, does not significantly affect the common-mode signals recognised by {\sc SysRem}. We refer the readers to previous studies for the technical details of the application of {\sc SysRem} in removing high-resolution telluric and stellar lines \citep[e.g.][]{Birkby2017, Nugroho2017, Cabot2019}. 

\begin{figure}
    \centering
    \includegraphics[width=0.9\linewidth]{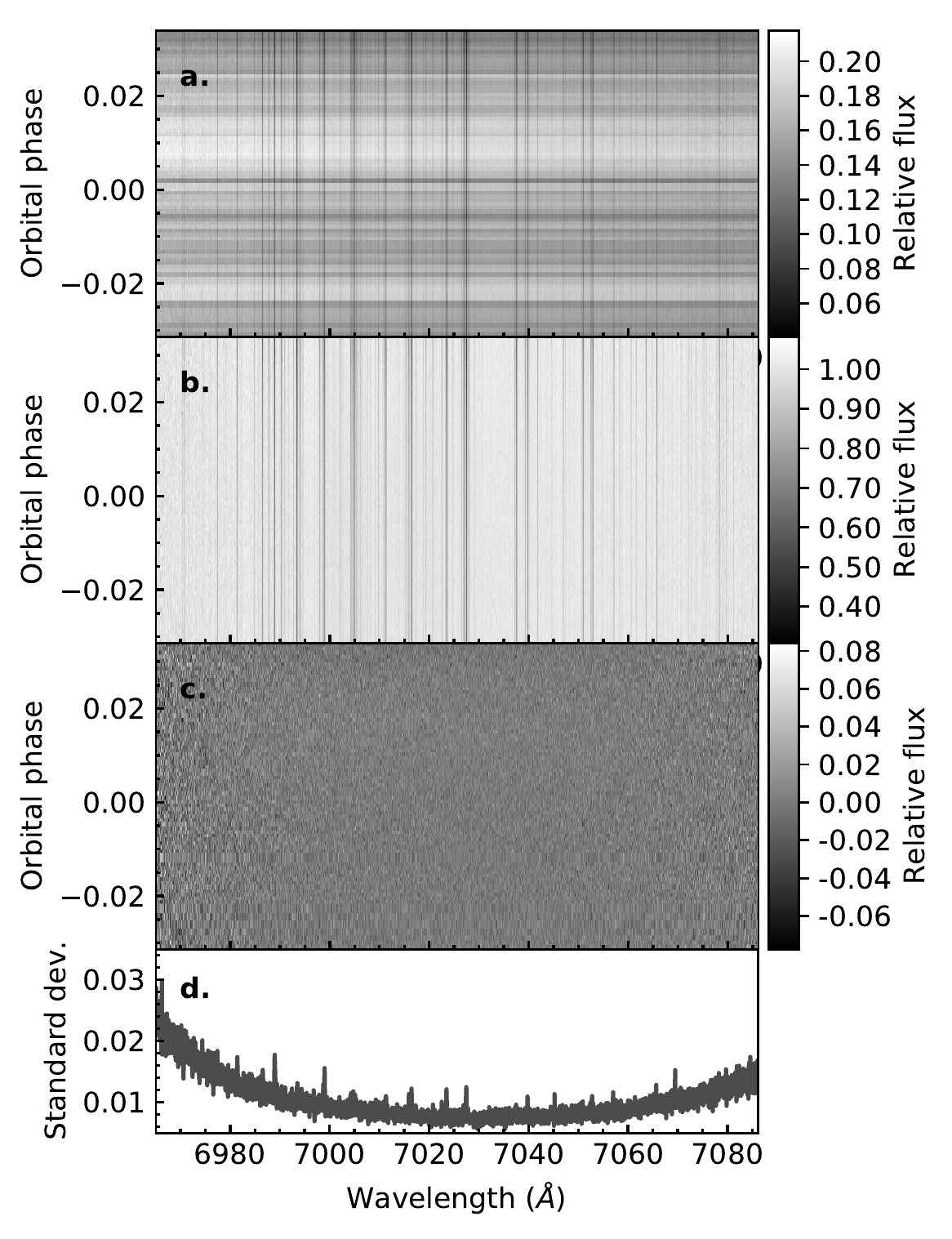}
    \caption{Example of telluric and stellar line removal using {\sc SysRem} for CARMENES data order 32. \textbf{(a.)} The stacked spectra after extraction from the .fits file. \textbf{(b.)} After correcting the blaze function and normalising to the continuum. \textbf{(c.)} The residual spectra after running {\sc SysRem} one iteration. \textbf{(d.)} The standard deviation of each wavelength bin in the residual spectra.
    \label{fig:sysrem}}
\end{figure}

To take into account the possibility that each order has different systematics, and to increase computational efficiency, we performed {\sc SysRem} order-by-order independently after removing the Doppler shadow. The error per pixel is estimated by taking the outer product of the standard deviation of each wavelength bin and each exposure before the normalisation. For each order, we run {\sc SysRem} for 10 iterations. Figure \ref{fig:sysrem} shows each step of the stellar and telluric line removal for CARMENES data order 32. 

\begin{figure}
    \centering
    \includegraphics[width=\linewidth]{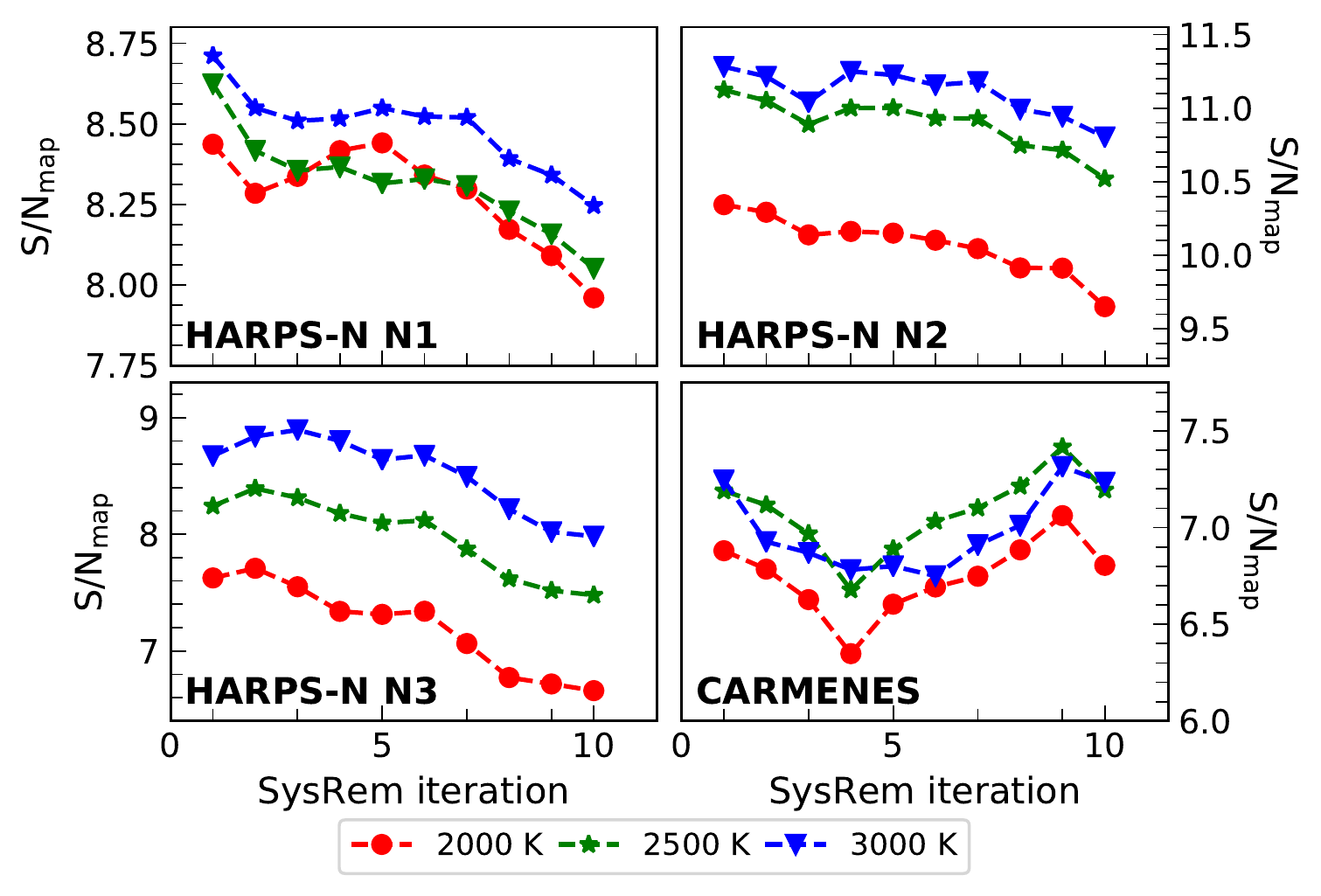}
    \caption{The S/N$_{\mathrm{map}}$ of the Fe\,{\sc i} signal in HARPSN (a, b and c) and CARMENES data-sets (d) as a function of {\sc SysRem} iterations for various temperatures of the atmospheric model. Here zero iterations corresponds to signal retrieval performed after the normalisation of each wavelength bin by its mean value before applying {\sc SysRem}. 
    \label{fig:FeIsysrem}}
\end{figure}

{\sc SysRem} iteratively removes the most dominant linear magnitude variations as a function of time. However, as the number of iterations increases, it will also remove the planetary signal. Generally, iterations can be stopped when the S/N of the detection is at a maximum before it begins to decrease as the algorithm starts to remove the planetary signal. As there could be different systematics between orders, the optimum number of iterations might differ. The optimisation could be performed by injecting weak artificial planetary signals at the expected velocity and choosing the number of iterations that give the highest S/N for each order \citep[e.g.][]{Brogi2013, Birkby2013, Birkby2017, Nugroho2017, Sanchez2019}. However, this could potentially bias the optimisation to recover the injected signal only at a specific velocity and/or for a specific model, as \citet{Cabot2019} have shown. To avoid this, we chose to use the same number of {\sc SysRem} iterations for all orders that maximises the S/N of the detected signal \citep[e.g.][]{deKok2013, Birkby2017, Turner2020}. This method would not result in the true optimised number of {\sc SysRem} iterations and might introduce additional noise from the order that has not been `completely cleaned' by {\sc SysRem} as has been pointed out by \citealt{Alonso-Floriano2019}), but it would not affect the result significantly if the S/N of the detection is strong, to begin with. As can be seen in Figure \ref{fig:atomic-ionic-spectrum}, the absorption features of Fe\,{\sc i} are found across a wide wavelength range for both HARPSN and CARMENES data-sets; therefore, in principle, the `optimum' number of {\sc SysRem} iterations is when the S/N of Fe\,{\sc i} is at maximum, as the S/N of the detection is defined by the ratio between the signal of the planet and the noise. We found that the S/N of Fe\,{\sc i} is maximised after one {\sc SysRem} iteration for HARPSN N1 and HARPSN N2 data-sets, two iterations for HARPSN N3 data-sets and nine iterations for CARMENES data-sets (see Figure \ref{fig:FeIsysrem}). The difference in the number of iterations between data-sets was expected because the telluric contamination across the wavelength range of the HARPSN data-sets is minimal, while the CARMENES data-sets cover a much redder wavelength range and hence contain many more telluric features, especially water lines. From this point, we used this number of iterations to search for all considered atomic/molecular elements in each data-set.

\section{Searching for atmospheric signals} \label{sec:planetsignal}
There are several possible methods which can be used to search for planetary atomic or molecular signatures in high-resolution spectroscopic data after removal of telluric and stellar lines. One could stack the in-transit spectra at the planetary rest frame and search for absorption features at specific wavelengths \citep[e.g.][]{Redfield2008, Snellen2008, Wood2011,Zhou2012, Wyttencbach2015,Wyttenbach2017, Yan2018,Casasayas-Barris2018, Cauley2019, Casasayas-Barris2019, Yan2019,Turner2020}. This method works well mostly for strong absorption lines (e.g. Na\,{\sc i} D, K\,{\sc i}, Ca\,{\sc ii} H$\&$K, Ca\,{\sc ii} IRT). However, for weaker absorption features with many lines in noisy data, cross-correlation analysis is superior, as it effectively sums over many lines weighted by their line strengths that are shifted at a certain velocity as a function of time, thus providing a robust detection.

\subsection{Modeling the transmission spectrum} \label{subsec:modeling}
In this paper, the focus of our analysis is finding the signature of thermal inversion agents in the transmission spectrum of KELT-20b and confirming the previous detections in \citet{Casasayas-Barris2019}. First, we calculated the cross-section of the atomic species (Fe\,{\sc i}, Fe\,{\sc ii}, Ca\,{\sc ii}, Na\,{\sc i}, Ti\,{\sc i}, Ti\,{\sc ii}, V\,{\sc i}, V\,{\sc ii}) using PYthon for Computational ATmospheric Spectroscopy (line-by-line), {\sc Py4CAtS} \citep{Schreier2019}. The line-by-line database was taken from \citet{Kurucz2018} and extracted into {\sc Py4CAtS}-supported format. Partition functions of each species were taken from \citet{Barklem2016} which were then spline-interpolated to estimate the partition function value at a certain temperature. We considered the absorption line as a Voigt profile with natural broadening, which is provided in \citet{Kurucz2018} for each species, and we consider thermal broadening only. Since we are probing the atmosphere of the planet at relatively low pressures, the effect of pressure broadening will be less significant than other broadening mechanisms. The strength of each line is calculated using Equation 1 in \citet{Sharp2007}. Instead of defining the spectral resolution at an absolute value, it can be defined as the number of grid-points per mean value of Half Width of Half Maximum (HWHM) of the absorption lines in the considered wavelength range. We set the absolute line wing cut off to 100 cm$^{-1}$ following \citet{Sharp2007} and the resolution to 5 grid points.

For molecular species, we extracted the line-by-line database from ExoMol, and consider only the main isotope species (except for TiO), in particular $^{27}$Al$^{16}$O \citep{Patrascu2015}, $^{40}$Ca$^{16}$O \citep{Yurchenko2016}, $^{56}$Fe$^{1}$H \citep{Wende2010}, $^{24}$Mg$^{1}$H \citep{Gharib2013}, $^{23}$Na$^{1}$H \citep{Rivlin2015}, $^{32}$S$^{1}$H \citep{Gorman2019}, $^{52}$V$^{16}$O \citep{McKemmish2016}. For TiO, we used the line list from \citet{Plez1998} (hereafter Plez '98), the updated version using lab measurements (Plez 2012, here after Plez '12) and from ExoMol \citep[hereafter TiO-ToTo, ][]{McKemmish2019} for the five most stable isotopes ($^{46}$Ti$^{16}$O, $^{47}$Ti$^{16}$O, $^{48}$Ti$^{16}$O, $^{49}$Ti$^{16}$O, $^{50}$Ti$^{16}$O). The line list from Plez was converted into {\sc HELIOS-K} binary format using a custom-built Python script (Grimm, priv. communication) and the partition function from TiO-ToTo. For $^{24}$Mg$^{1}$H, we used the partition function provided in \citet{Barklem2016}. We calculated the cross-section using {\sc HELIOS-K} \citep{Grimm2015} at a resolution of 0.01 cm$^{-1}$ with a full Voigt profile (technically the line-wing cut off was set to 10$^{30}$ times the Lorentz line widths). Similar to the calculation of atomic cross-sections using {\sc Py4CAtS}, we considered natural and thermal broadening only. All of the cross-sections were calculated at temperatures of 2000 K, 2500 K and 3000 K. The $T_{\mathrm{eq}}$ of the planet assuming zero Bond albedo and instantaneous re-radiation is about 2900 K.

The atmosphere of KELT-20b was modelled using the physical parameters of the planet in Table \ref{table1} (for the planetary mass we adopted the upper limit) assuming a 1D plane-parallel atmosphere divided into a hundred evenly-spaced layers in log-pressure from 10 bar to 10$^{-15}$ bar. We used {\sc FastChem}\footnote{with an update to include all additional elements similar to \citet{Hoeijmakers2019}, Kitzmann priv. communication} \citep{Stock2018} to calculate the abundance of each species (volume mixing ratio, VMR) at each layer assuming chemical equilibrium, solar metallicity and an isothermal temperature profile. For FeH, since it is not covered by {\sc FastChem}, we assumed a constant abundance of 10$^{-8}$ at all altitudes for all temperatures. The resulting abundances of the considered species can be seen in Figure 4. We included Rayleigh scattering by H$_{2}$ and bound-free continuum absorption of H$^{-}$ calculated using the formula from \citet{John1988}. Following \citet{Brown2001}, we assumed the stellar light propagates in parallel through the so-called transit chord lines, the transmission spectrum of the planet is 
\begin{equation}
    Tr (\lambda)= 1 - \left(\frac{R_{\mathrm{p}}(\lambda)}{R_{\mathrm{s}}}\right)^{2},
\end{equation}{}
where $R_{\mathrm{s}}$ is the radius of the star, and $R_{\mathrm{p}}(\lambda)$ is the radius of the planet as a function of wavelength. The radius of the planet as a function of wavelength can be calculated by
\begin{equation}
    R_{\mathrm{p}}^{2}(\lambda)= R_{\mathrm{p_{0}}}^{2} + 2\int_{R_{\mathrm{p_{0}}}}^{R_{\mathrm{p_{0}}}+H_{\mathrm{a}}} (1-e^{-\tau_{\mathrm{chord}}(\lambda,r)}) \ r dr,
\end{equation}
where $R_{\mathrm{p_{0}}}$ is the white light planet radius (taken from Table \ref{table1}), $H_{\mathrm{a}}$ is the assumed maximum height of the planet atmosphere, and $\tau_{\mathrm{chord}}(\lambda,r)$ is the integrated optical depth of the transit chord at radius $r$ and wavelength $\lambda$ \citep[e.g.][]{Brown2001}. The resulting transmission spectrum was then convolved with a Gaussian kernel to the spectral resolution of HARPSN and CARMENES and normalised to its continuum, which is determined by the combination of the Rayleigh scattering and bound-free H$^{-}$ continuum. The result of the normalisation is the negative of the transmission spectrum ($\Delta F$): see Figures \ref{fig:atomic-ionic-spectrum} and \ref{fig:molecular-spectrum}. The spectral transmission models used in this analysis may be provided upon request.

\begin{figure*}
    \subfigure{\centering\includegraphics[width=0.8\linewidth]{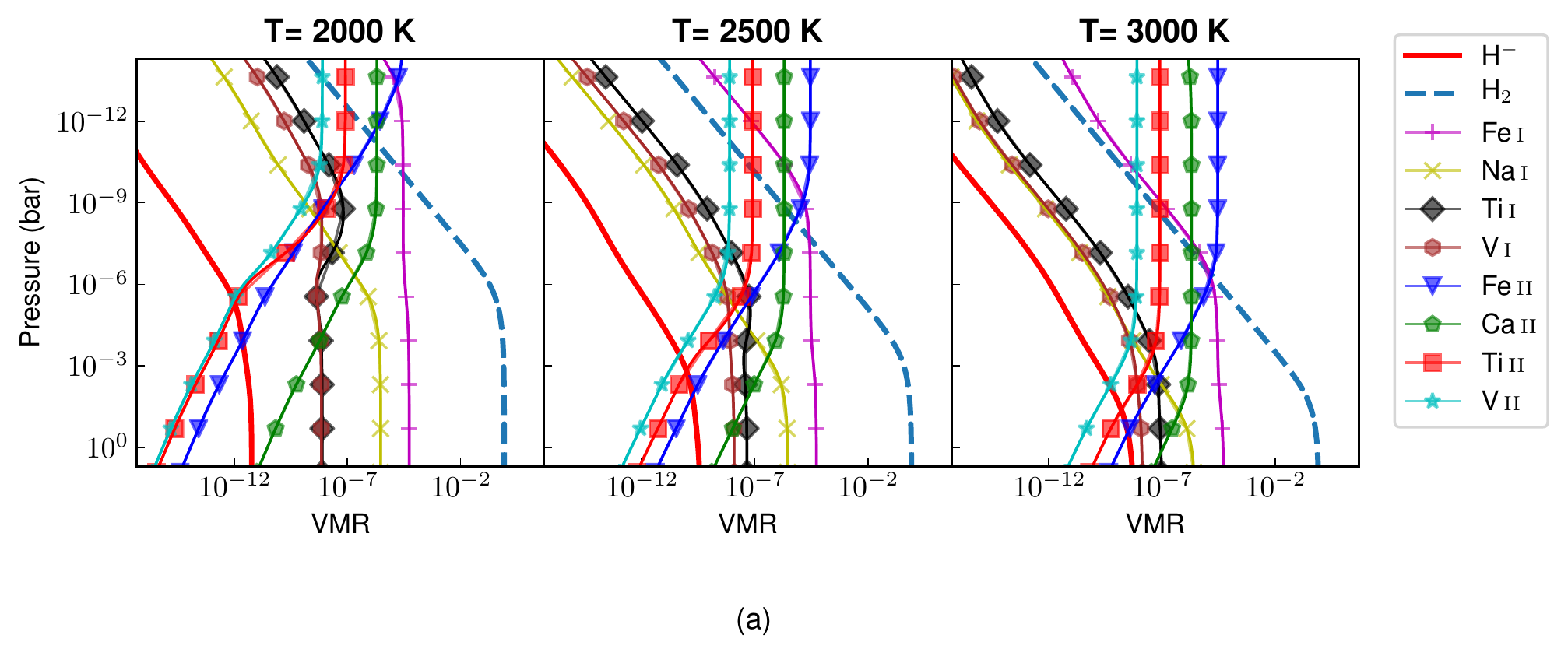}\label{fig:atomic-ionic-fastchem}}

    \subfigure{\centering\includegraphics[width=0.8\linewidth]{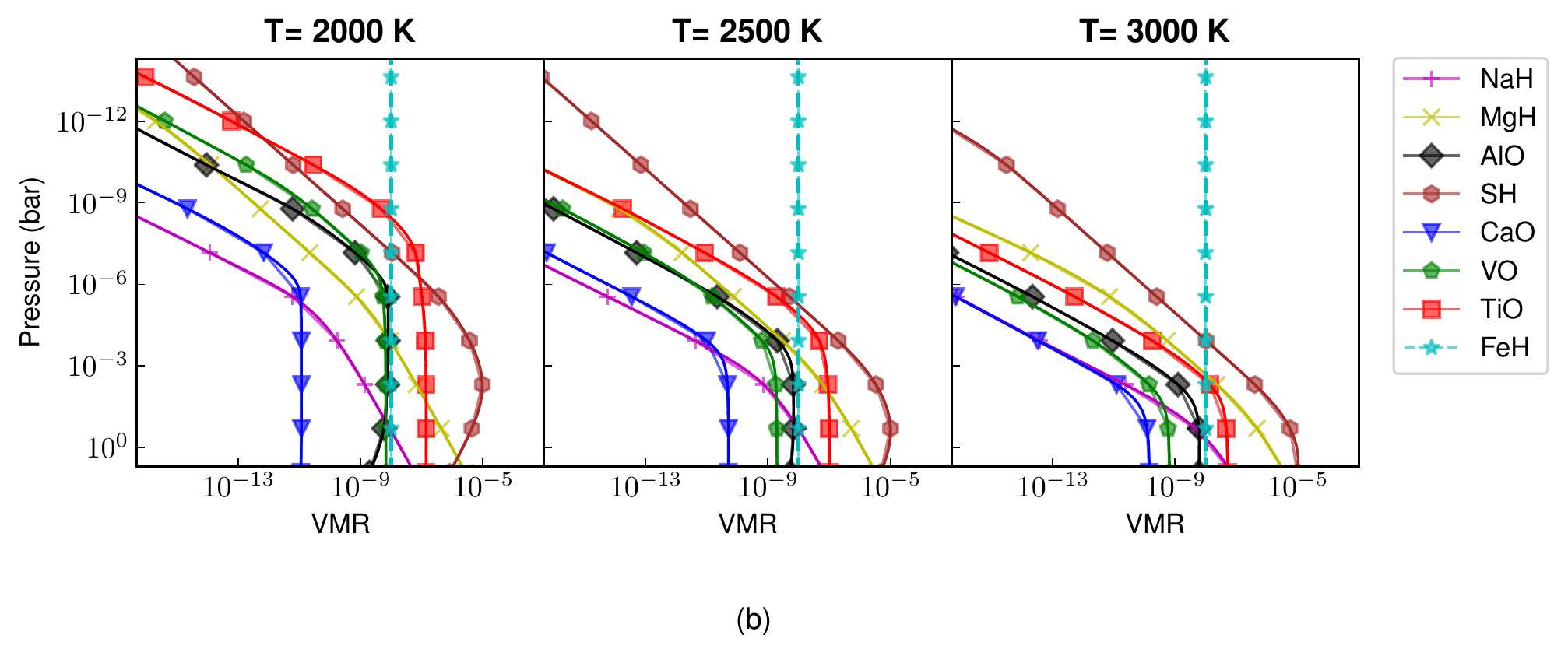}\label{fig:molecular-fastchem}}
    \caption{The abundances of atomic/ionic (a) and molecular (b) species at T of 2000 K, 2500 K and 3000 K calculated using {\sc FastChem}, with the exception of FeH which was fixed at 10$^{-8}$ at all altitudes for all temperatures.}
\end{figure*} 

\begin{figure*}
    \subfigure{\centering\includegraphics[width=.47\linewidth]{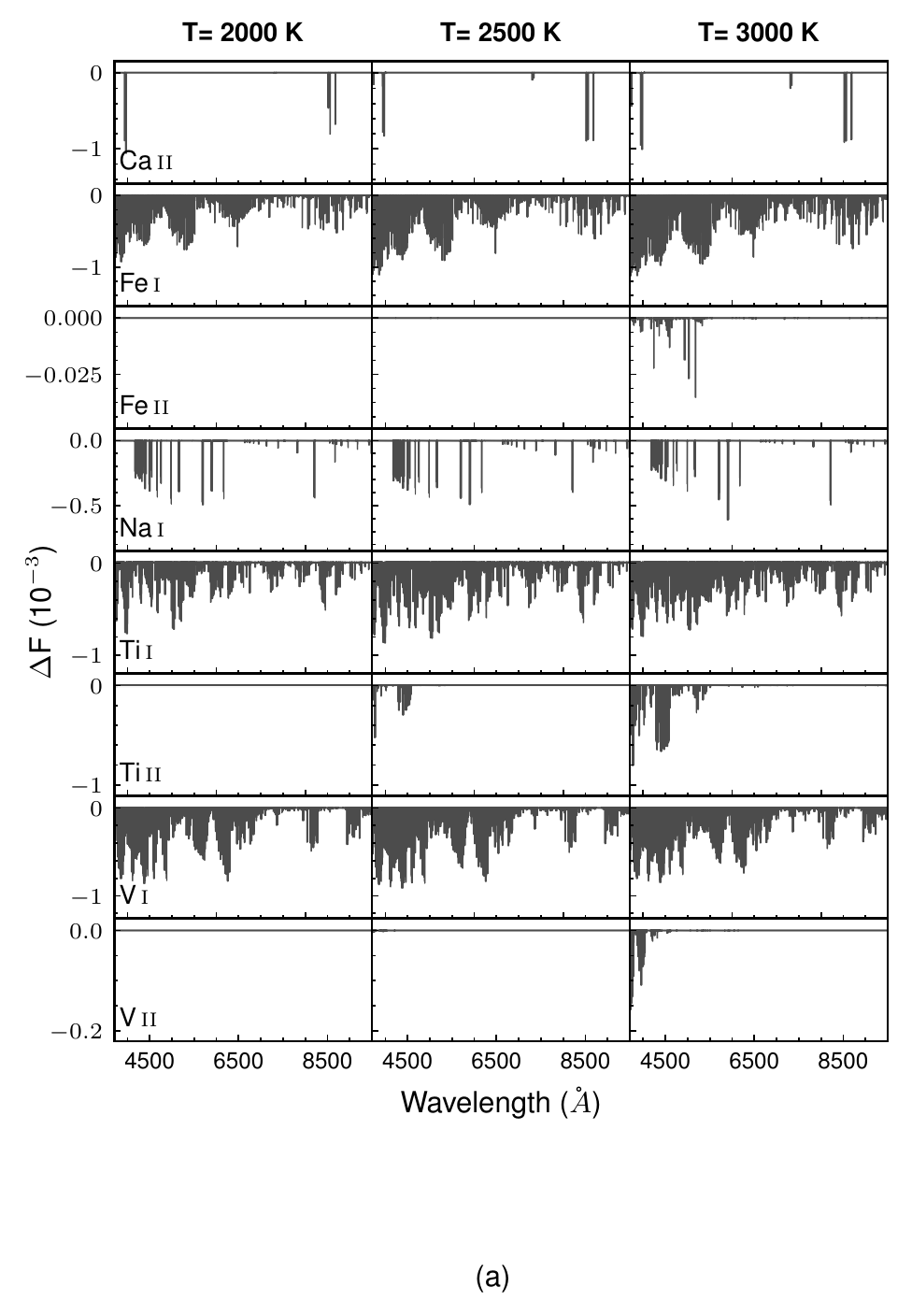}\label{fig:atomic-ionic-spectrum}}
    \subfigure{\centering\includegraphics[width=.46\linewidth]{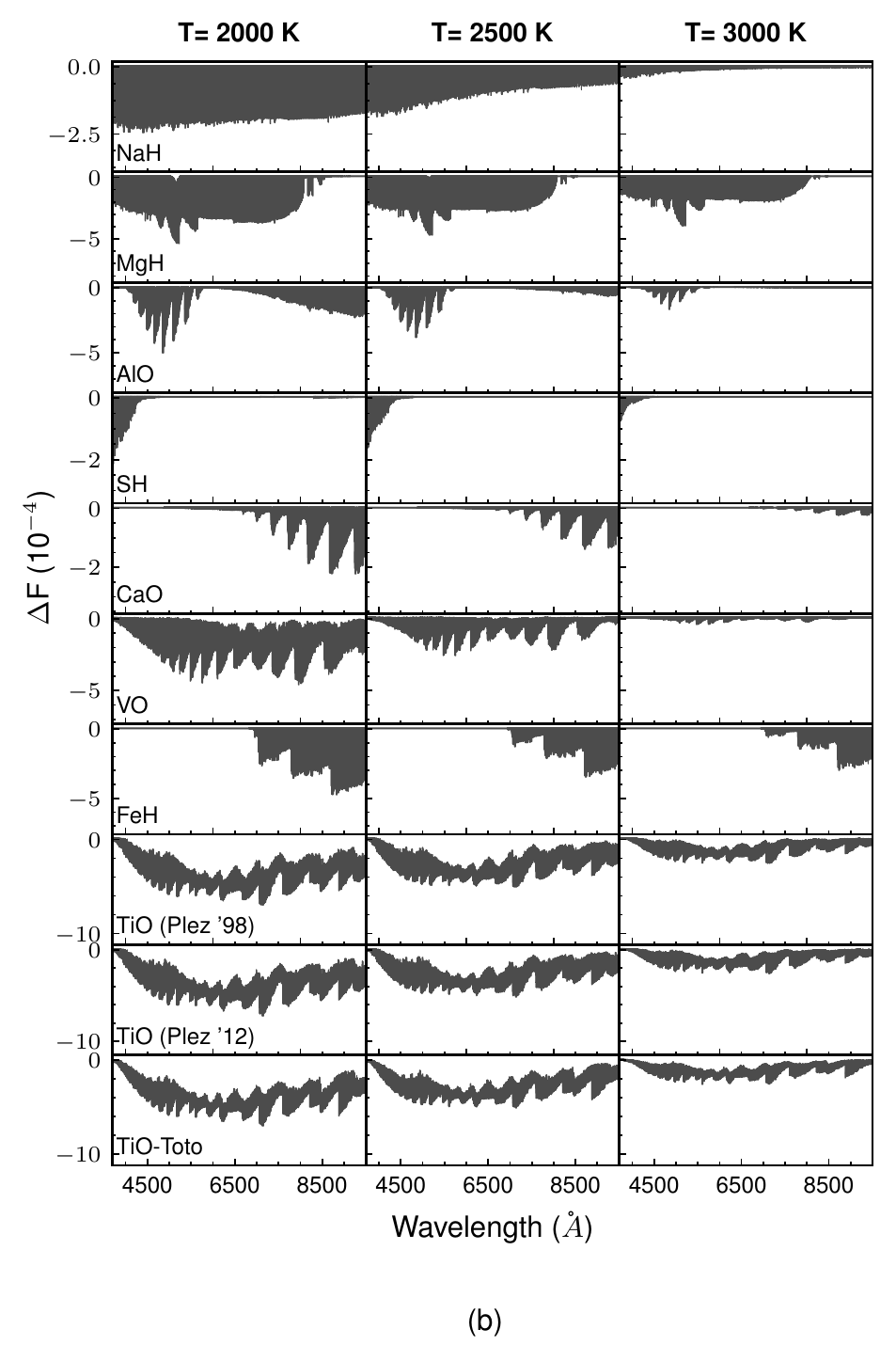}\label{fig:molecular-spectrum}}
    \caption{The normalised transmission spectrum model for atomic (a) and molecular (b) species in the wavelength range of HARPSN and CARMENES at T of 2000 K, 2500 K and 3000 K}
\end{figure*}

\subsection{Systemic velocity of KELT-20/MASCARA-2} \label{subsec:systemic}
There are two constraints on the systemic velocity in the literature, which are inconsistent with each other (see Table \ref{table1}). The difference of $\approx$2\,km s$^{-1}$ will not affect the robustness of the detection: however, it still might affect the interpretation of the detected signal, e.g. in detecting a global blue-shift caused by day-night winds. Thus, we measured the systemic velocity of KELT-20 by cross-correlating N1 and CARMENES data with the stellar spectrum model. The cross-correlation function (CCF) is defined as 
\begin{equation}
    \mathrm{CCF}= \sum \frac{f_{i}m_{i}}{\sigma_{i}^{2}},
\end{equation}
where $f_{i}$ is the data, $m_{i}$ is the Doppler-shifted spectrum model, and $\sigma_{i}$ is the error per pixel $i$.

\begin{figure*}
\subfigure{\centering\includegraphics[width=.48\linewidth]{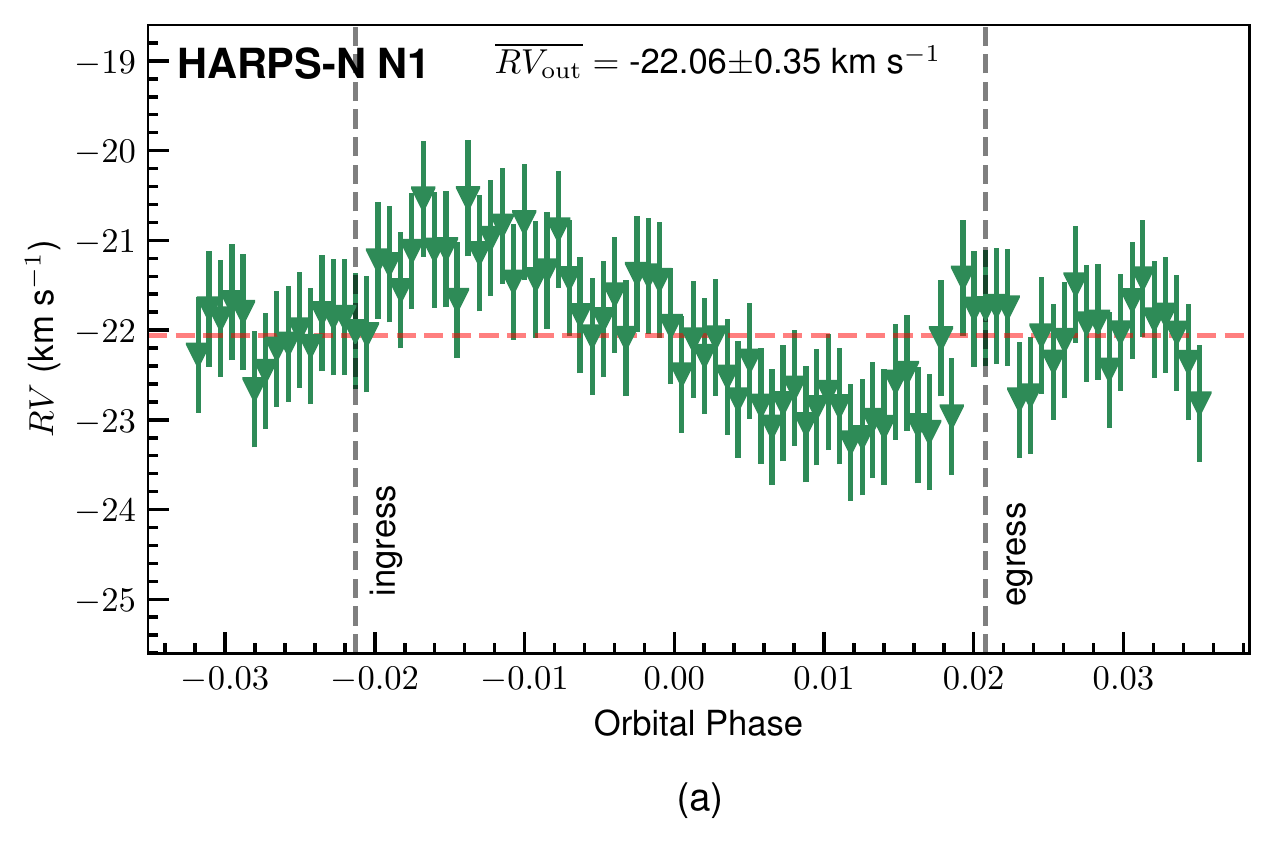}}
\subfigure{\centering\includegraphics[width=.48\linewidth]{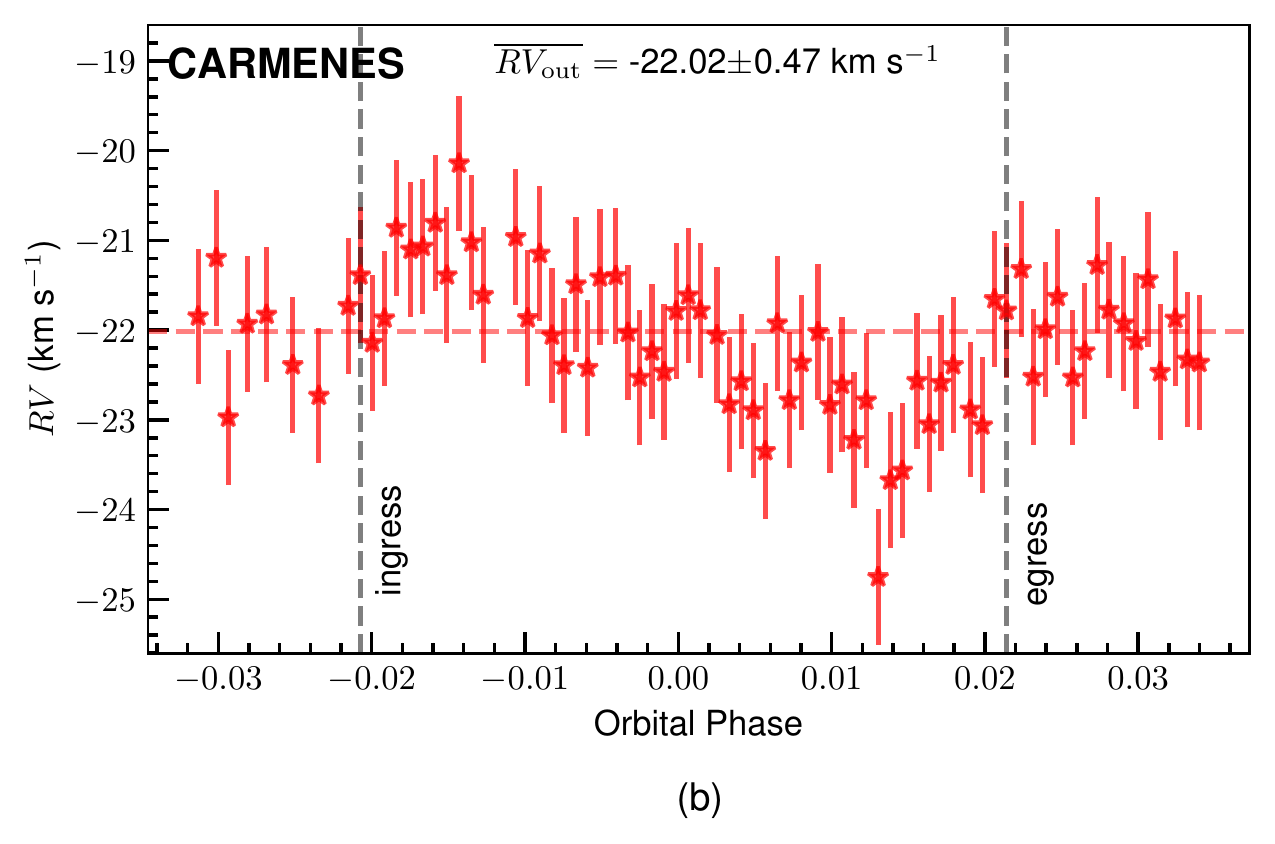}}
\caption{The systemic velocity of KELT-20/MASCARA-2 measured using HARPSN (N1) data (a) and CARMENES data (b). The black dashed lines mark the beginning of the ingress and the ending of the egress, while the red dashed lines mark the mean RV value for each data-sets. The mean RV value is calculated by only considering the RV outside the transit to neglect the R-M effect on the observed RV. \label{fig:systemic-rv}}
\end{figure*}

Figure \ref{fig:systemic-rv} shows the RV of the star in the heliocentric rest frame. The systemic velocity and the error-bars are calculated by taking the mean value and the standard deviation of the out-of-transit RV, respectively. The measurements from both data-sets agreed with each other: however, these values differ by at least 1-$\sigma$ from the available values in the literature. Note that we do not correct the measured RV with the reflex motion from the host star due to the planet since the value is insignificant during the observation ($\approx$0.07\,km s$^{-1}$). It is reasonable, however, to use these measured values for further analysis since they are measured from the same data-sets: they can, therefore, act as a reference systemic velocity to constrain any velocity deviation.

\subsection{Cross-correlation with the transmission spectrum model} \label{subsec:crossc}
After the telluric and stellar lines were removed, the residuals were then cross-correlated with the Doppler-shifted transmission spectrum model order-by-order from $-$984\,km s$^{-1}$ to $+$840\,km s$^{-1}$ in 1\,km s$^{-1}$ steps. During the removal of the telluric and stellar lines, any broad features in the planetary spectrum will be removed as well: therefore, we applied a high-pass filter to the transmission model to remove any broad features, especially for the Na\,{\sc i}, Ca\,{\sc ii}, VO and TiO spectrum models. 

The CCF of each exposure was then stacked into a matrix with the velocity lag as the column and frame number as the row. We then summed the CCF of the ``good-orders" to calculate the total CCF. The spectral order is included into the ``good-order" list if the strongest line strength in the transmission spectrum model within the specific order is larger than 0.1 per cent of the strongest line strength in the transmission spectrum model within the whole wavelength range of each data-sets. We also excluded the orders that were most heavily contaminated with telluric lines in both HARPSN (order 69, oxygen lines) and CARMENES data (order 39, oxygen lines). For Na\,{\sc i}, we only considered the order that contains the Na\,{\sc i} doublet (Na\,{\sc i} D) at around 5900 \AA, while for Ca II, we only considered the order that contains Ca\,{\sc ii} H$\&$K for HARPSN data-sets and Ca\,{\sc ii} IRT for CARMENES data-sets. For TiO, the ``good-orders" were also determined by the accuracy of the line-list. For the transmission spectrum model using Plez '98, we only consider wavelengths longer than 6300 \AA \ \citep{Nugroho2017}, and for Plez '12 and TiO-ToTo, we refer to Figure 15 in \citet{McKemmish2019}.

To search for the planet signal and possible spurious signals, we integrated the in-transit CCF at the rest frame of the planet assuming orbital velocities ($K_{\mathrm{p}}$) of $-$300\,km s$^{-1}$ to $+$300\,km s$^{-1}$ and delta velocities ($\Delta V$) of $-$500\,km s$^{-1}$ to $+$500\,km s$^{-1}$ in 0.5\,km s$^{-1}$ steps. For the integration, the CCF was weighted by the transit light-curve model of KELT-20b calculated using BAsic Transit Model cAlculatioN \citep[{\sc BATMAN},][]{Kreidberg2015} assuming linear limb darkening. The limb darkening coefficient was taken from the V band R-M linear limb darkening coefficient in \citet{Lund_2017}. Assuming a circular orbit, the planetary radial velocity at a given time ($\mathrm{RV}_{\mathrm{p}}$ ($t$)) is defined as
\begin{equation}
\mathrm{RV}_{\mathrm{p}} (t)= K_{\mathrm{p}} \sin (2\pi\phi(t)) + v_{\mathrm{sys}} + v_{\mathrm{bary}}+\Delta V
\end{equation}
where $\phi(t)$ is the orbital phase of the planet at $t$ time calculated using the parameter in Table \ref{table1}, $v_{\mathrm{sys}}$ is the systemic velocity taken from the measured value in Section \ref{subsec:systemic} for each data-set, and $v_{\mathrm{bary}}$ is the barycentric correction taken from the header of the data, and $\Delta V$ is the velocity shift (e.g. due to day-night wind). The result is the integrated CCF at various combination of $K_{\mathrm{p}}$ and $\Delta V$ values ($K_{\mathrm{p}}-\Delta V$ map). The $K_{\mathrm{p}}-\Delta V$ map of all HARPSN data-sets was then summed up to obtain the total $K_{\mathrm{p}}-\Delta V$ map. The S/N map was calculated by taking the standard deviation of the $K_{\mathrm{p}}-\Delta V$ map -- avoiding the planetary signal by masking the CCF value within $K_{\mathrm{p}}$ of 0-300\,km s$^{-1}$ and $\Delta V$ of $\pm$50\,km s$^{-1}$ -- and divided out from the map. Note that if the $K_{\mathrm{p}}-\Delta V$ map covers too narrow a velocity range, then the S/N of the possible detected signal using this method might not represent the true S/N. 

We also generated a likelihood map for each of the detected species by using the new likelihood-based approach outlined by \citet[][see also \citealt{Brogi2019} for a related approach]{gibson2020}. The log-likelihood ($\ln \mathcal{L}$) is defined as:
\begin{equation}\label{eq:lnlikelihood}
    \ln \mathcal{L}= -\frac{N_{\mathrm{eff}}}{2} \ln \left[\frac{1}{N_{\mathrm{eff}}} \left( \sum\frac{f_{i}^{2}}{\sigma_{i}^{2}} + \alpha^{2}\sum \frac{m_{i}^{2}}{\sigma_{i}^{2}} -2\alpha \sum \frac{f_{i}m_{i}}{\sigma_{i}^{2}}\right) \right],
\end{equation}
where $N_{\mathrm{eff}}$ is the number of pixels used weighted by the transit light curve model, and $\alpha$ is a set of scale factors. Note that the last term in equation \ref{eq:lnlikelihood} is the (scaled) CCF, which can also be represented by the $K_{\mathrm{p}}-\Delta V$ map. The result is a 3-dimensional data cube with $K_{\mathrm{p}}$, $\Delta V$ and $\alpha$ as the axes. The likelihood was then obtained by subtracting the maximum value from the data-cube (effectively normalising the likelihood) before computing the exponential.
To constrain $K_{\mathrm{p}}$, $\Delta V$, and $\alpha$, we took a slice through the maximum value of the data cube and fit the conditioned likelihood with a Gaussian function. The error was estimated by the standard deviation of the best-fit Gaussian function for each value. The parameter of $\alpha$ tells us how well the average line-contrast of the spectrum model matches the average line-contrast of the observed planetary spectrum. An $\alpha$ of 1 means the average line contrast of the observed planetary spectrum is perfectly represented by the spectrum model, while an $\alpha$ of zero means that spectrum model is perfectly matched with a zero value: or, in other words, no detection. Therefore the significance of the detection can be calculated by dividing the median value of the conditioned likelihood of $\alpha$ by its standard deviation.

\section{Result and discussion}\label{sec:resanddis}
\subsection{Detection of neutral iron and other atomic species}\label{subsec:FeIdet}

\begin{figure*}
    \subfigure{\centering\includegraphics[width=0.49\linewidth]{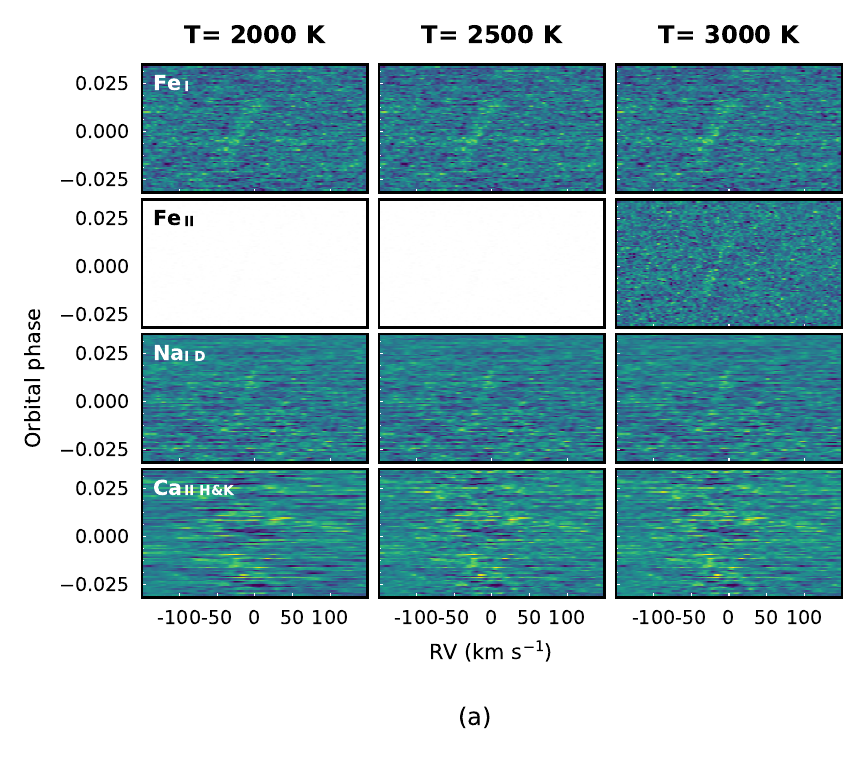}\label{fig:ccmat-resultA}}
    \subfigure{\centering\includegraphics[width=0.49\linewidth]{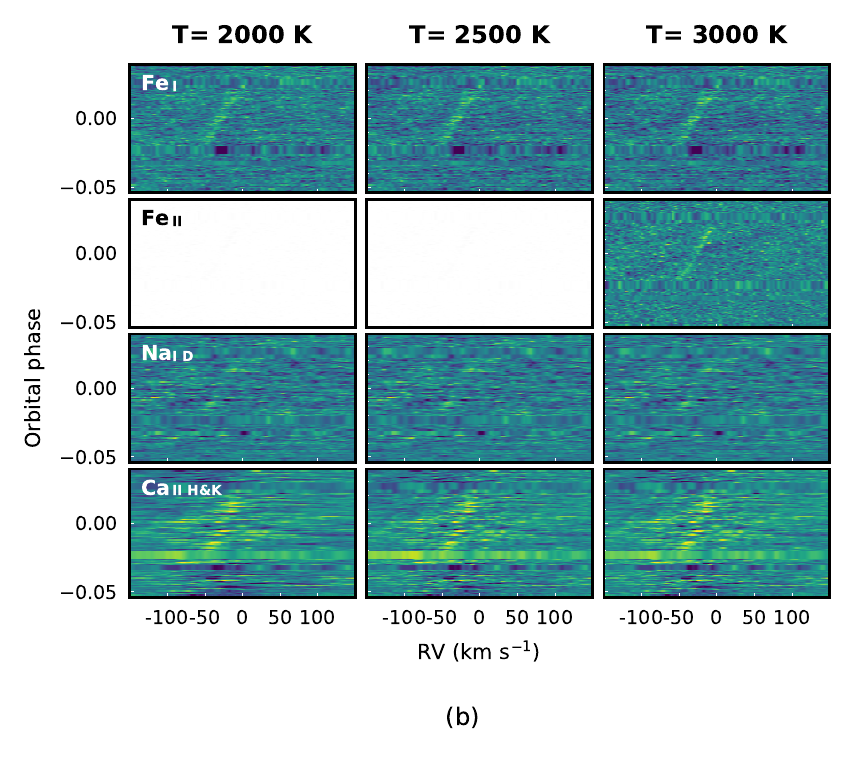}\label{fig:ccmat-resultB}}
    
    \subfigure{\centering\includegraphics[width=0.49\linewidth]{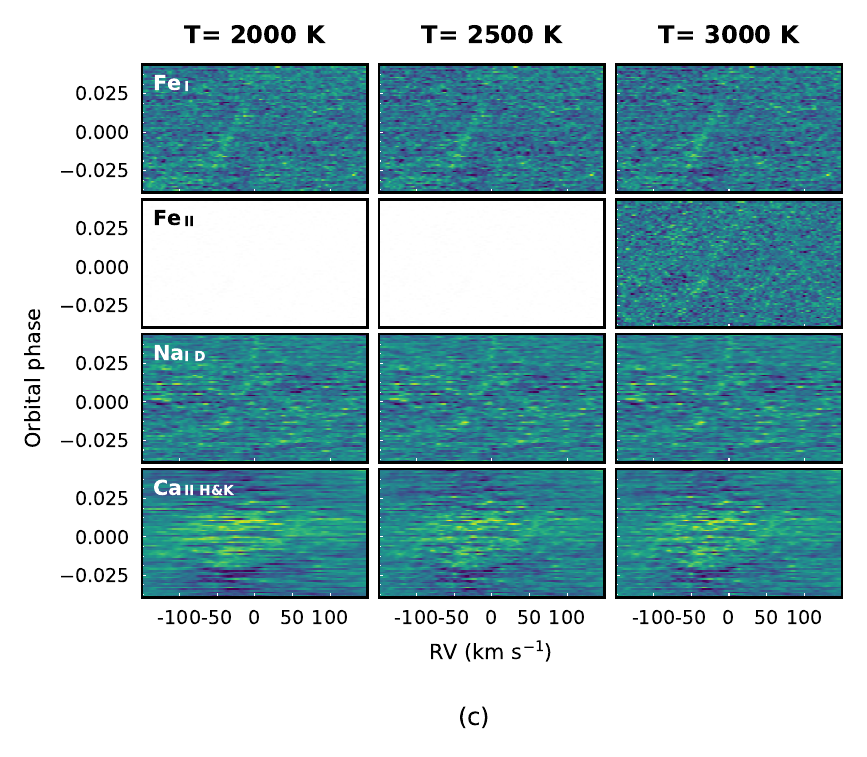}\label{fig:ccmat-resultC}}
    \subfigure{\centering\includegraphics[width=0.49\linewidth]{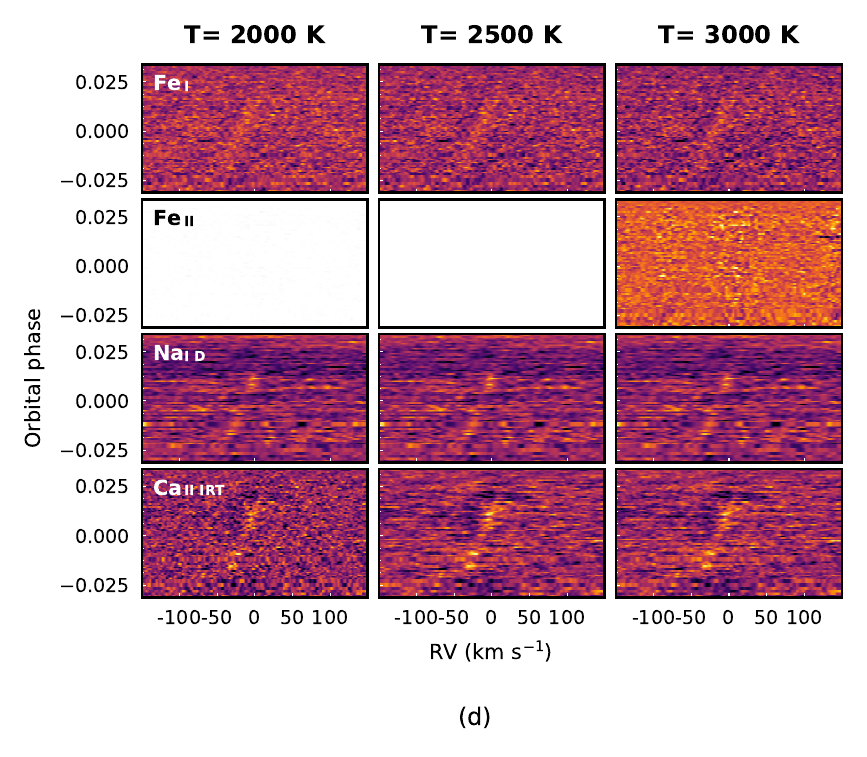}\label{fig:ccmat-resultD}}
    \caption{The Doppler-shadow-removed cross-correlation map of the detected atomic species (Fe\,{\sc ii}, Fe\,{\sc ii}, Ca\,{\sc ii} H$\&$K, Ca\,{\sc ii} IRT, and Na\,{\sc i} D) using the residuals after one {\sc SysRem} iteration for  N1 (a), N2 (b), N3 (c) and CARMENES (d). The planetary signal can be seen as a bright streak from an orbital phase of -0.02 to 0.02 and from an RV of $\approx$-50\,km s$^{-1}$ to 0\,km s$^{-1}$. Each column in each panel represents the cross-correlation result using a transmission spectrum model with different temperature. The dark green band in (b) around orbital phase of -0.03, -0.02 and 0.03 represents either a gap in the observation or masked frames due to low S/N.}
    \label{fig:ccmat-result}
\end{figure*}

\begin{figure*}
    \subfigure{\centering\includegraphics[width=0.499\linewidth]{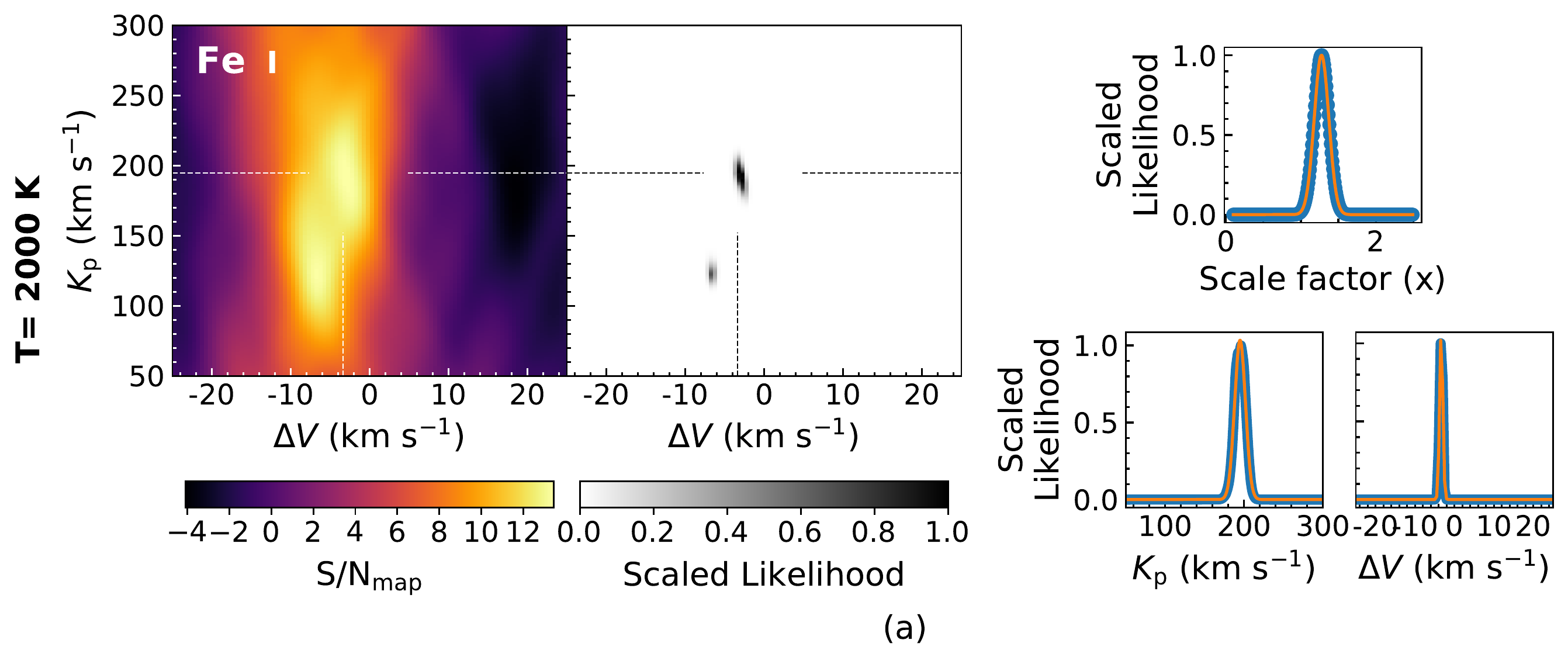}\label{fig:kpdv-FeIresultA}}
    \subfigure{\centering\includegraphics[width=0.495\linewidth]{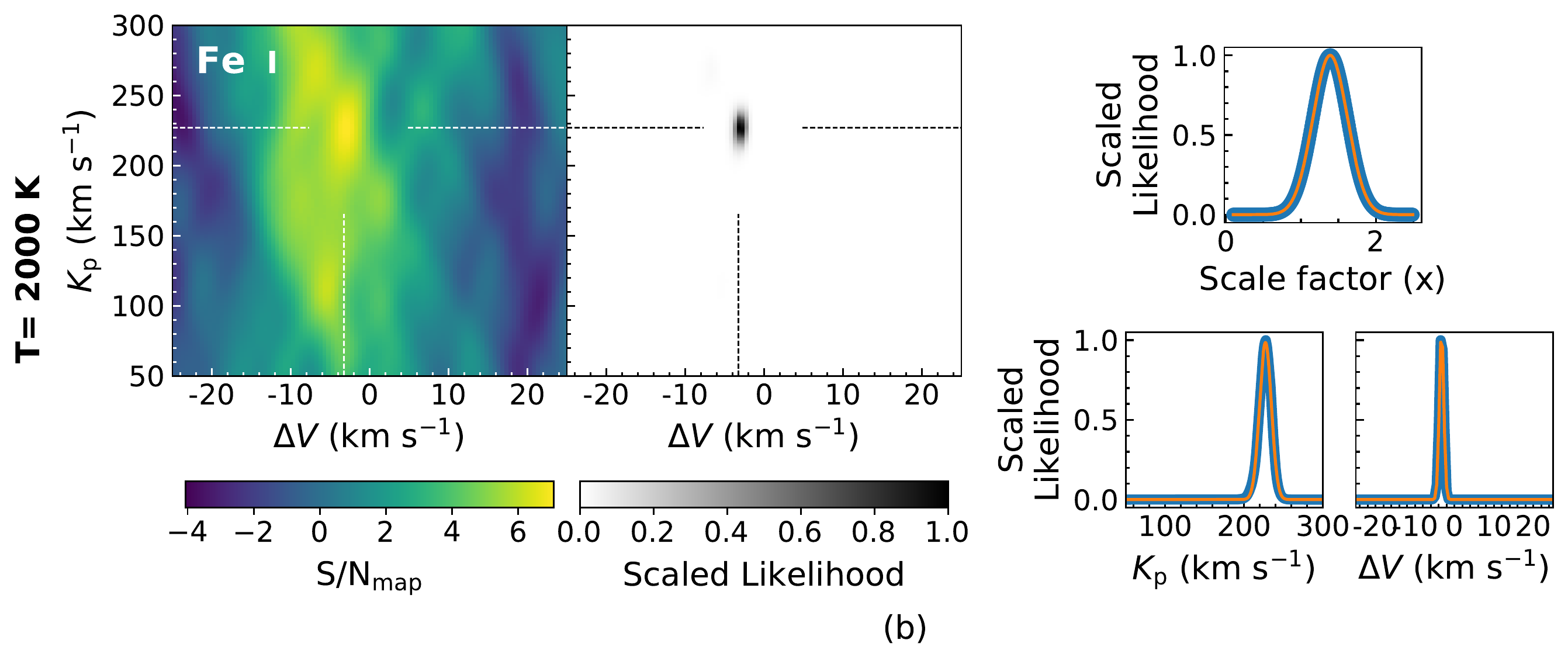}\label{fig:kpdv-FeIresultB}}
    
    \subfigure{\centering\includegraphics[width=0.499\linewidth]{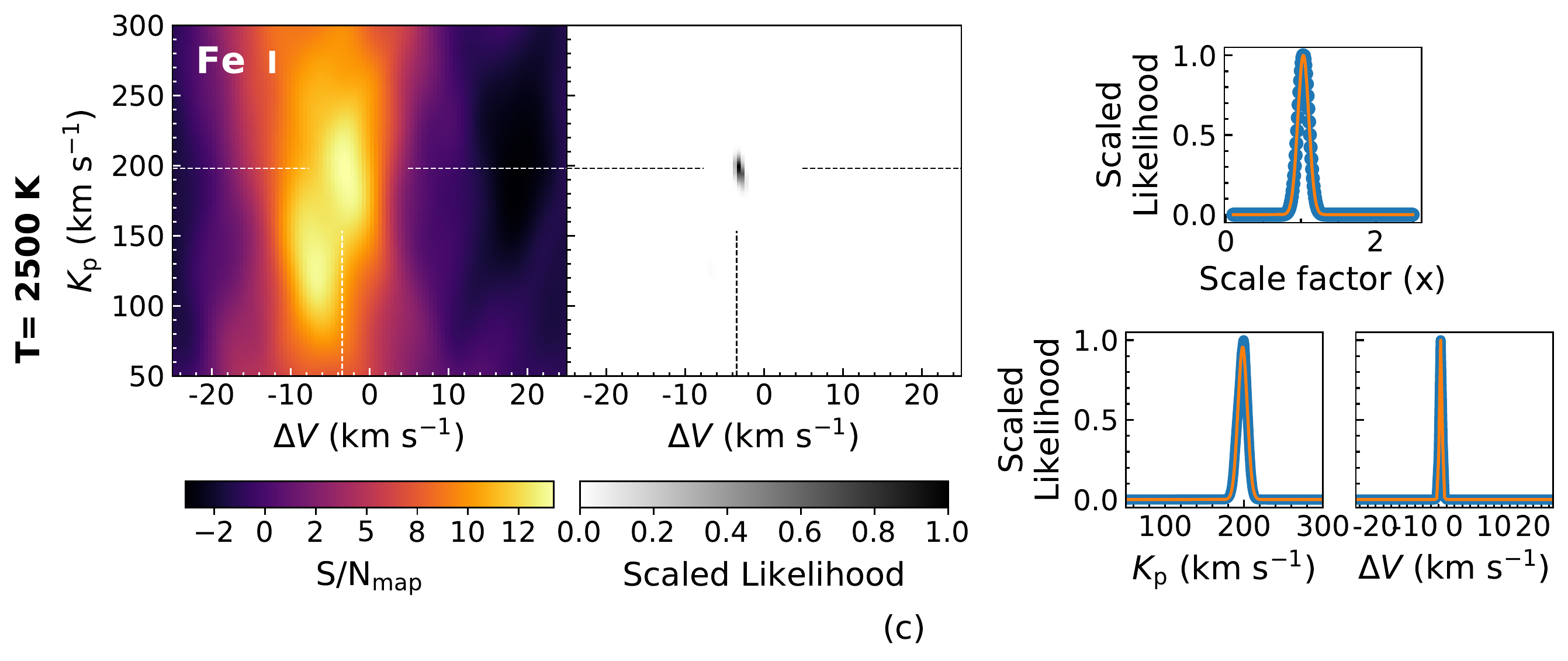}\label{fig:kpdv-FeIresultC}}
    \subfigure{\centering\includegraphics[width=0.495\linewidth]{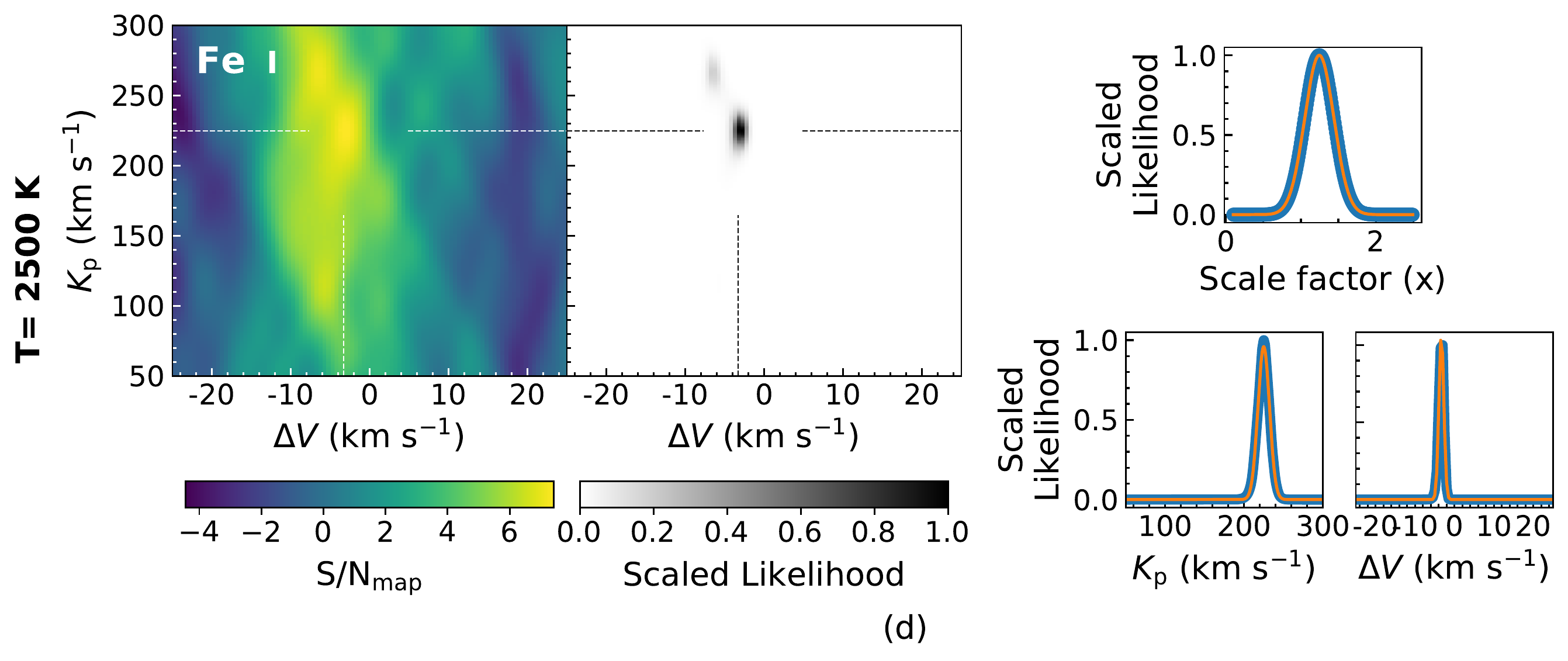}\label{fig:kpdv-FeIresultD}}
    
    \subfigure{\centering\includegraphics[width=0.499\linewidth]{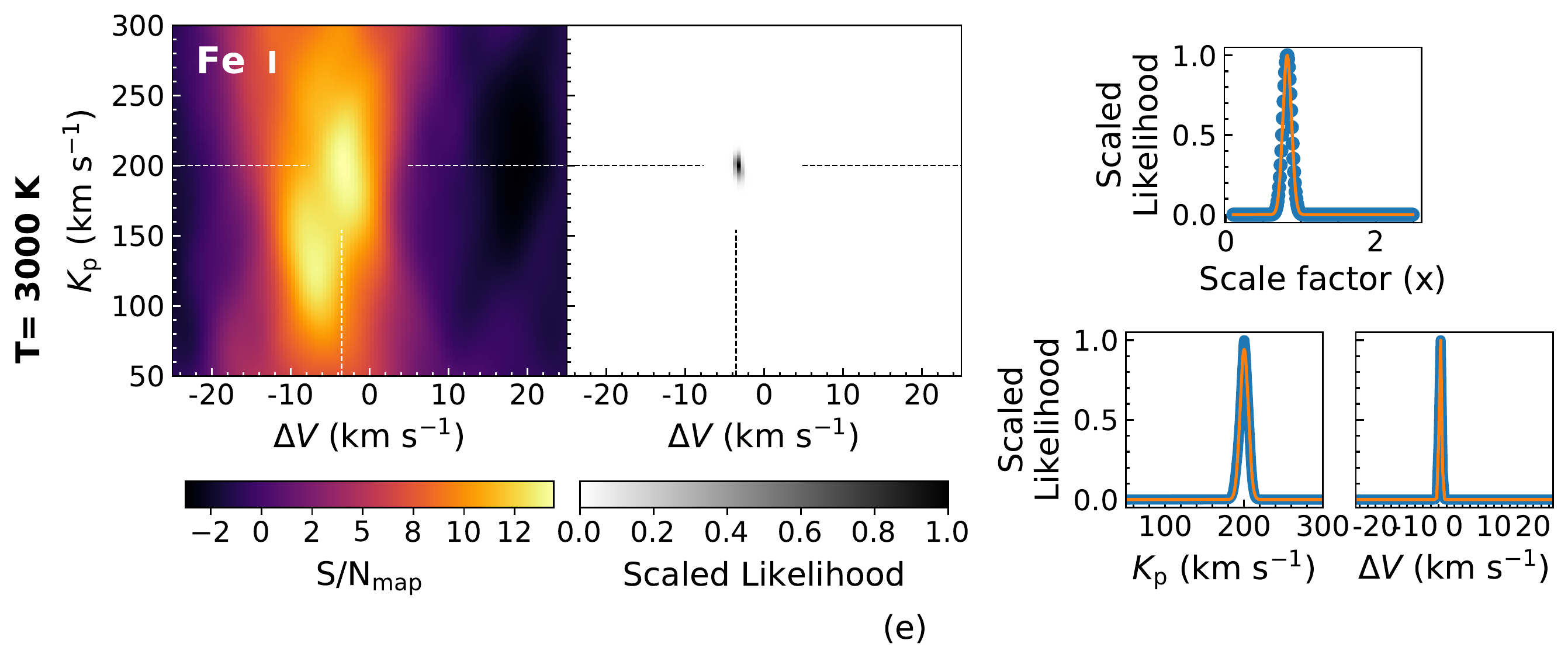}\label{fig:kpdv-FeIresultE}}
    \subfigure{\centering\includegraphics[width=0.495\linewidth]{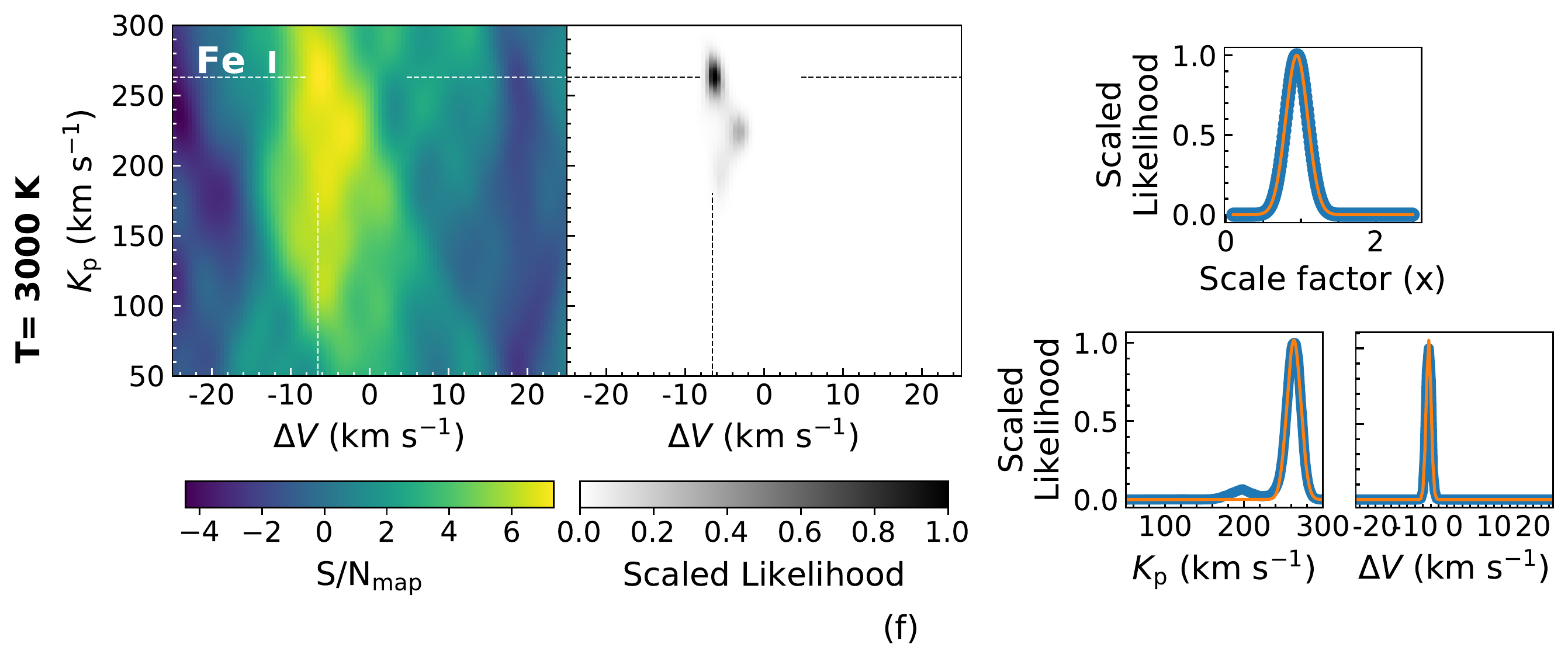}\label{fig:kpdv-FeIresultF}}  
    
    \caption{The orbital velocity ($K_{\mathrm{p}}$)-velocity offset($\Delta V$) map of Fe\,{\sc i} for the combined HARPSN-N (\textit{left column}) and CARMENES (\textit{right column}) data-sets. Each row shows the result of a specific atmospheric temperature. Each panel consists of an S/N map (\textit{left subpanel}), a likelihood map (\textit{middle subpanel}), and the conditioned distributions (\textit{right subpanel}). The dotted line shows the highest peak in the map. The colour bars in the S/N map and likelihood map represents the S/N$_{\mathrm{map}}$ and the scaled likelihood respectively. The conditioned distributions subpanel consists of three distributions: the scale factor (\textit{upper}), the $K_{\mathrm{p}}$ (\textit{lower left}), and the $\Delta V$ (\textit{lower right}). The blue dot is the data, while the orange line is the best-fit Gaussian function.}
    \label{fig:kpdv-FeIresult}
\end{figure*}

\begin{figure*}
    \subfigure{\centering\includegraphics[width=0.499\linewidth]{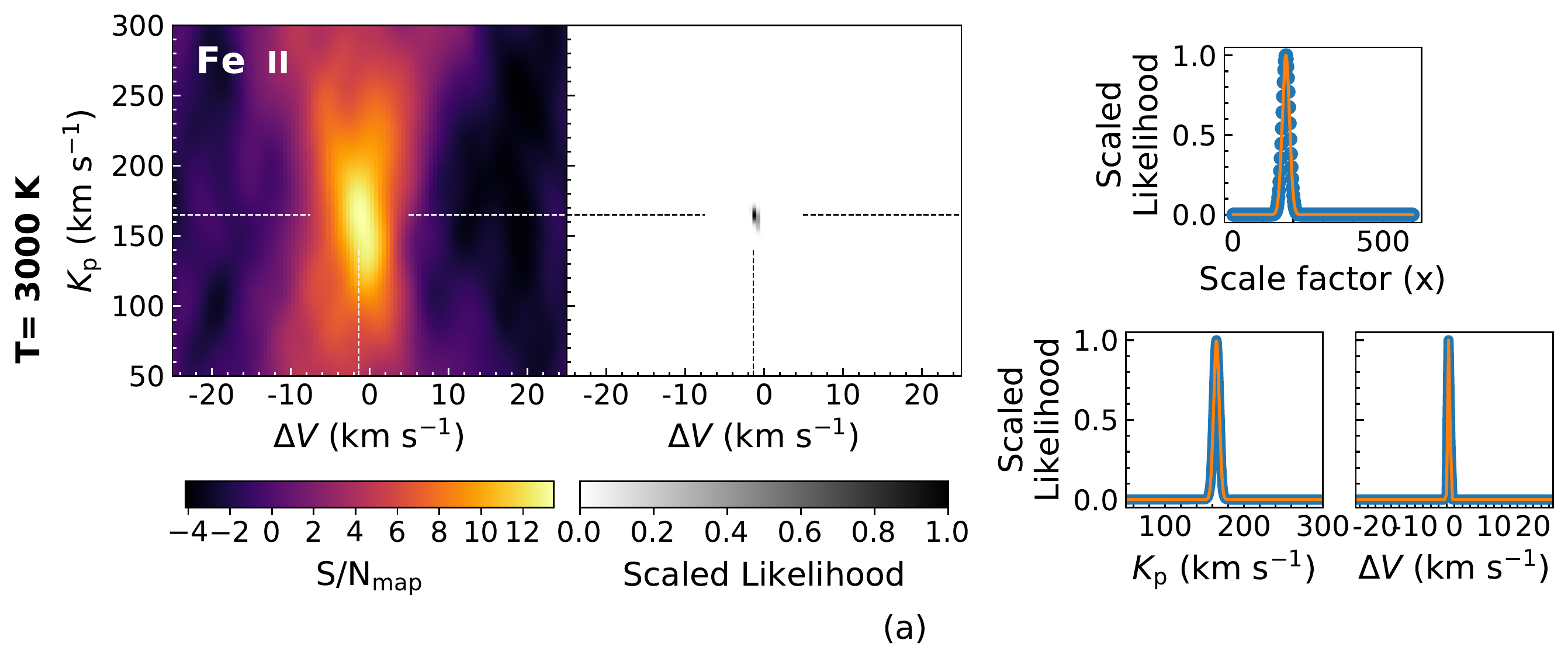}\label{fig:kpdv-FeIIresultA}}
    \subfigure{\centering\includegraphics[width=0.495\linewidth]{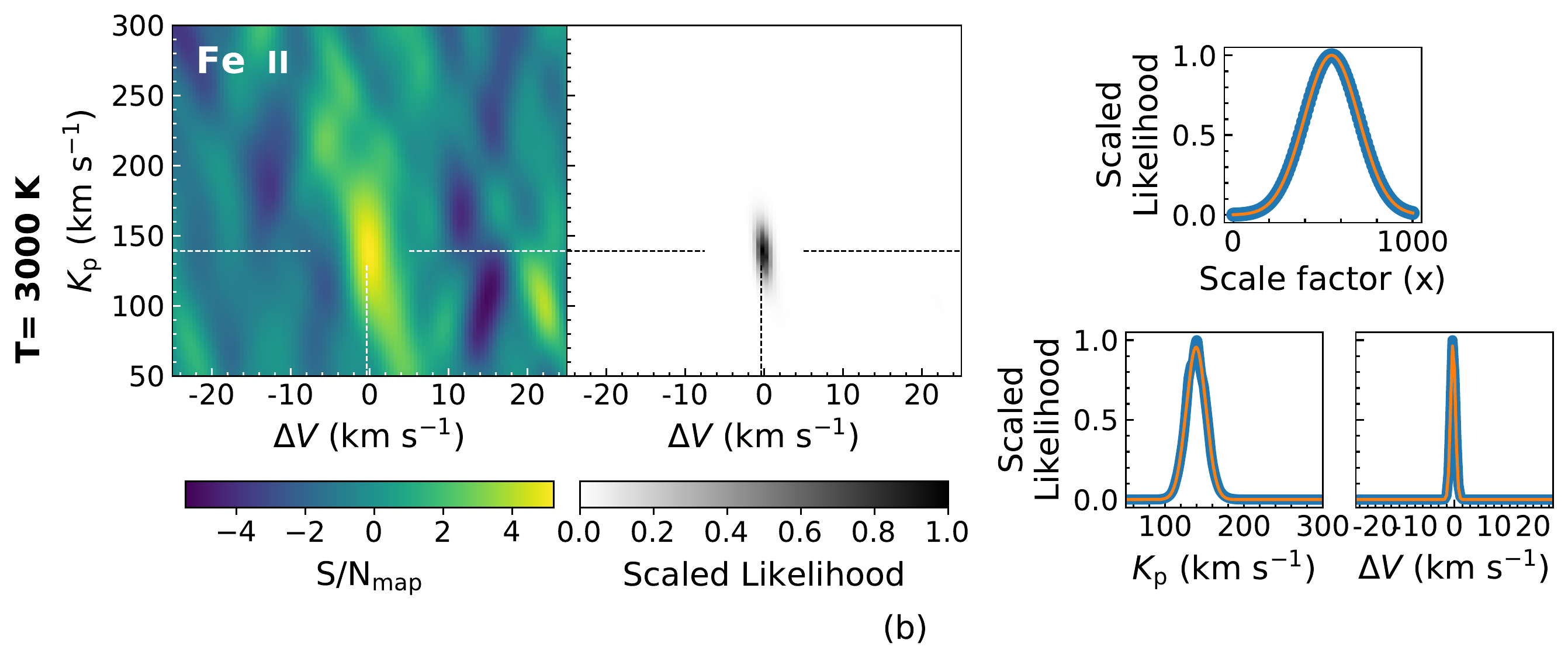}\label{fig:kpdv-FeIIresultB}}  
    
    \caption{Similar to Figure \ref{fig:kpdv-FeIresult} but for Fe\,{\sc ii}.}
    \label{fig:kpdv-FeIIresult}
\end{figure*}

\begin{figure*}
    \subfigure{\centering\includegraphics[width=0.499\linewidth]{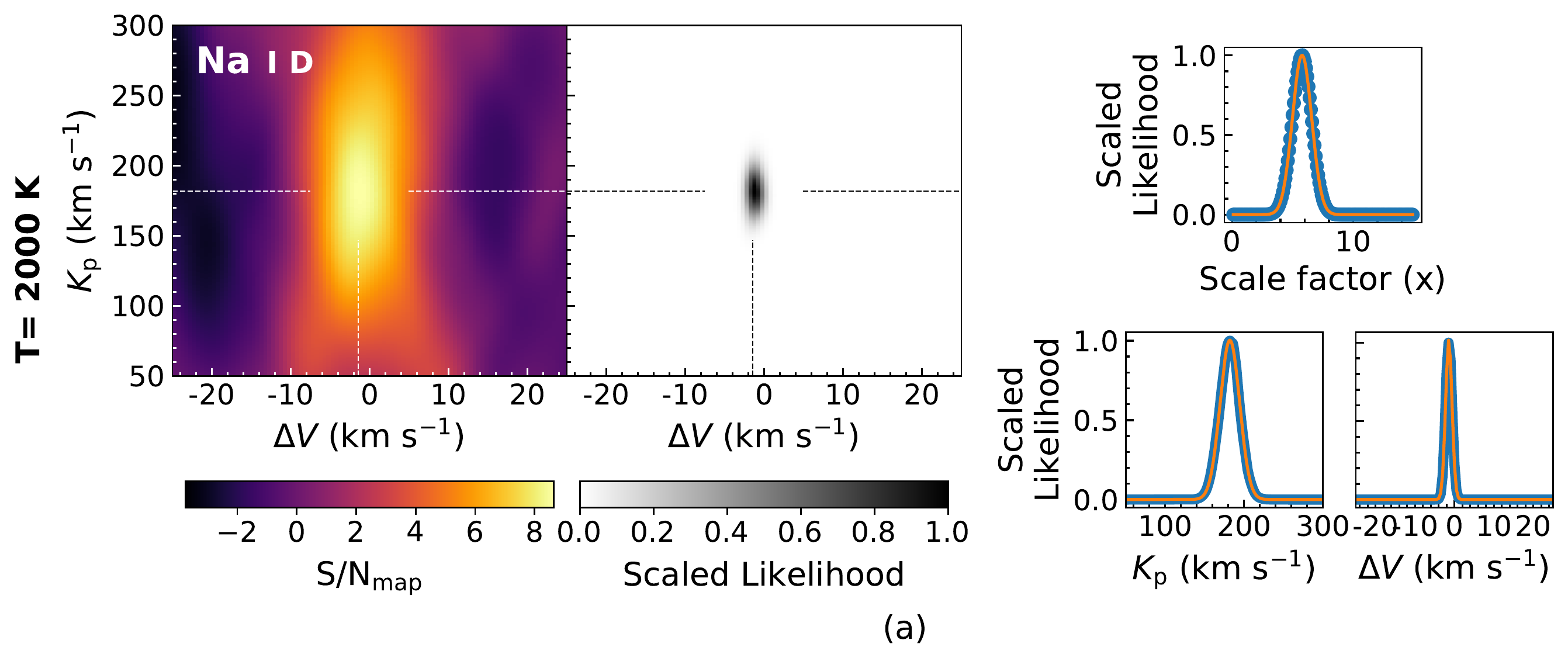}\label{fig:kpdv-NaIresultA}}
    \subfigure{\centering\includegraphics[width=0.495\linewidth]{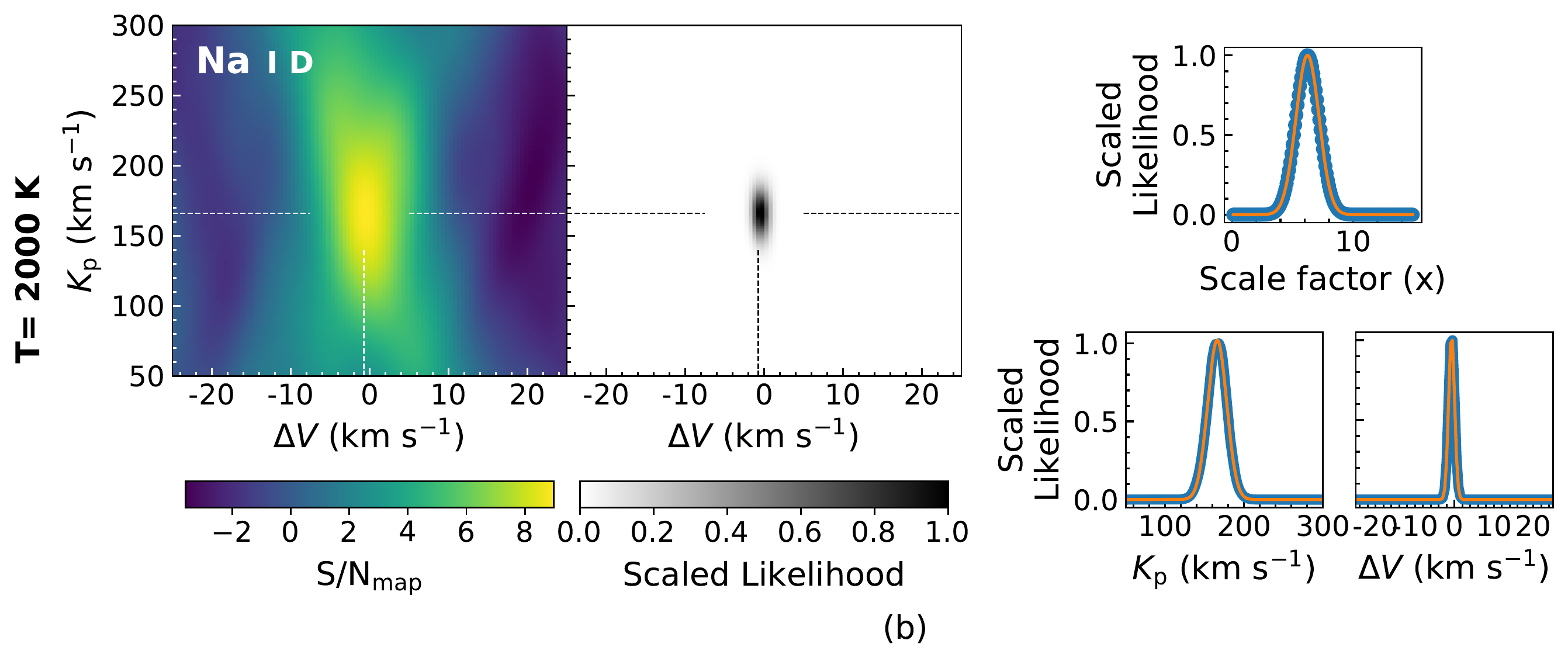}\label{fig:kpdv-NaIresultB}}
    
    \subfigure{\centering\includegraphics[width=0.499\linewidth]{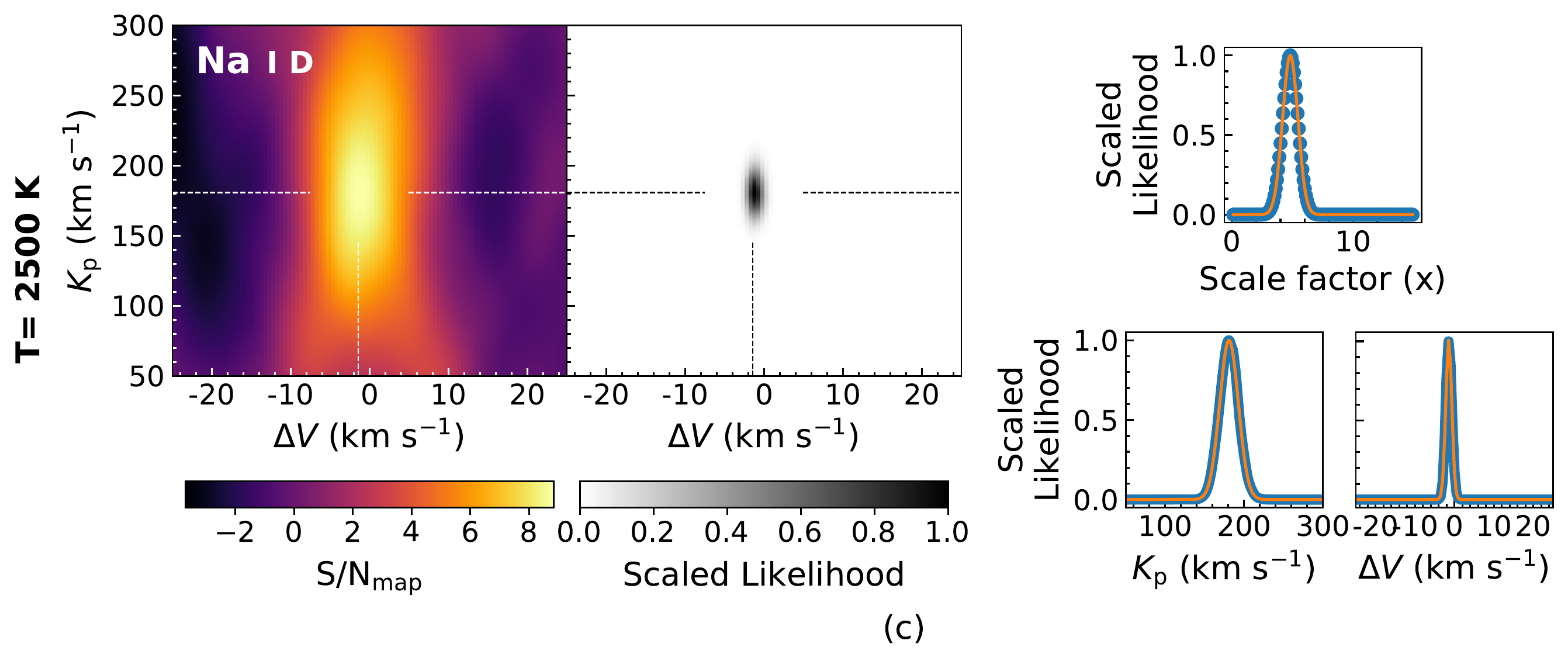}\label{fig:kpdv-NaIresultC}}
    \subfigure{\centering\includegraphics[width=0.495\linewidth]{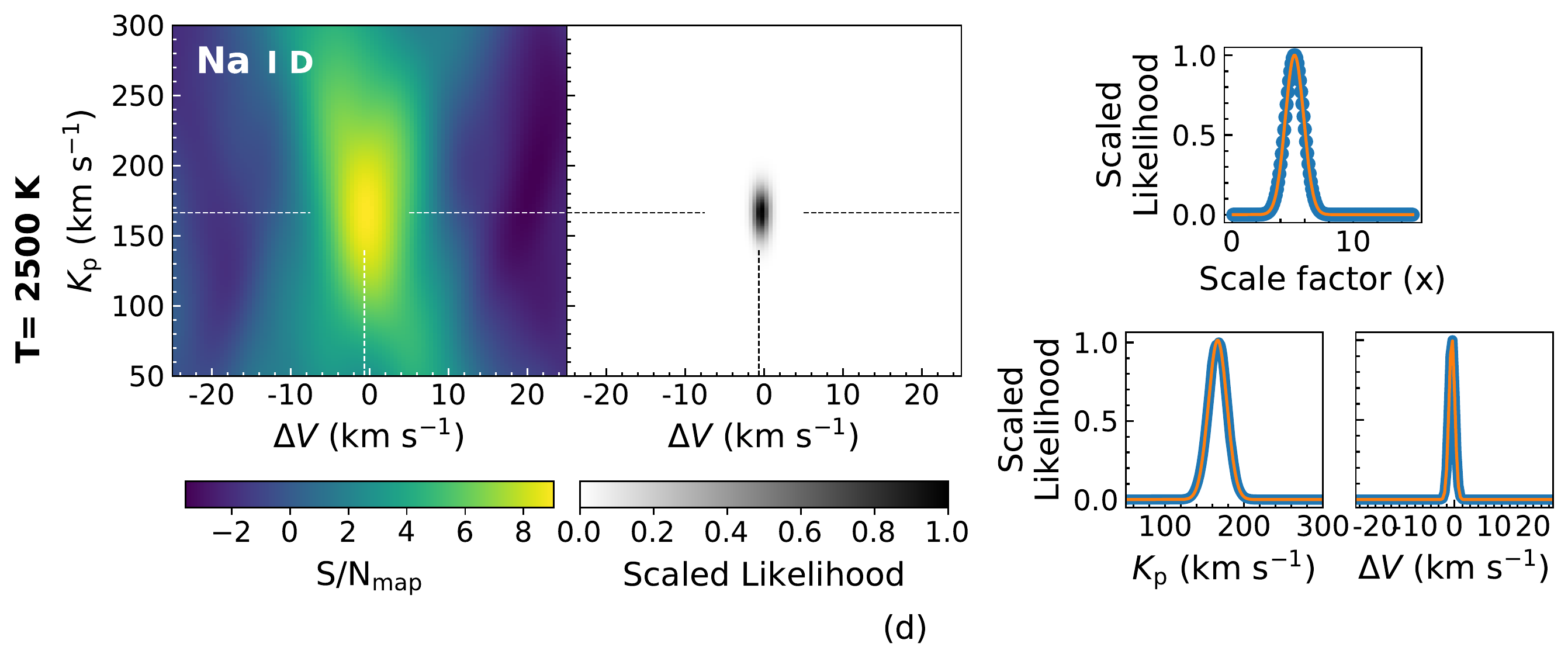}\label{fig:kpdv-NaIresultD}}
    
    \subfigure{\centering\includegraphics[width=0.499\linewidth]{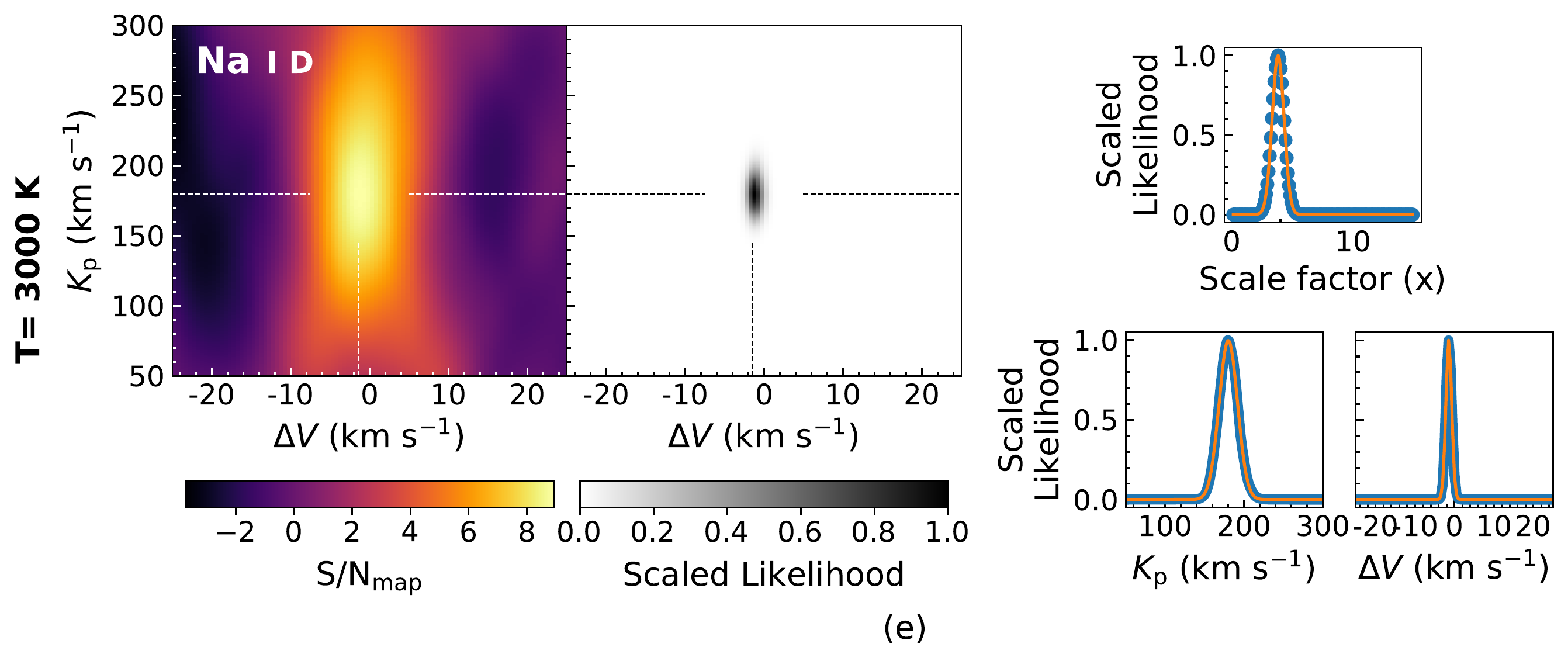}\label{fig:kpdv-NaIresultE}}
    \subfigure{\centering\includegraphics[width=0.495\linewidth]{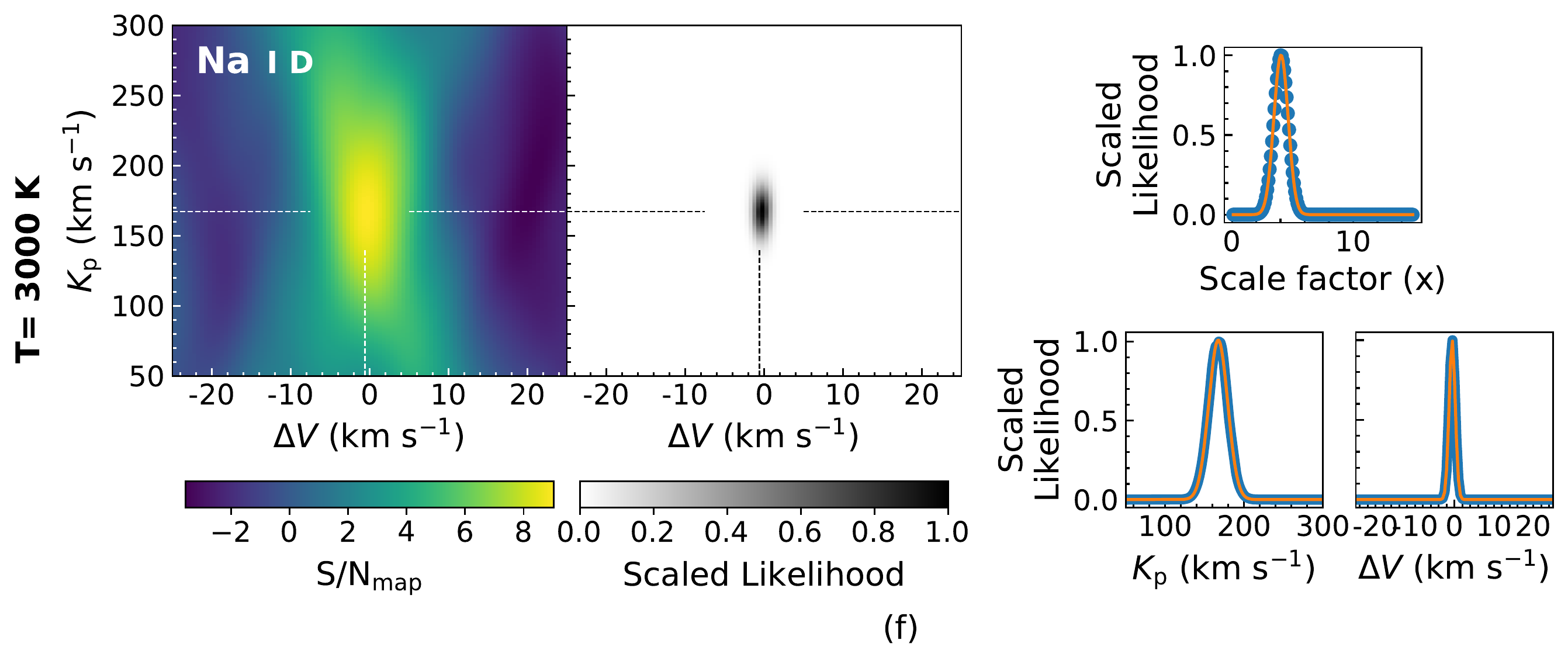}\label{fig:kpdv-NaIresultF}}  
    
    \caption{Similar to Figure \ref{fig:kpdv-FeIresult} but for Na\,{\sc i} D.}
    \label{fig:kpdv-NaIresult}
\end{figure*}

\begin{figure*}
    \subfigure{\centering\includegraphics[width=0.499\linewidth]{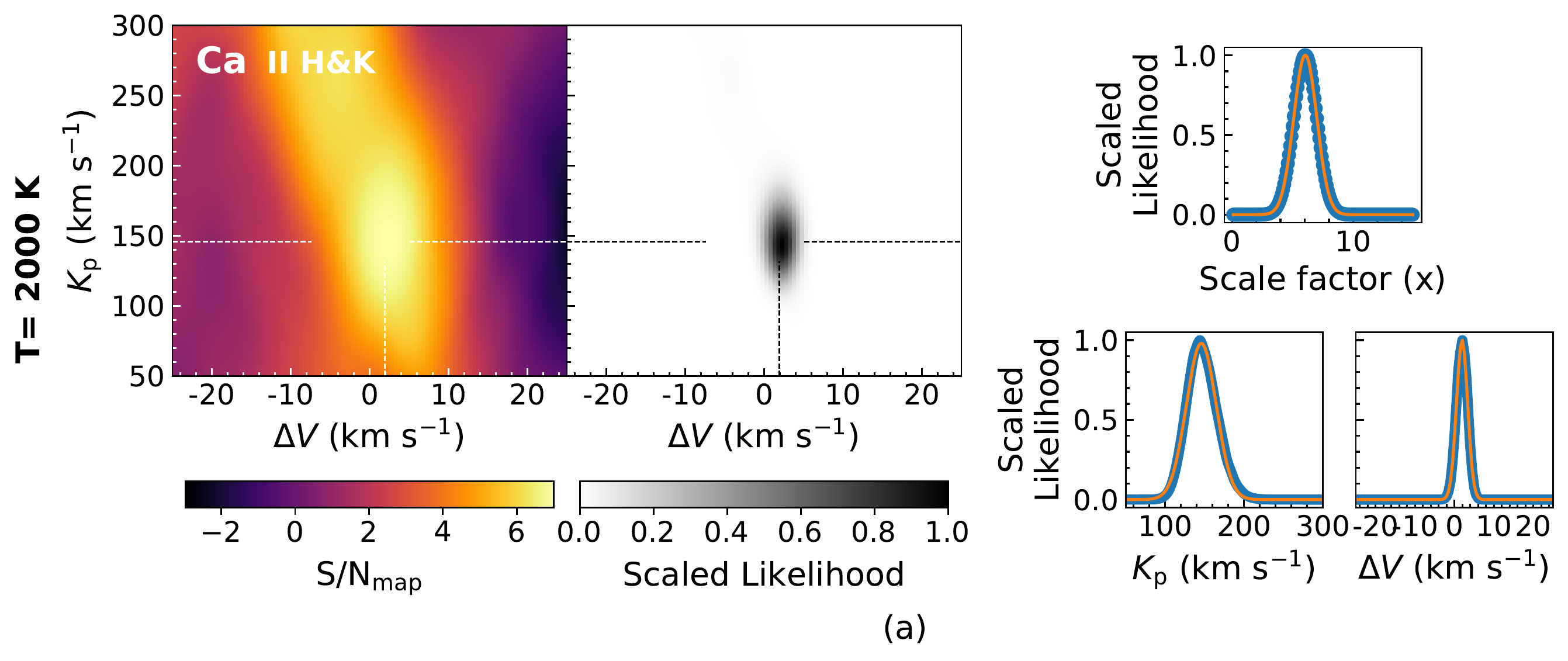}\label{fig:kpdv-CaIIresultA}}
    \subfigure{\centering\includegraphics[width=0.495\linewidth]{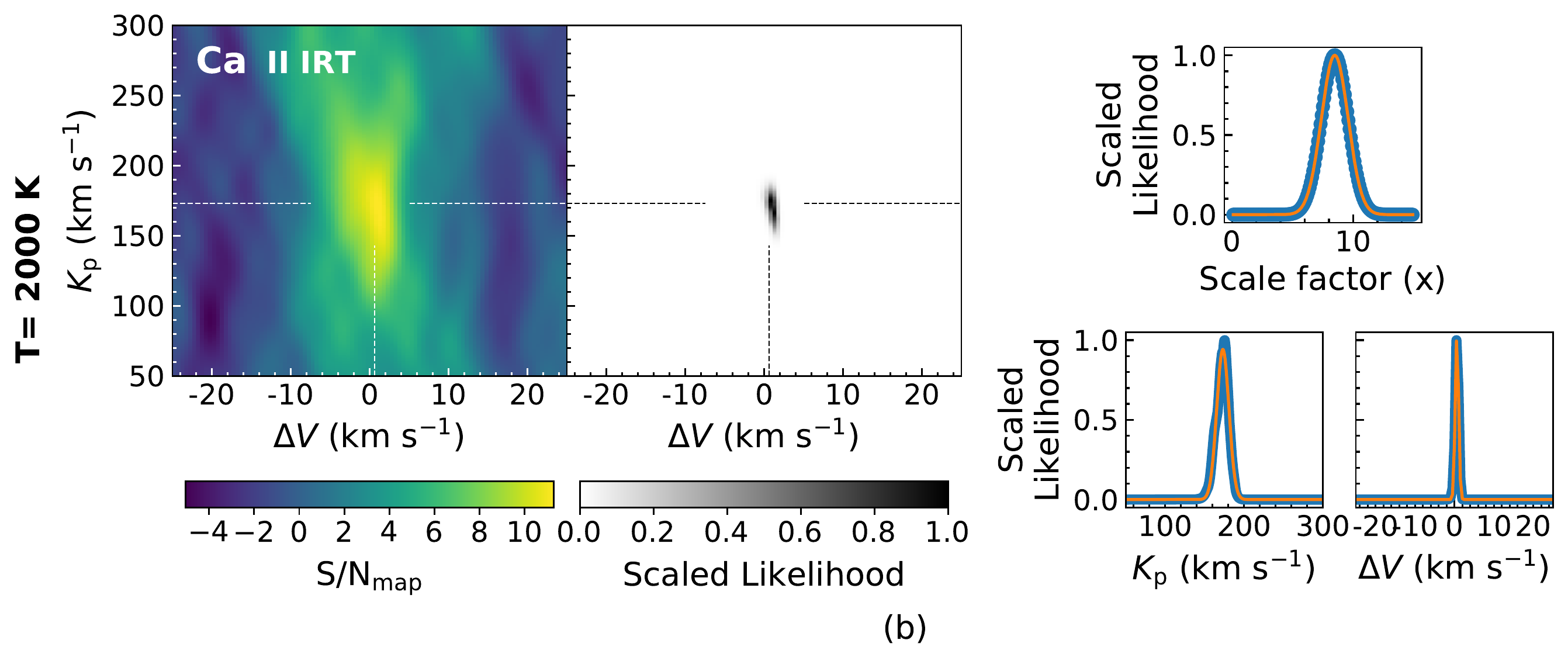}\label{fig:kpdv-CaIIresultB}}
    
    \subfigure{\centering\includegraphics[width=0.499\linewidth]{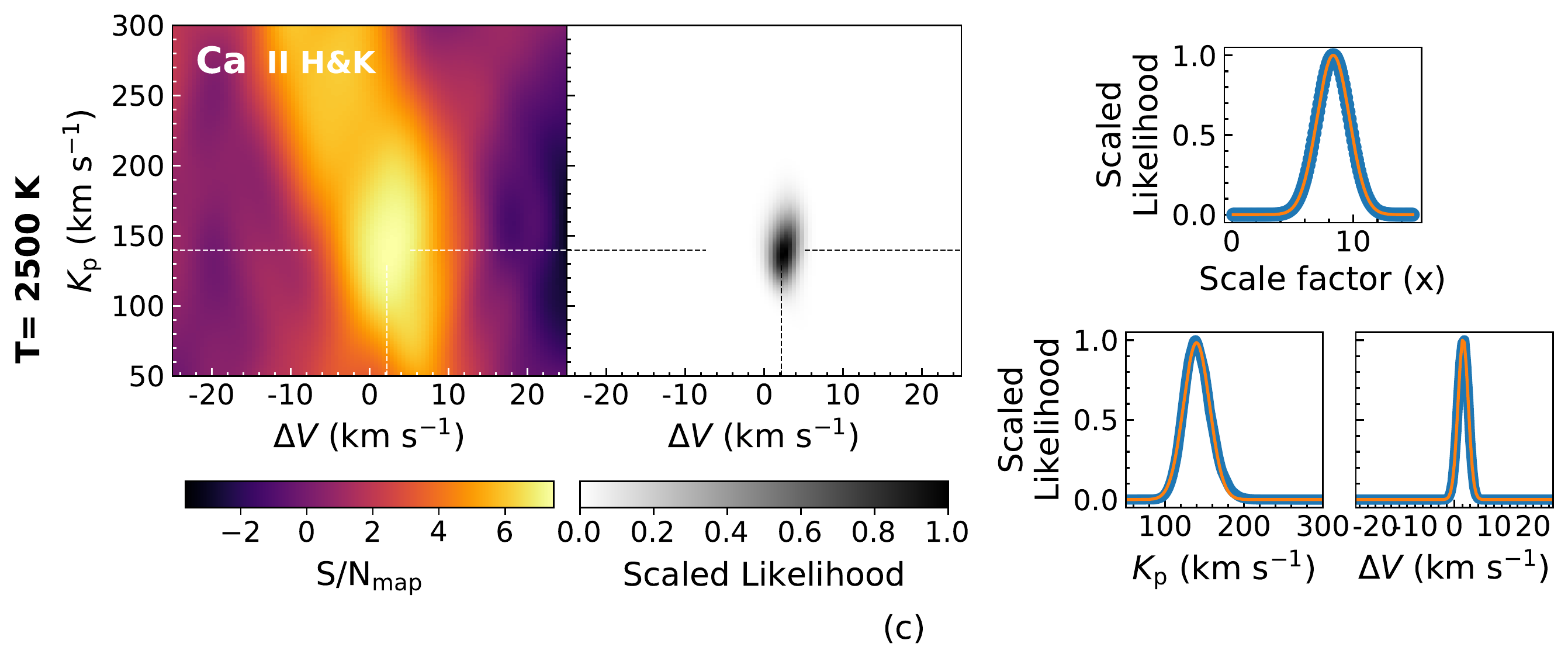}\label{fig:kpdv-CaIIresultC}}
    \subfigure{\centering\includegraphics[width=0.495\linewidth]{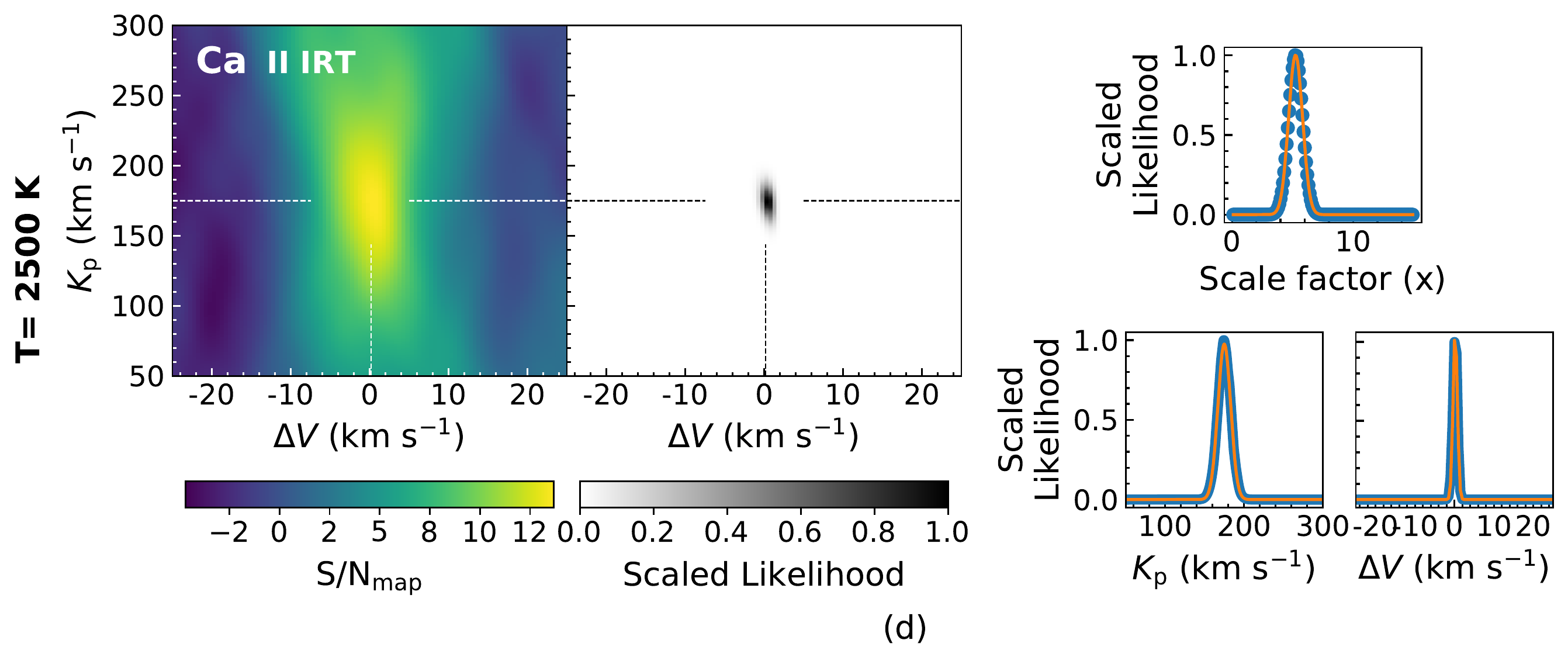}\label{fig:kpdv-CaIIresultD}}
    
    \subfigure{\centering\includegraphics[width=0.499\linewidth]{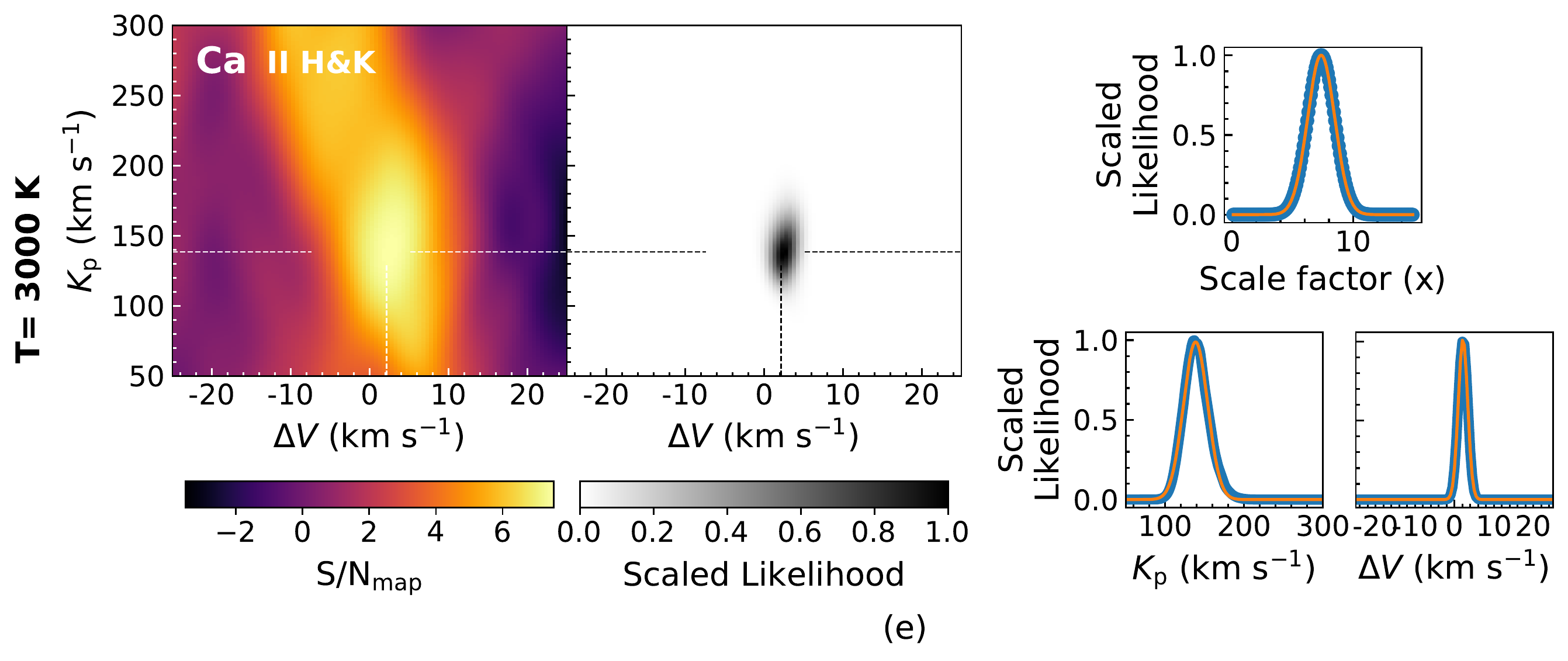}\label{fig:kpdv-CaIIresultE}}
    \subfigure{\centering\includegraphics[width=0.495\linewidth]{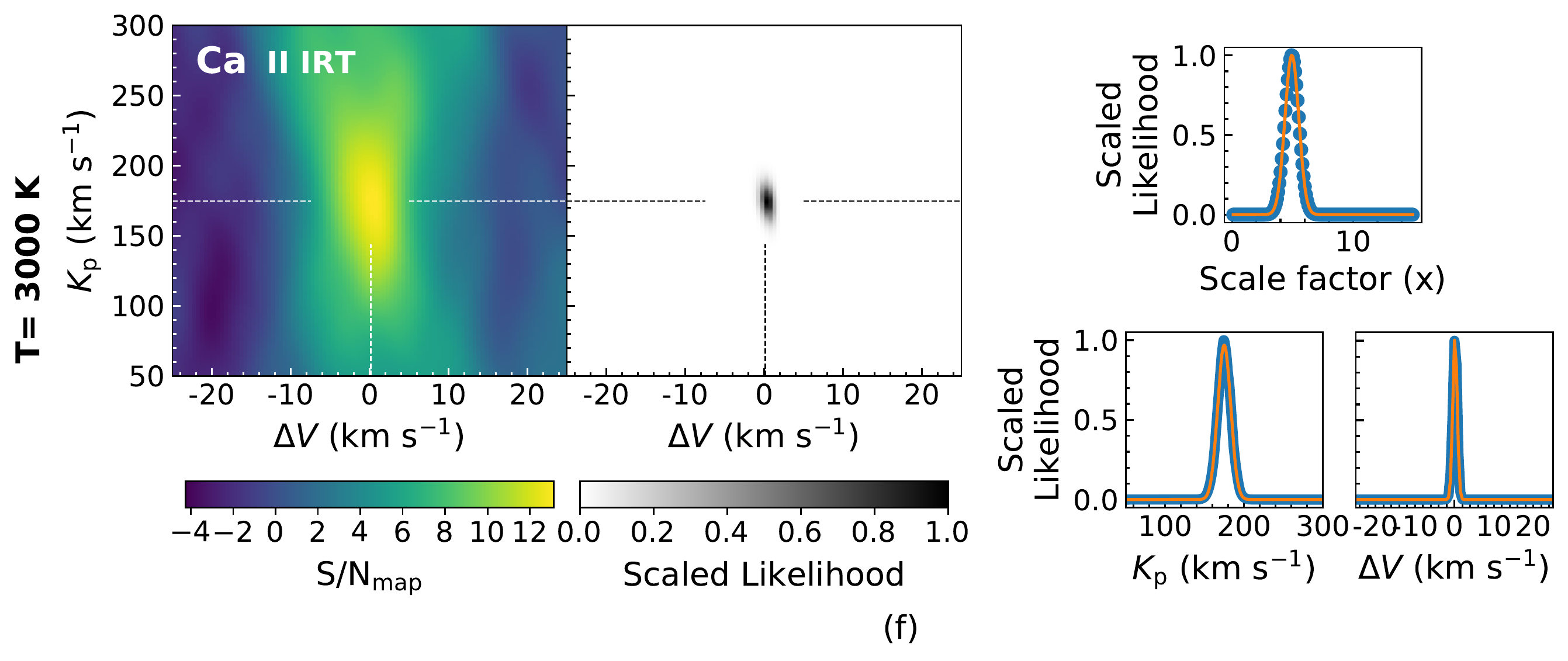}\label{fig:kpdv-CaIIresultF}}  
    
    \caption{Similar to Figure \ref{fig:kpdv-FeIresult} but for Ca\,{\sc ii} H$\&$K from the combined HARPSN data-sets (\textit{left column}), and Ca\,{\sc ii} IRT from the CARMENES data-sets (\textit{right column}).}
    \label{fig:kpdv-CaIIresult}
\end{figure*}

We detected Fe\,{\sc i} and Ca\,{\sc ii} H$\&$K in the transmission spectrum of KELT-20b in the combined HARPSN data-sets and Fe\,{\sc i} in the CARMENES data-sets. We also confirmed the previous detection of Fe\,{\sc ii}, Ca\,{\sc ii} IRT, Na\,{\sc i} D. Figure \ref{fig:ccmat-result} shows the CCF maps of the detected species in the HARPSN and CARMENES data-sets. Most of the Doppler shadow can be removed from the data; however, some residuals remained, especially in the CCF map of Fe\,{\sc ii} at T = 3000 K of the HARPSN-N2 data-sets which can be seen in Figure \ref{fig:ccmat-resultB}. These residuals might have affected the estimation of the S/N of the detected signals. However, as the bright CCF trail during the in-transit phase ($-$0.02 to $+$0.02) of each detected species can be seen visually in the total CCF map (see Figure \ref{fig:ccmat-result}), the robustness of the detection is not affected. The S/N map and the likelihood analysis for all of the detected species are shown in Figure \ref{fig:kpdv-FeIresult}, \ref{fig:kpdv-FeIIresult}, \ref{fig:kpdv-NaIresult}, and \ref{fig:kpdv-CaIIresult}, while the conditioned distributions of each parameter are summarised in Table \ref{table2} and \ref{table3} for the combined HARPSN and CARMENES data respectively. 

\begin{table*}
\caption{Summary of the conditioned distribution of parameters of the detected species in the combined HARPSN data-set}\label{table2}
\begin{tabular}{l|c|c|c|c|c|}

Species & $K_{\mathrm{p}}$  & $\Delta V$ & $\alpha$ &S/N$_{\mathrm{map}}$& Significance \\
                            &(km s$^{-1}$) &(km s$^{-1}$) &($\times$)&&($\sigma$)\\
\hline
\hline
Fe\,{\sc i}   (2000 K) &194.9$\pm$6.9& -3.4$\pm$0.4& 1.28$\pm$0.10& 13.45& 13.05\\
Fe\,{\sc i}   (2500 K) &198.2$\pm$5.9& -3.5$\pm$0.3& 1.03$\pm$0.07& 14.22& 14.05\\
Fe\,{\sc i}   (3000 K) &200.1$\pm$5.2& -3.6$\pm$0.3& 0.82$\pm$0.06& 14.43& 14.30\\
\hline                  
Fe\,{\sc ii}  (3000 K) &165.0$\pm$3.5& -1.4$\pm$0.2& 175.42$\pm$12.00& 13.47& 14.61\\
\hline
Na\,{\sc i} D (2000 K) &182.0$\pm$12.4& -1.4$\pm$0.8& 5.77$\pm$0.80& 8.64& 7.25\\
Na\,{\sc i} D (2500 K) &180.9$\pm$11.9& -1.4$\pm$0.8& 4.80$\pm$0.63& 8.84& 7.61\\
Na\,{\sc i} D (3000 K) &180.0$\pm$11.8& -1.4$\pm$0.7& 3.79$\pm$0.49& 8.92& 7.72\\
\hline
Ca\,{\sc ii} H$\&$K (2000 K) &145.8$\pm$18.9& 1.9$\pm$1.4& 6.04$\pm$0.96& 7.00& 6.32\\
Ca\,{\sc ii} H$\&$K (2500 K) &139.7$\pm$16.5& 2.2$\pm$1.3& 8.35$\pm$1.34& 7.47& 6.24\\
Ca\,{\sc ii} H$\&$K (3000 K) &138.5$\pm$15.3& 2.2$\pm$1.2& 7.35$\pm$1.12& 7.53& 6.57\\
\hline
\end{tabular}
\end{table*}

\begin{table*}
\caption{Summary of the conditioned distribution of parameters of the detected species in the CARMENES data-set}\label{table3}
\begin{tabular}{l|c|c|c|c|c|}
Species               & $K_{\mathrm{p}}$  & $\Delta V$ & $\alpha$ &S/N$_{\mathrm{map}}$& Significance \\
                            &(km s$^{-1}$) &(km s$^{-1}$) &($\times$)&&($\sigma$)\\
\hline
\hline
Fe\,{\sc i}   (2000 K) &227.1$\pm$7.5& -3.3$\pm$0.6& 1.39$\pm$0.23& 7.07& 6.04\\
Fe\,{\sc i}   (2500 K) &224.9$\pm$7.3& -3.3$\pm$0.6& 1.25$\pm$0.19& 7.42& 6.42\\
Fe\,{\sc i}   (3000 K) &263.1$\pm$8.8& -6.5$\pm$0.6& 0.95$\pm$0.15& 7.34& 6.44\\
\hline
Fe\,{\sc ii}  (3000 K) &139.2$\pm$12.5& -0.4$\pm$0.6& 548.06$\pm$152.20& 5.20& 3.60\\
\hline

Na\,{\sc i} D (2000 K) &166.0$\pm$12.1& -0.7$\pm$0.8& 6.23$\pm$0.96& 8.98& 6.51\\
Na\,{\sc i} D (2500 K) &166.5$\pm$11.7& -0.7$\pm$0.7& 5.15$\pm$0.76& 9.04& 6.80\\
Na\,{\sc i} D (3000 K) &167.3$\pm$12.1& -0.6$\pm$0.7& 4.04$\pm$0.59& 9.03& 6.83\\
\hline
Ca\,{\sc ii} IRT (2000 K) &173.1$\pm$7.7& 0.6$\pm$0.4& 8.47$\pm$1.10& 11.29& 7.69\\
Ca\,{\sc ii} IRT (2500 K) &174.9$\pm$8.1& 0.2$\pm$0.5& 5.24$\pm$0.58& 13.68& 9.06\\
Ca\,{\sc ii} IRT (3000 K) &174.8$\pm$8.2& 0.1$\pm$0.6& 4.93$\pm$0.57& 13.13& 8.60\\
\hline
\end{tabular}
\end{table*}

As can be seen in Figure \ref{fig:kpdv-FeIresult}, the signal for the detected species manifest as a single peak except for the signal of Fe\,{\sc i}, which has a double-peak structure, especially for the combined HARPSN dataset. The primary feature of Fe\,{\sc i} (the peak that has the highest S/N, which is marked by the white dashed line) has a $K_{\mathrm{p}}$ value consistent or close to the expected value in all data-sets. The secondary feature is weaker and more blue-shifted at around $K_{\mathrm{p}}$ of 125\,km s$^{-1}$ and $\Delta V$ of -8\,km s$^{-1}$. However, in the CARMENES data-sets, the Fe\,{\sc i} signal is blurrier which might have caused the  $K_{\mathrm{p}}$ slightly different than the expected value. In this section, we focus our analysis by considering the primary Fe\,{\sc i} signal only to provide a consistent interpretation of the state of the planetary atmosphere. The possible scenario of the double-peak structure will be discussed in detail in Section \ref{subsec:pecFeI}.

From Table \ref{table2} and \ref{table3}, the S/N$_{\mathrm{map}}$ and significance of the detected signal are mostly consistent with each other, although S/N$_{\mathrm{map}}$ might be over-estimated when there are strong residuals in the CCF map. This is especially the case for the species which have a broad absorption feature like Na\,{\sc i} D, Ca\,{\sc ii} IRT, and Ca\,{\sc ii} H$\&$K which makes the R-M+CLV modelling difficult. The $K_{\mathrm{p}}$ of most of the detected species is consistent with the expected value within 1 or 2-$\sigma$ which is due to a relatively small change of the planetary radial velocity during the transit. The signal of Fe\,{\sc i} is blue-shifted by $>$ 3 km s$^{-1}$ at $>$ 5.35-$\sigma$. Assuming that the planet is tidally-locked, the (equatorial) rotational velocity is $\approx -2.6$ km s$^{-1}$, while the equatorial day-night wind velocity for a tidally locked hot Jupiter can be up to $\approx$5\,km s$^{-1}$ \citep[e.g.][]{2016Kataria}, combining both effects would explain the blue-shifted signal of Fe\,{\sc i} in our result. In contrast with the result from \citet{Casasayas-Barris2019}, we did not detect a significant shift from the planetary rest-frame within 1-$\sigma$ in the signal of all of the other detected species, which is due to the different systemic velocity that we adopt. We should note that if we use a similar systemic velocity value to \citet{Casasayas-Barris2019}, our results are consistent with theirs. We note that in Figure \ref{fig:atomic-ionic-spectrum} there are no significant absorption lines for Fe\,{\sc ii} at temperatures of 2000 K and 2500 K that can be used for the cross-correlation, resulting in the detection of Fe\,{\sc ii} at 3000 K only. It might be that we are probing a hotter atmospheric layer than we have assumed. For KELT-9b, \citet{Hoeijmakers2019} argued that the thermal reaction only is enough to ionise almost all of the Fe\,{\sc i} for pressures lower than 3.5 mbar, but this might not be the case for KELT-20b as its $T_{\mathrm{eq}}$ is much lower. As can be seen in Figure \ref{fig:atomic-ionic-fastchem}, for T = 3000 K the VMR of Fe\,{\sc ii} begins to saturate at pressures of 10$^{-3}$ mbar while for cooler atmospheres saturation happens at much lower pressures. Another possibility is that this might also because we ignored the effect of photo-ionisation. If photo-ionisation was considered, it could ionise Fe\,{\sc i} and increase the VMR of Fe\,{\sc ii} at higher pressure,allowing it to be visible in the transmission spectrum of the planet even at T $<$ 3000 K, if it was considered. However, we do not attempt to constrain the VMR of the detected species as it is also degenerate with the assumed temperature and the continuum level set by the Rayleigh scattering and H$^{-}$. 

\begin{figure*}
    \subfigure{\centering\includegraphics[width=0.489\linewidth]{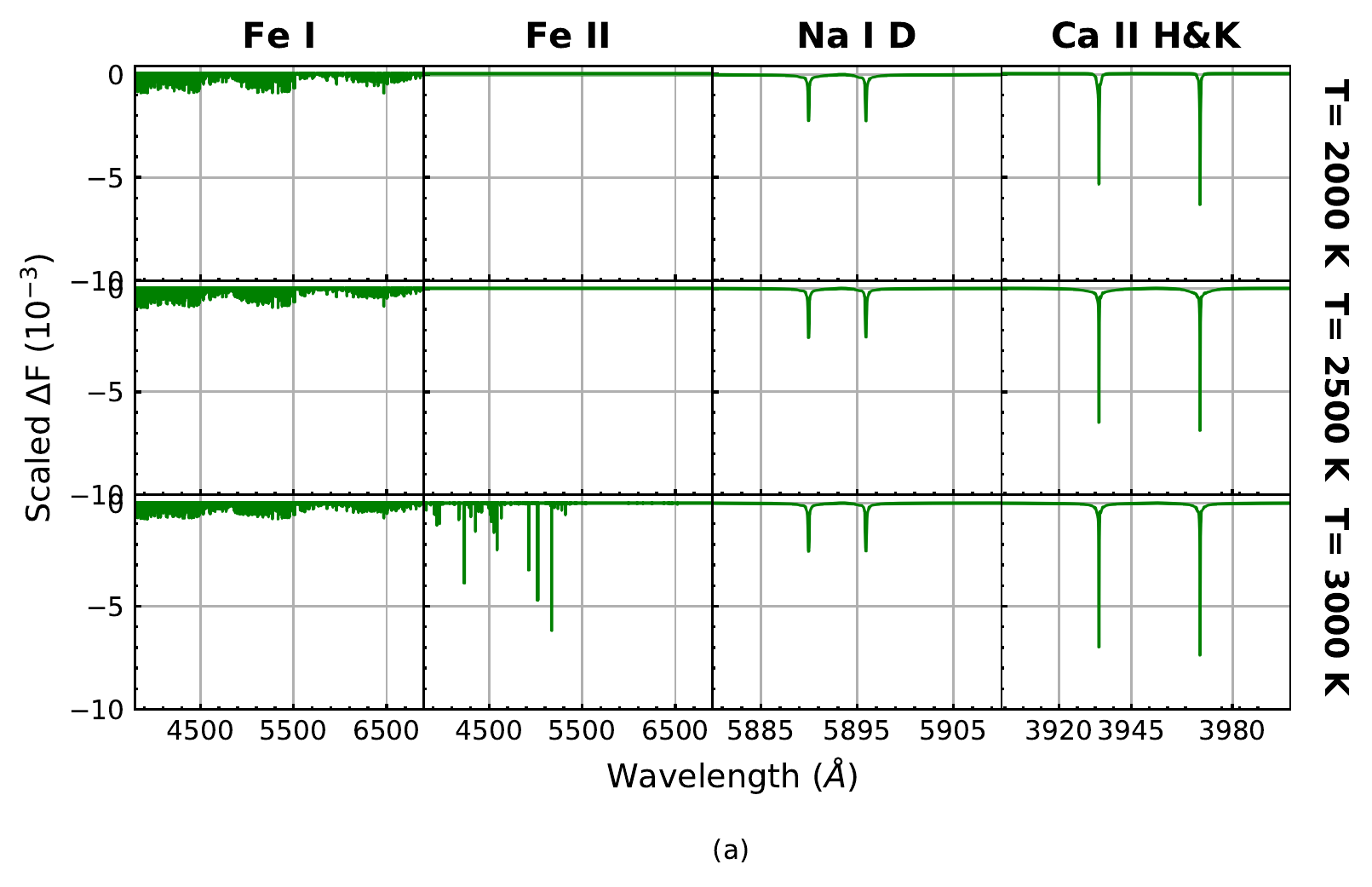}\label{fig:har-bestfitspec}}
    \subfigure{\centering\includegraphics[width=0.49\linewidth]{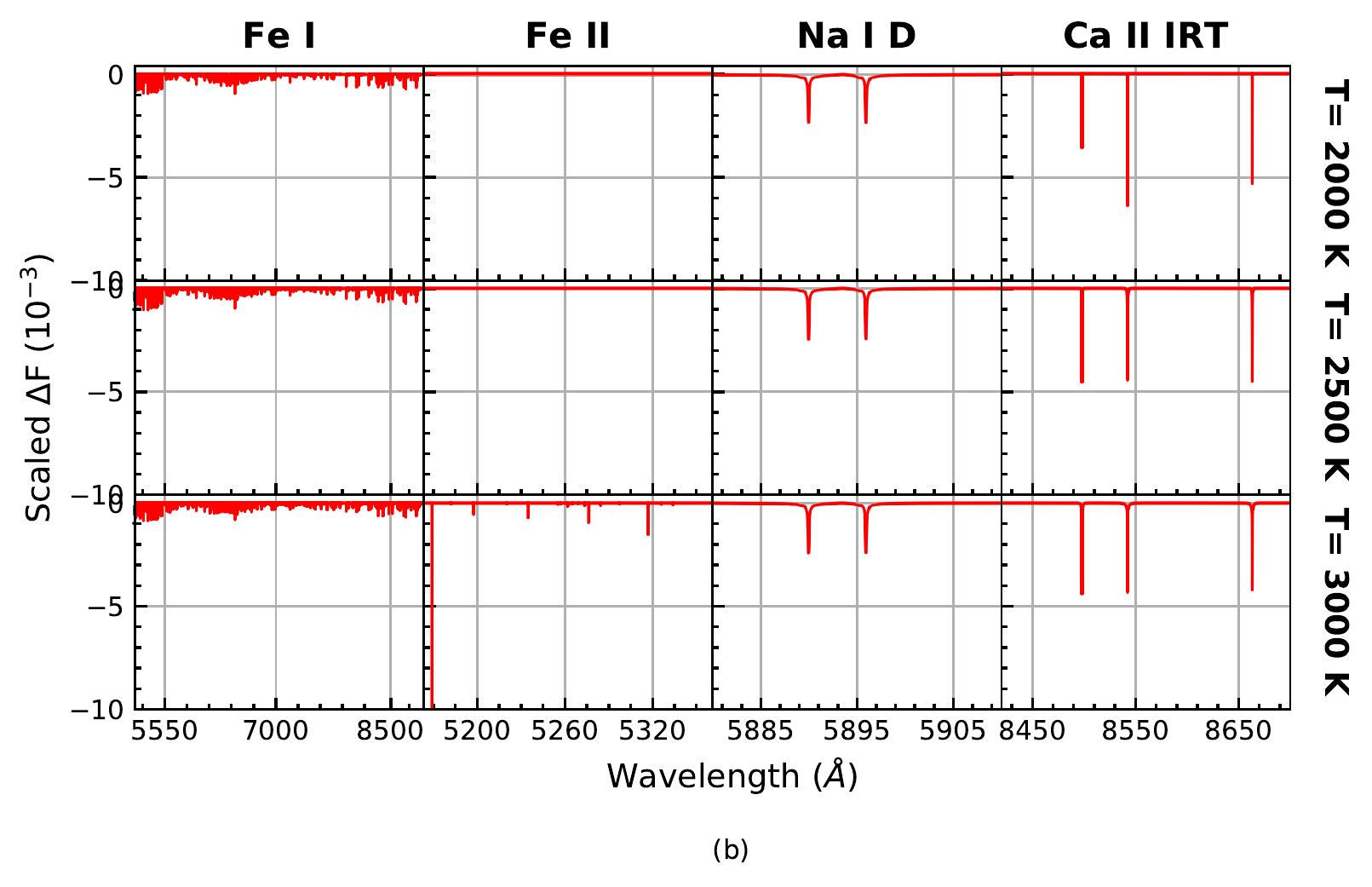}\label{fig:car-bestfitspec}}
    \caption{The best-fit spectrum model of the detected species at three different temperature for the combined HARPSN-N (a) and CARMENES (b) database.\label{fig:bestfitspec}}
\end{figure*}

One advantage of using the new likelihood mapping developed by \citet{gibson2020} is that we do not need injection tests to constrain the $\alpha$ value. The deviation from our chemical equilibrium assumption is also reflected in the $\alpha$ value of each detected species. As summarised in Table \ref{table2} and \ref{table3}, Fe\,{\sc i} has an $\alpha$ value of 1.03 at T = 2500 K, while at T = 3000 K the $\alpha$ value has decreased to 0.82. This means that our Fe\,{\sc i} model at T = 2500 K has an average line-contrast similar to the observed signal. Meanwhile, the other species need an $\alpha$ value $>$ 3, especially for Fe\,{\sc ii} which has an $\alpha$ value $>$ 175. The $\alpha$ value of Fe\,{\sc ii} from the combined HARPSN data-sets might be closer to the real value than the CARMENES data-sets, which cover far fewer Fe\,{\sc ii} lines, resulting in a significance of 3.60 $\sigma$ only. To make the constrained $\alpha$ value easier to interpret, we plotted our transmission spectrum model multiplied by its $\alpha$ value. The result is shown in Figure \ref{fig:bestfitspec}. It is clear that the absorption feature of Fe\,{\sc ii}, Na\,{\sc i} D, Ca\,{\sc ii} H$\&$K and Ca\,{\sc ii} IRT extend to a relatively higher altitude than Fe\,{\sc i}. This might also indicate that the atmospheric layers above the Fe\,{\sc i} altitude have either higher VMR for the detected species other than Fe\,{\sc i} or a higher temperature, which might hint at the existence of an inversion layer. 

\subsection{Non-detection of other thermal inversion agents} \label{sec:nondetec}
We were unable to find any significant signal from any of the possible molecular thermal inversion agents (NaH, MgH, AlO, SH, CaO, VO, FeH and TiO) or Ti\,{\sc i}, Ti\,{\sc ii}, V\,{\sc i}, and V\,{\sc ii} at the expected planetary position (see Figures \ref{fig:nondet-mol-resultA} and \ref{fig:nondet-mol-resultC}). To assess our detection capability, we injected artificial planetary signals of all of the non-detected species into the data and tried to recover them. First, the spectrum is Doppler-shifted to the expected planetary radial velocity for each frame at $K_{\mathrm{p}}$ of 173\,km s$^{-1}$ (assuming the parameters in Table \ref{table1}) and $V_{\mathrm{sys}}$ of -22.05\,km s$^{-1}$ (from our measurements). The spectrum was then convolved with a rotational kernel\footnote{Using \textsc{pyasl.fastRotBroad}} to take into account the broadening due to the planetary rotation by 2.6\,km s$^{-1}$ (assuming a tidally-locked planet and a linear limb-darkening coefficient of 0), then convolved with a Gaussian function to the instrumental resolution of HARPSN and CARMENES\footnote{Using \textsc{pyasl.instrBroadGaussFast}}. The exposure time of each frame is short enough to have a negligible smearing effect on the data so we did not perform any further broadening to the artificial signal. The broadened and Doppler-shifted artificial spectrum was then injected into all data-sets after the blaze function correction. We then performed {\sc SysRem}, cross-correlated with the Doppler-shifted spectrum model, then shifted all of the transit light-curve-weighted CCFs to the planetary rest frame before phase-folding them. We then calculated the 1-$\sigma$ and 3-$\sigma$ detection limits using the phase-shuffling method \citep[e.g.][]{Esteves2017,Deibert2019}. This is done by assigning random in-transit phases to the total CCF during the transit then integrating them at the planetary rest-frame by adopting the $K_{\mathrm{p}}$ of the detected signal. This process was repeated 10000 times, then the noise level was estimated by taking the standard deviation at each $\Delta V$ value. Finally, we combined all of the HARPSN CCF and propagated the error estimations from the phase-shuffling.

\begin{figure*}
    \subfigure{\centering\includegraphics[width=0.51\linewidth]{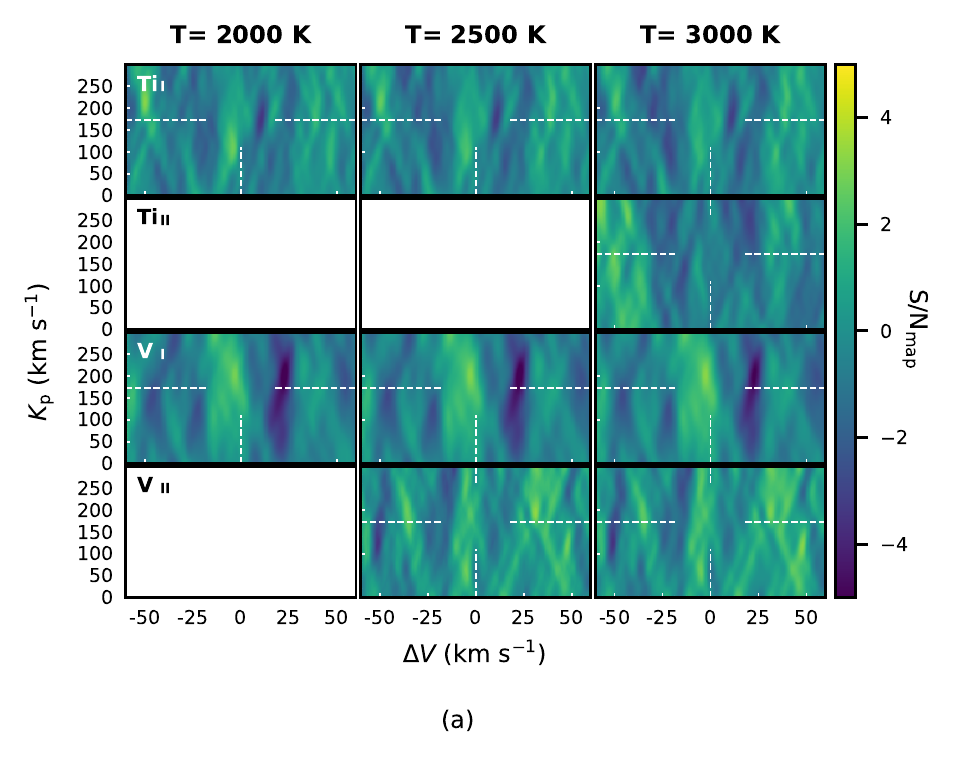}\label{fig:nondet-at-resultA}}
    \subfigure{\centering\includegraphics[width=0.449\linewidth]{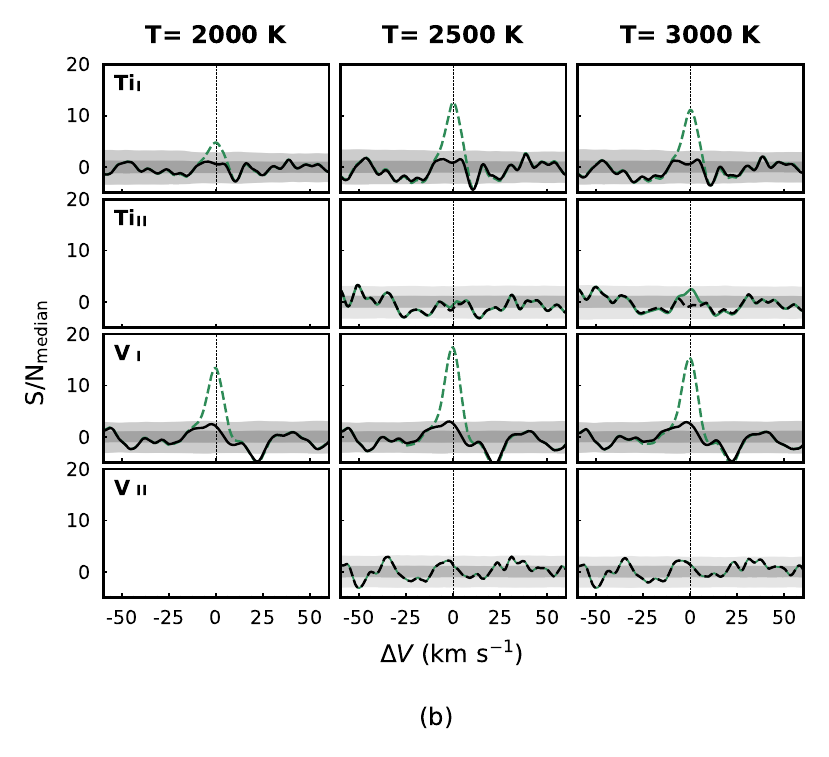}\label{fig:nondet-at-resultB}}
    
    \subfigure{\centering\includegraphics[width=0.51\linewidth]{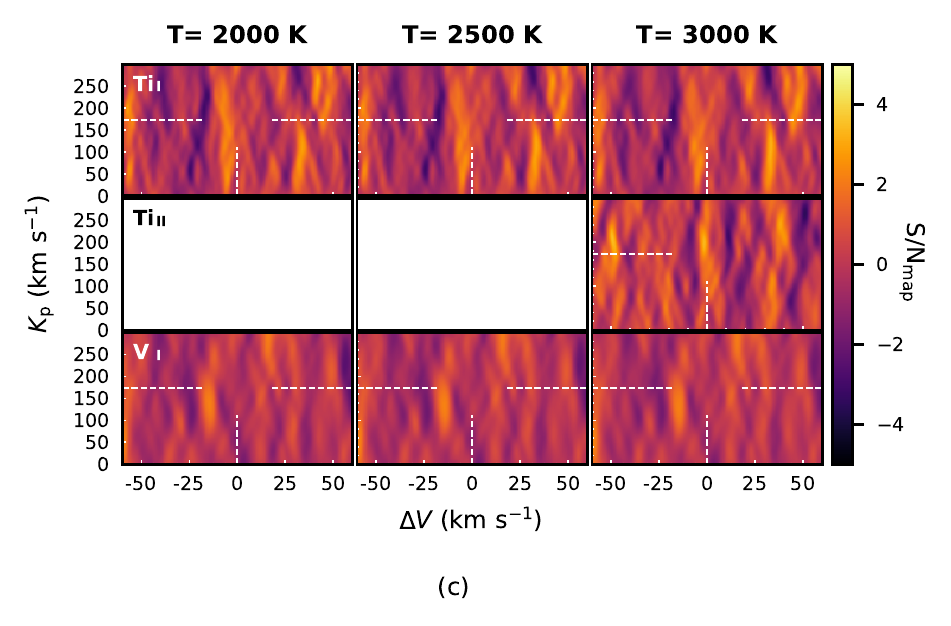}\label{fig:nondet-at-resultC}}
    \subfigure{\centering\includegraphics[width=0.449\linewidth]{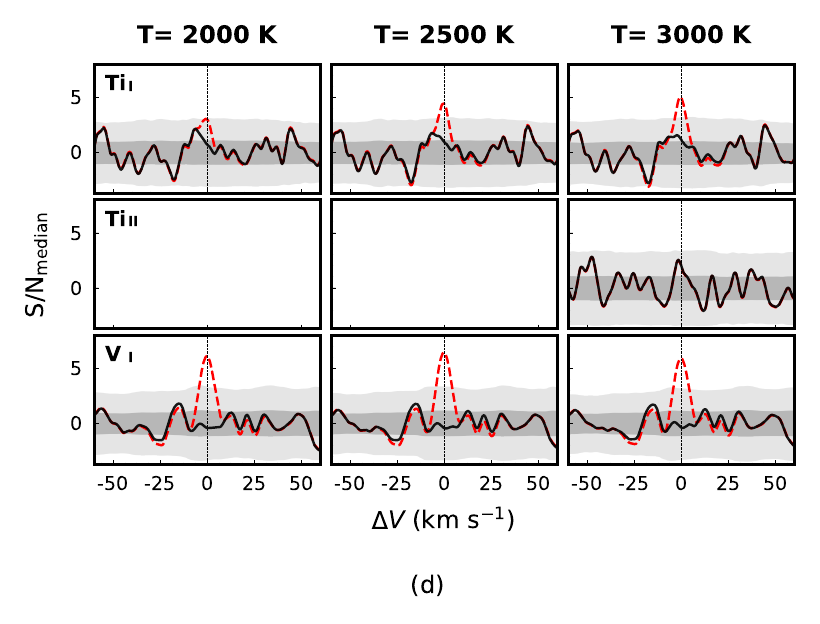}\label{fig:nondet-at-resultD}}
    \caption{The non-detection results for the considered neutral and ionic species. (a) and (c) shows the $K_{\mathrm{p}}-\Delta V$ map for combined HARPSN and CARMENES data-sets respectively. The white dashed line marks the location of the expected planetary signal if the atomic species was detected. The colour bar represents the S/N$_{\mathrm{map}}$. (b) and (c) show the 1-D cross-section along the expected $K_{\mathrm{p}}$ with the solid black line showing the real data and the green dashed line (HARPSN) or the red dashed line (CARMENES) showing the recovered injected planetary signal. The dark shade represents the 1-$\sigma$ detection limit, while the bright shade represents the 3-$\sigma$ detection limit. The vertical black dashed line marks $\Delta V$= 0\,km s$^{-1}$.}
    \label{fig:nondet-at-result}
\end{figure*}

\begin{figure*}
    \subfigure{\centering\includegraphics[width=0.447\linewidth]{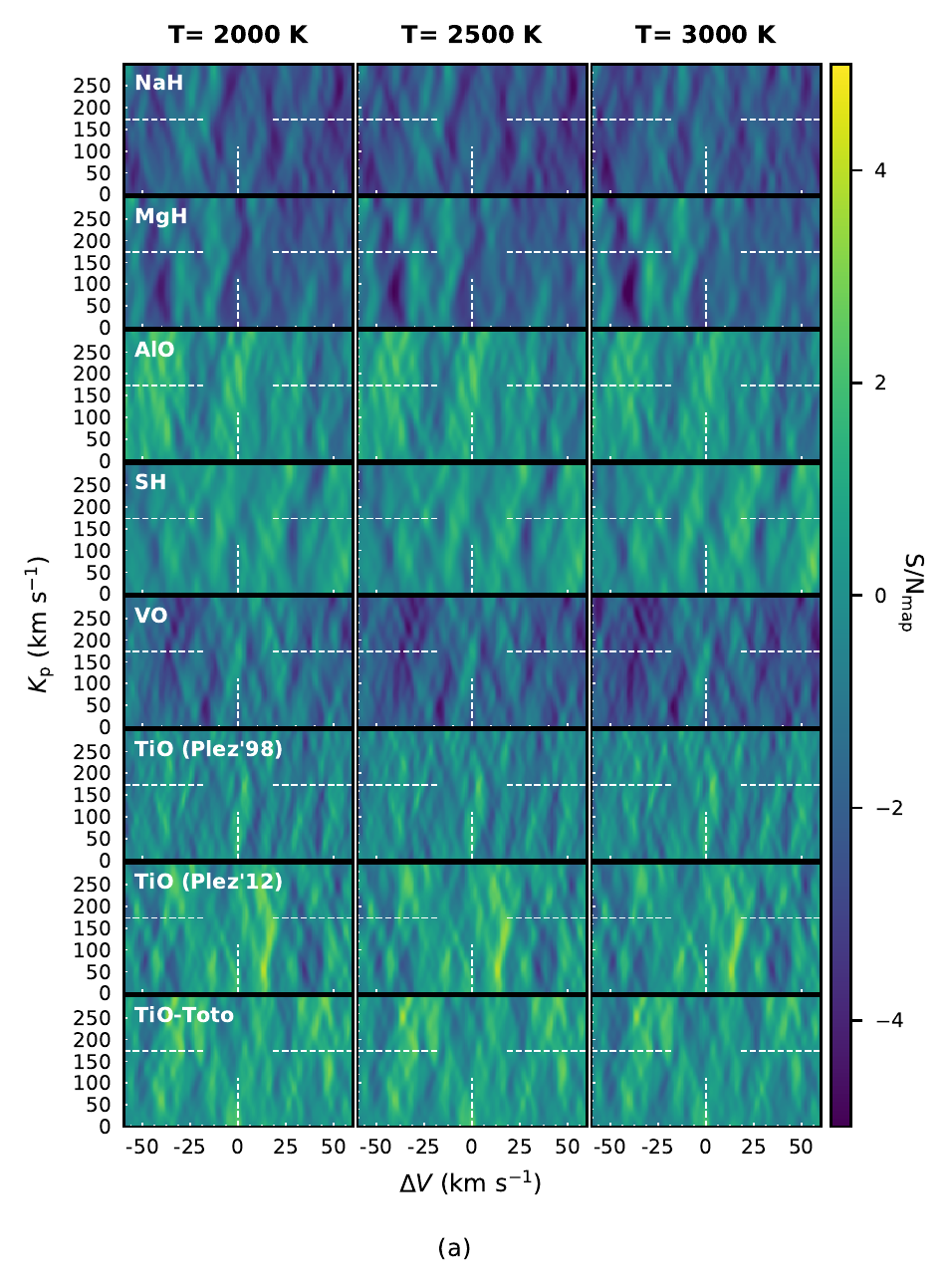}\label{fig:nondet-mol-resultA}}
    \subfigure{\centering\includegraphics[width=0.39\linewidth]{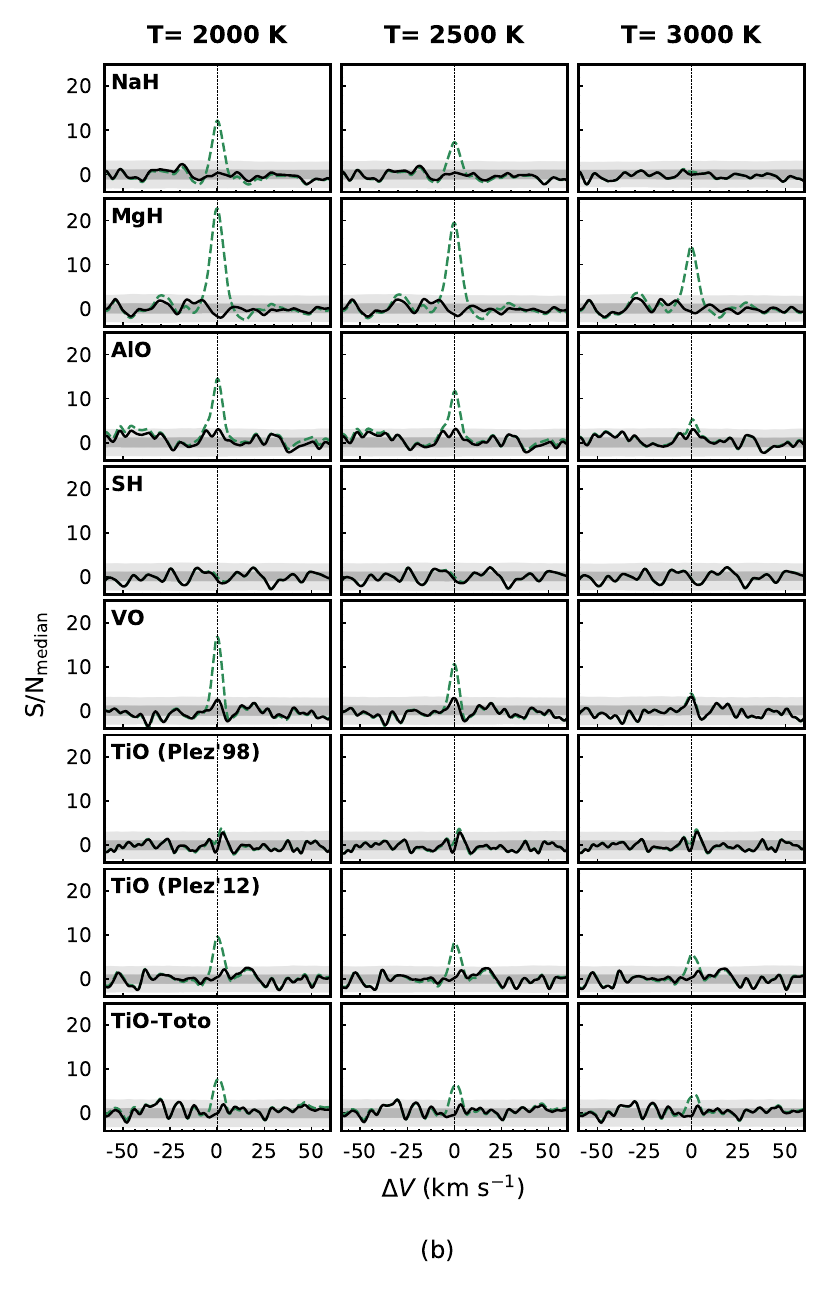}\label{fig:nondet-mol-resultB}}
    
    \subfigure{\centering\includegraphics[width=0.44\linewidth]{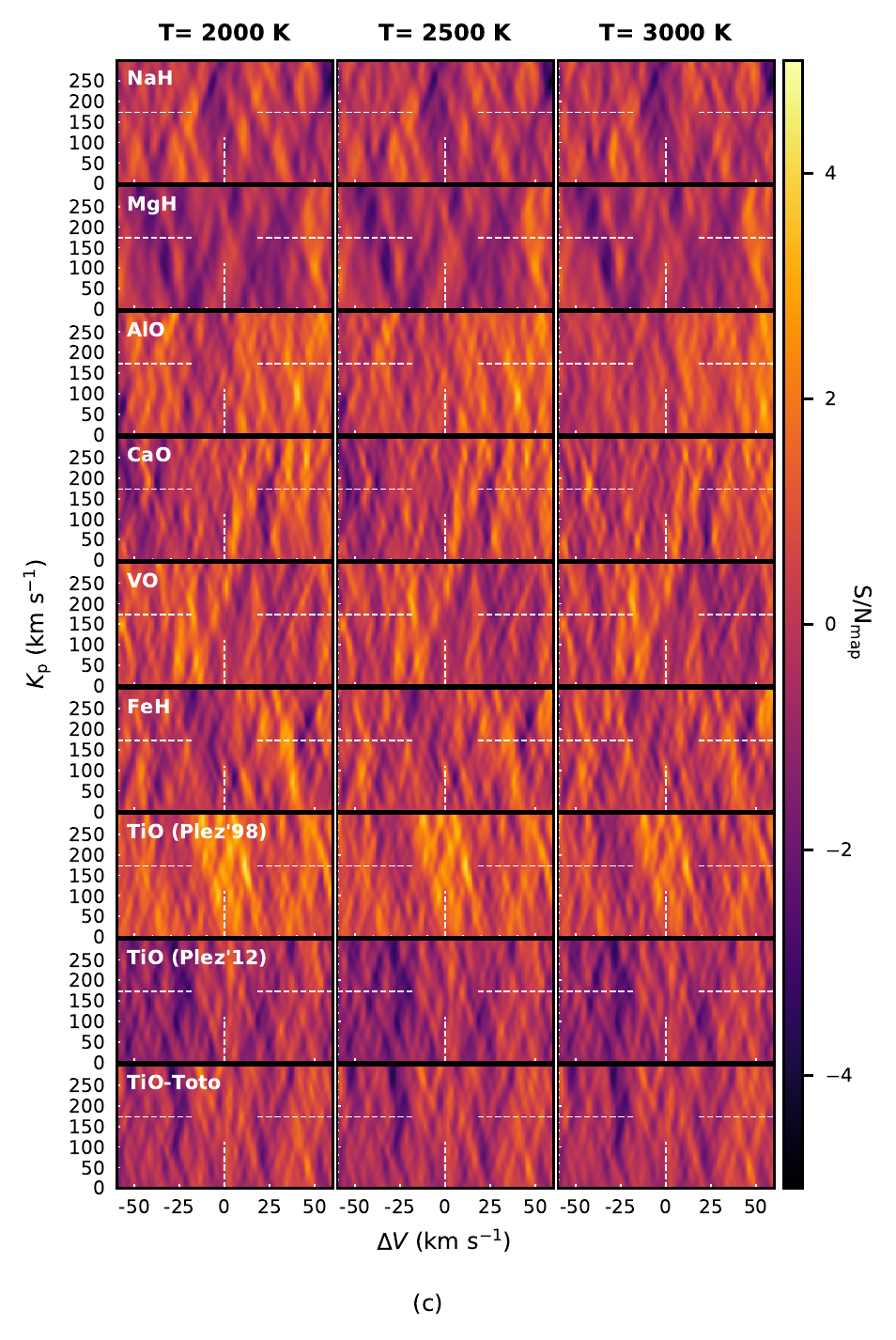}\label{fig:nondet-mol-resultC}}
    \subfigure{\centering\includegraphics[width=0.385\linewidth]{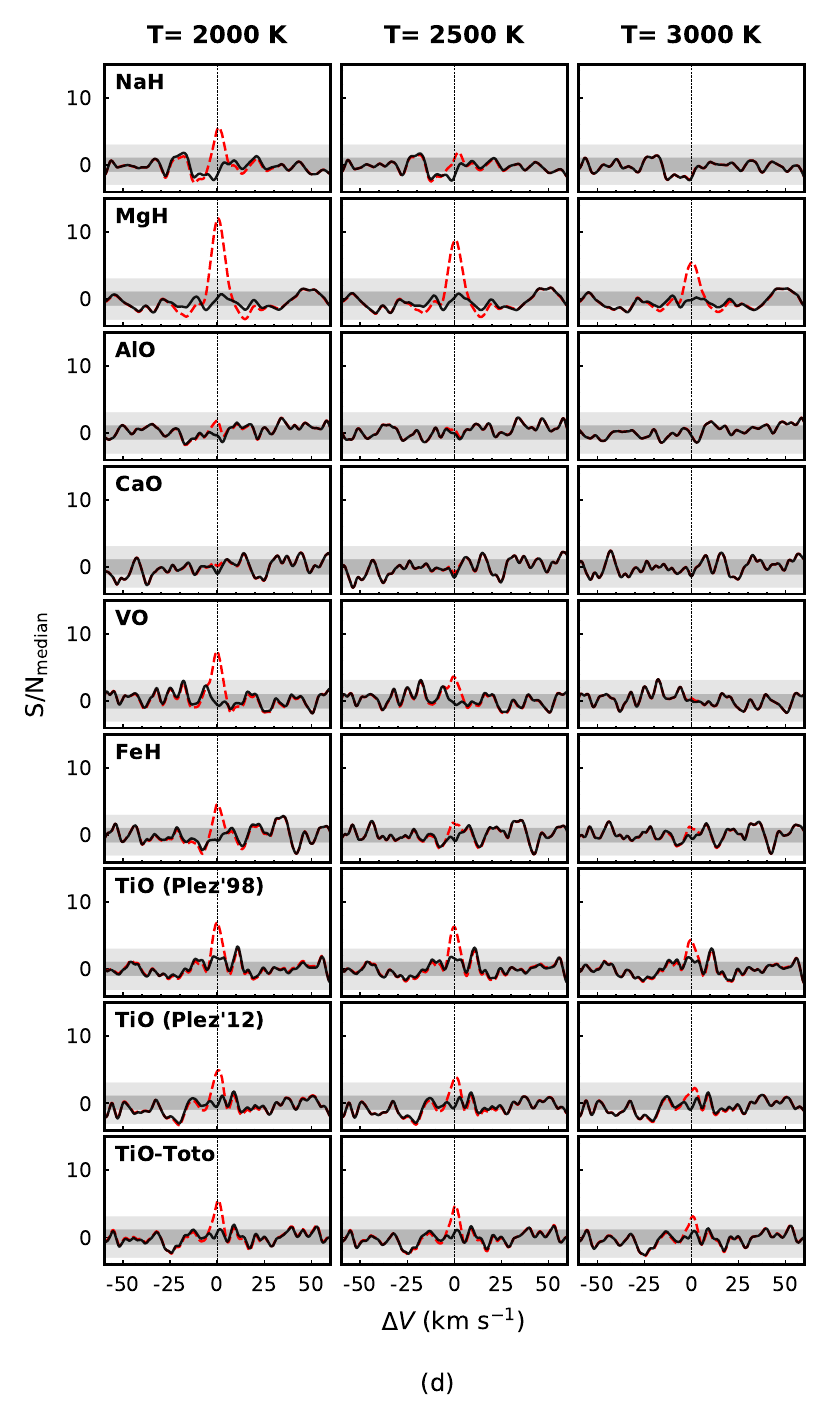}\label{fig:nondet-mol-resultD}}
    \caption{Similar to Figure \ref{fig:nondet-at-result} but for the considered molecular species}
    \label{fig:nondet-mol-result}
\end{figure*}

The result can be seen in Figures \ref{fig:nondet-at-resultB} and \ref{fig:nondet-at-resultD}. For a visual purpose, the y-axis represents the S/N of the recovered signal (S/N$_{median}$) calculated by dividing the CCF with the median of the noise of all $\Delta V$ value. The injected signal of Ti\,{\sc i} and V\,{\sc i} can be recovered at all temperature regimes that we assumed by more than 3-$\sigma$ in the combined HARPSN and CARMENES data-sets. In contrast, we were unable to recover the injected signals of Ti\,{\sc ii} and V\,{\sc ii}. As Figure \ref{fig:atomic-ionic-spectrum} shows, at T = 2000 K the spectrum of Ti\,{\sc ii} and V\,{\sc ii} (and at T = 2500 K) have no absorption lines at all, so the non-detection of these species at these temperatures was expected. However, at T = 3000 K, where both species have many lines, we were still unable to detect them. To investigate if the atmosphere that we are probing for these two species has a higher temperature than assumed, we cross-correlated the data with Ti\,{\sc ii} and V\,{\sc ii} spectrum template at T = 3500 K and 4000 K, but we could not detect any signal. 

For the molecular species at low temperature (T = 2000 K), most of the injected signals can be recovered both in HARPSN and CARMENES data-sets, except for SH and CaO. For higher temperatures, these molecular species will be less abundant due to thermal dissociation (see Figure \ref{fig:molecular-fastchem}) which makes the S/N$_{\mathrm{shf}}$ of the recovered signal lower than at T = 2000 K. This can be seen in Figures \ref{fig:nondet-mol-resultB} and \ref{fig:nondet-mol-resultD}. High-resolution analysis using cross-correlation is highly dependent on the accuracy of the line list. Among the considered molecular species the only line list that has been proven to be accurate enough for high-resolution analysis is the TiO line list \citep{Hoeijmakers2015, Nugroho2017, McKemmish2019}. In the combined HARPSN data-sets, the S/N$_{\mathrm{shf}}$ of TiO recovered using TiO (Plez '12) and TiO-Toto line list is higher than using TiO (Plez '98), while in the CARMENES data-set the S/N$_{\mathrm{shf}}$ of the recovered signals are comparable. This is because the TiO (Plez '98) line-list is only accurate for wavelengths longer than $\approx$6300 \AA, while the other two line lists are accurate enough for wavelengths longer than $\approx$4500 \AA. For the HARPSN data-sets, we can cover many more lines using the TiO (Plez '12) and TiO-Toto line-lists than the TiO (Plez '98) line-list, while for CARMENES data-sets the line-list coverage is more comparable. Therefore, assuming that all of the line lists that we used for these molecular species are perfectly accurate at the wavelength range that we considered, these results rule out the presence of NaH, MgH, AlO, FeH, VO and TiO in the atmosphere of KELT-20b for an isothermal atmosphere (T = 2000 K) in chemical equilibrium at solar metallicity. If the atmospheric temperature is 2500 K, we can rule out NaH, MgH, VO and TiO, and for T = 3000 K we can only rule out the presence of MgH by > 5$\sigma$ and TiO by > 4$\sigma$. However, we did not further our analysis to constrain the upper limit of the detection because, unlike KELT-9b, the $T_{\mathrm{eq}}$ of KELT-20b is low enough to allow clouds to be formed, therefore introducing a degeneracy between cloud altitude and the chemical abundance of the considered species which could not be broken with this kind of analysis alone.

\subsection[Peculiar double-peak structure in Fe I signal] {Peculiar double-peak structure in Fe\,{\sc i} signal} \label{subsec:pecFeI}

In this section, we showed that the double-peak feature of Fe\,{\sc i} could arise from the planetary atmosphere and not an artifact, we checked this by various methods. First of all, stellar pulsation can cause the stellar line profile to change as a function of time and might produce a spurious signal in the $K_{\mathrm{p}}-\Delta V$ map. However, there is no evidence that KELT-20 is pulsating \citep{Lund_2017}. Moreover, de-trending algorithms like {\sc SysRem} would not be able to remove the pulsation signature, since the pulsation effect mimics the planetary signal, which also moves in wavelength as a function of time. Therefore, if KELT-20 is pulsating, we would have seen its signature in Figure \ref{fig:ccmat-result}, similar to $\beta$ Pictoris \citep[Figure 4 in ][]{Koen2003}, WASP-33 \citep[Figure 2 in ][]{Johnson2015}, and KELT-13 \citep[Figure 4 in ][]{Temple2017}. Moreover, stellar pulsation would have affected all of the lines in the stellar spectrum, thus the possibility that this double-peak feature is caused by the stellar pulsation is very low. Second, as can be seen in Image G2 in Figure 5 of \citet{Watson2019}, a ring-like feature in the $K_{\mathrm{p}}-\Delta V$ map (or equivalent with K$_{p}-V_{sys}$ map) can be created if there is a phase offset. However, as there is no similar structure in the $K_{\mathrm{p}}-\Delta V$ map of other detected species this is very unlikely. Third, we masked the overlapped planetary signal with the Doppler shadow (orbital phase of -0.006 to 0.003) after Doppler shadow removal and recalculated $K_{\mathrm{p}}-\Delta V$ map, but the double-peak structures persisted. We also tried to remove the Doppler shadow using a similar method to \citet{Hoeijmakers2019}, and even tried to not to remove the Doppler shadow at all, but no difference was observed. Therefore it is unlikely that this structure is due to some residual from the Doppler shadow removal. Fourthly, there might be some remaining systematic that affected the data and produced spurious noise splitting the planetary signal, so we applied different algorithms to remove the telluric and stellar lines (using air-mass de-trending, e.g. \citealt[][]{Brogi2018}, and in-out transit spectrum division, e.g. \citealt[][]{Wyttencbach2015}) and even by only normalising each wavelength bin by its mean value and repeating the cross-correlation steps. However, all of these methods had minimal or no effect on the structure of Fe\,{\sc i} in the $K_{\mathrm{p}}-\Delta V$ map, thus suggesting that the structure is independent of the telluric removal algorithm that we used. Fifthly, Figure \ref{fig:wavecalib} shows that the wavelength solutions were stable enough over the course of the observations, and this is also clearly represented by the one peak signal of Fe\,{\sc ii} in the $K_{\mathrm{p}}-\Delta V$ map, since the absorption lines of Fe\,{\sc ii} at 3000 K exist across wide wavelength range ($<$5500 \AA). To check if the structure is caused by a spurious signal from some spectral order we masked several orders of the data alternately and recalculated the $K_{\mathrm{p}}-\Delta V$ map of Fe\,{\sc i}. The S/N of both peaks changed but in general, the structure did not change significantly, suggesting that whatever the cause of these structures, it exists in all orders. This could also be caused by the Fe\,{\sc i} template that we used. We check this by cross-correlating our Fe\,{\sc i} template with the reduced HARPSN data of KELT-9b that was used in \cite{Hoeijmakers2018}. However, we found only a single peak in the Fe\,{\sc i} signal. We note that the exposure time that was used in KELT-9 data was 600 s per frame. By taking into account the orbital velocity of KELT-9b from \citet{Hoeijmakers2019}, this is equivalent to $\approx$4.5-7.5\,km s$^{-1}$ smearing in the planetary signal; therefore, the double-peak feature might have been smeared out. We also checked the accuracy of the Fe\,{\sc i} line-list by cross-correlating it with the stellar spectrum of HD 209458 taken using HARPS (PI: Mayor, PID: 60.A-9036(A)). We found that the CCF has a single peak and its width is comparable to the width of Fe\,{\sc i} KELT-20b's CCF showing that the double-peak feature was not caused by the line-list itself. Finally, we investigated the impact of the atmospheric models used by calculating the cross-sections of Fe\,{\sc i} using HELIOS-K, but we found no difference.

\begin{figure}
    \subfigure{\centering\includegraphics[width=1\linewidth]{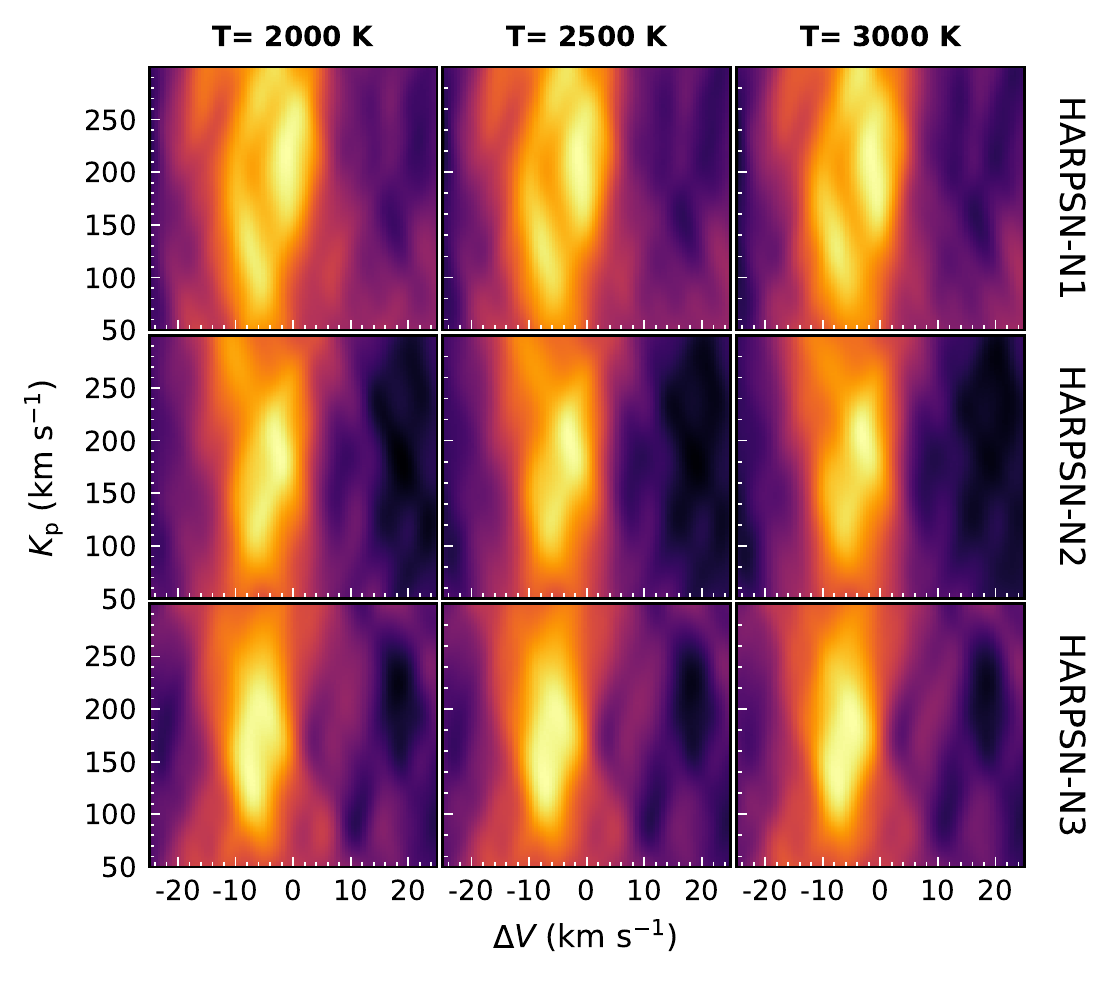}}
    \caption{$K_{\mathrm{p}}-\Delta V$ map of Fe\,{\sc i} for HARPSN-N1 (\textit{first row}), HARPSN-N2 (\textit{second row}) and HARPSN-N3 (\textit{third row}) data-sets.}
    \label{fig:FeIallHARPSN}
\end{figure}

This structure presents in all of the data-sets that were taken at different times with two different facilities (see Figure \ref{fig:FeIallHARPSN}) even after performing these tests, therefore it is very unlikely that the structure was caused by any noise or residual from the data reduction that we know of, or by the use of an incorrect template in the cross-correlation analysis. Assuming that the signals are real and originated from the atmosphere of the planet, we tried to replicate the structure. In Figure \ref{fig:kpdv-FeIresult} and \ref{fig:FeIallHARPSN}, the detected signal of Fe\,{\sc i} manifests in different shapes. Most of them are asymmetrical except in the CARMENES data which has a diamond-like shape. This symmetrical feature in the $K_{\mathrm{p}}-\Delta V$ map can be explained if there are two resolved planetary signals with similar $K_{\mathrm{p}}$ but different $\Delta V$. The evidence of this can be seen in Figure \ref{fig:ccmat-result}, as the width of the Fe\,{\sc i} signal looks wider than the signal of either Ca\,{\sc ii} IRT or Na\,{\sc i} D. From the $K_{\mathrm{p}}-\Delta V$ map of Fe\,{\sc i} in the CARMENES data (see Figure \ref{fig:kpdv-FeIresult}), the central-width of the diamond-like shape signal is $\approx$10 km s$^{-1}$. Based on these, we simulated two Fe\,{\sc i} signals (T = 2500 K) Doppler shifted with $K_{\mathrm{p}}$ of 200 km s$^{-1}$, V$_{\mathrm{sys}}$ of -22 km s$^{-1}$ and $\Delta V$ of -3.5 km s$^{-1}$ and -13.5 km s$^{-1}$ for the primary and secondary signal respectively in the in-transit phase. Then we added Gaussian noise, cross-correlated the signal with the model template, and calculated the $K_{\mathrm{p}}-\Delta V$ map. For the other data-sets, the Fe\,{\sc i} signal in N1 looks wider from about the mid-transit point to the end of the transit than before the mid-transit (see Figure \ref{fig:ccmat-result}a). Therefore, we masked some of the early parts of the secondary signal and repeated the simulation.

\begin{figure}
    \centering\includegraphics[width=1\linewidth]{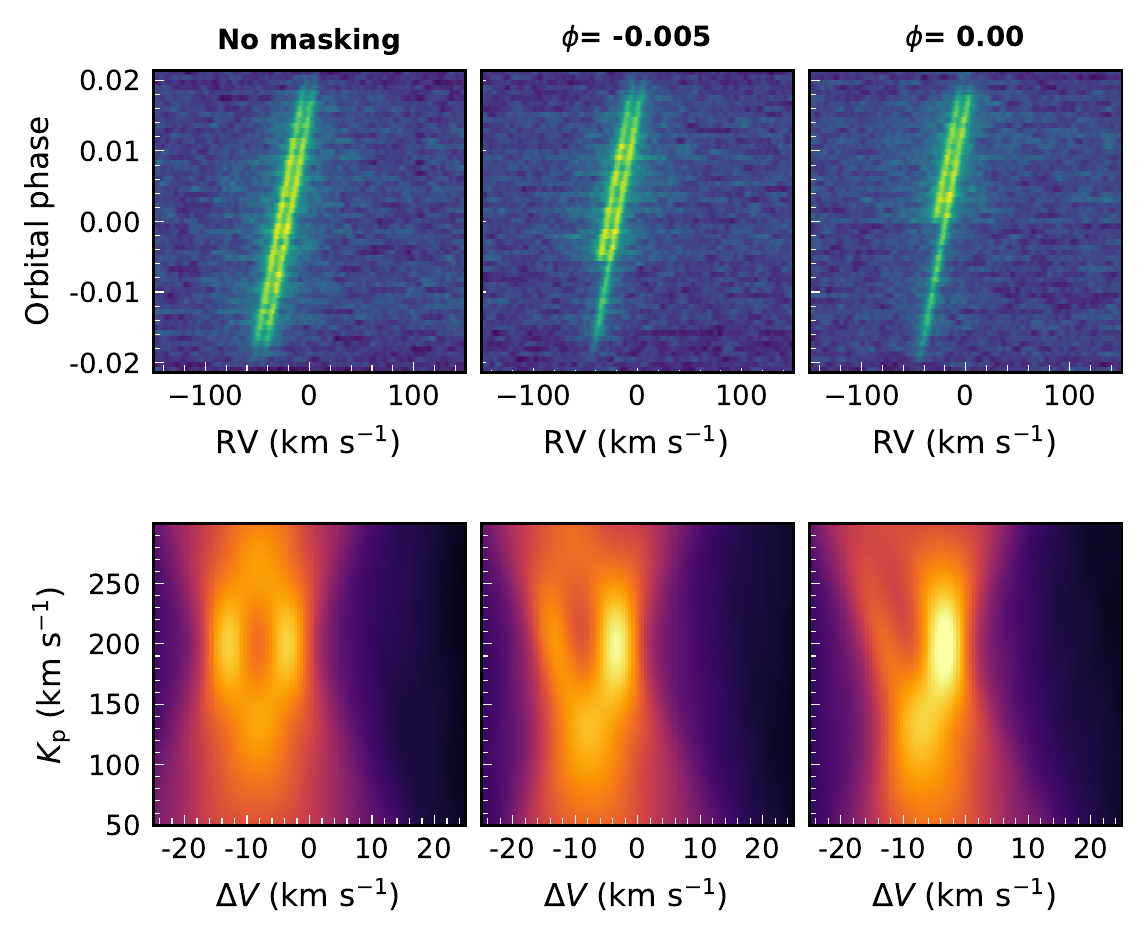}
    \caption{The cross-correlation (\textit{first row}) and the $K_{\mathrm{p}}-\Delta V$ maps (\textit{second row}) of the simulated two planetary signals with similar $K_{\mathrm{p}}$ but different $\Delta V$ without masking the signal (\textit{left column}), masking the signal from the beginning of the ingress to $\phi$ of -0.01 (\textit{middle column}), and $\phi$ of -0.016 (\textit{right column}). The white dashed line indicates the expected velocity of the injected primary signal.}
    \label{fig:ring_sim}
\end{figure} 

As can be seen in the bottom left panel of Figure \ref{fig:ring_sim}, the $K_{\mathrm{p}}-\Delta V$ map of the simulated data without masking the secondary signal resembles the signal of Fe\,{\sc i} in CARMENES data-set. We found that if the secondary signal begin to appear close to the mid-transit, $K_{\mathrm{p}}-\Delta V$ maps resemblance of the Fe\,{\sc i} structure in N1 and N2+N3. Interestingly, in contrast to the primary signal that was recovered at the injected location, the secondary signal of these model appears to be shifted to a smaller $K_{\mathrm{p}}$ and $\Delta V$. This result is consistent with the secondary signal in the $K_{\mathrm{p}}-\Delta V$ map of Fe\,{\sc i} for the combined HARPSN data-set (see Figure \ref{fig:kpdv-FeIresultA}, \ref{fig:kpdv-FeIresultC} and \ref{fig:kpdv-FeIresultE}). 

In general, the double-peak structure appears in the $K_{\mathrm{p}}-\Delta V$ map of all data-sets, which could be an indication that we are probing the Fe\,{\sc i} signal from two different atmospheric limbs of the planet. The resolved-signals might be due to different net Doppler shifts at each limb resulting from the combination of planetary rotation and atmospheric dynamics (e.g. equatorial jets, day-night wind). The net Doppler shift of the leading limb can be close to zero if the rotational velocity and the day-night wind velocity are comparable, while the trailing limb can be largely blue-shifted and dominated by the signal from the equator due to the combination of a strong day-night jet and the planetary rotation. This has been predicted for a highly irradiated and fast-rotating planet where the zonal winds peak in the equator, which might result in a double-peak feature in the high-resolution spectroscopy analysis \citep[e.g. ][]{Showman2011, Miller2012, Showman2013, Showman2015, Flowers2019} or even a single blue-shifted signal as it was observed in WASP-76b when there is not enough gas to absorb the stellar light in the morning terminator \citep{Ehrenreich2020}. The cause of 'delay' in the appearance of the weaker secondary signal is unclear and it is very unlikely that an unstable observational condition could cause this feature as it would affect all of the planetary signals. One of the possible explanation is it might be due to the atmospheric variability of the planet which affects only Fe\,{\sc i}. Indeed, this was hinted by the constrained $\alpha$ value of each detected species (see Figure \ref{fig:bestfitspec}), indicating that the detected Fe\,{\sc i} feature extends to a relatively different altitude than the rest of the detected species and potentially highly affected by the atmospheric dynamics of the planet which is showed by the blue-shifted signal of Fe\,{\sc i} while it is not the case for the other detected species. The first direct observational evidence of this equatorial wind in a hot Jupiter was seen in the high-resolution transmission spectroscopy analysis of HD 209458b using CRIRES, where the signal of CO was blue-shifted by 2$\pm$1 km s$^{-1}$ from the rest frame velocity of the planet \citep{Snellen2010}. \citet{Louden2015} spatially resolved the atmosphere of HD 189733b, comparing the shape of the Na\,{\sc i} D absorption line during the ingress and the egress and measuring the rotation of the planet and a strong eastward wind of -1.9 km s$^{-1}$. Recently, \citet{Brogi2016, Brogi2018} detected H$_{2}$O and CO in the near-infrared transmission spectrum of this planet and measured a similar wind velocity. In principle, the cause of this double-peak structure can be further confirmed using a similar technique as \citet{Louden2015}; however, further analysis is beyond the scope of this paper and will be subject to future study.

\subsection{Possible interpretation of the planetary atmospheric conditions}

At the temperature equilibrium of KELT-20b, most of the Ti-bearing species are present mostly as TiO \citep{Lodders2002} while mono-atomic Fe gas is the most dominant of the Fe-bearing species \citep{Visscher2010}. In this work, we show that we were able to detect Fe gas but not any Ti- or V-bearing species, especially atomic Ti, V, TiO and VO. Based on our injection tests, if TiO and VO were thermally dissociated, we should have been able to detect Ti\,{\sc i} and V\,{\sc i} in our data relatively straightforwardly, as has been shown by our injection and recovery tests.

The non-detection of Ti- and V- bearing species, therefore, more likely indicates the presence of a non-chemical equilibrium process such as a vertical and/or day-night cold-trap in the atmosphere of the planet. Vertical cold-traps exist at the specific pressure when the temperature of the atmosphere is lower than the condensation temperature of the gas. When the gas gravitationally settles beyond this pressure, it condenses and might be removed from the upper atmosphere depending on the strength of the vertical mixing. As \citet{Parmentier2013} showed, even if the vertical cold-trap is assumed to be inefficient, the tidally-locked rotation makes the temperature of the night side much colder than the day-side, creating a day-night cold-trap which should remove the gas phase of TiO/VO from the upper atmosphere of the planet. Based on the statistical study of the infrared phase curve of twelve hot Jupiters, \citet{Keating2019} found that the mean value of the night side temperature of hot Jupiters is around 1100 K, which is below the condensation temperature of TiO and VO \citep[e.g.][]{Lodders2002}. Depending on the ratio between the growth timescale of the Ti- or V-bearing condensate and the advective timescale, the gas phase of these molecules could still exist in the upper atmosphere of the planet.

Our results, however, seem to indicate that it is very unlikely that TiO/VO exists in the day-side of the planet. Therefore, the lack of the detection of any molecular thermal inversion agents might indicate that no inversion layers in the atmosphere of KELT-20b are driven by TiO/VO (if the line lists used in this analysis are accurate). Other mechanisms, e.g. absorption from metal atoms \citep{Lothringer2018}, can still provide enough opacity to create an observable inversion. This could be the case as it is supported by our detection of Fe\,{\sc i}. Our analysis showed that the other detected species are at a higher altitude than Fe\,{\sc i}, and the line-contrast is underestimated by our chemical equilibrium model which might indicate a higher temperature in the upper layer or, in other words, an inversion layer. Furthermore, this possibility has been supported observationally for WASP-121b, which shows a clear evidence of thermal inversion layer \citep{Evans2017} and strong absorption from Fe\,{\sc i} and Fe\,{\sc ii} \citep{gibson2020,Sing2019}, but has no TiO/VO as shown by \citet[]{Merrit2020} although again, subject to the accuracy of line lists. Further investigation should be done using the secondary eclipse technique or emission spectroscopy in the near-infrared to reveal the temperature structure on the day-side of the planet using either HST, JWST or ground-based facilities.

\section{Conclusions}\label{sec:concl}
We have searched for possible thermal inversion agents in the transmission spectrum of KELT-20b/MASCARA-2b in HARPSN and CARMENES data-sets. By combining all of the HARPSN data-sets, we were able to detect Fe\,{\sc i} at $>$ 13-$\sigma$ and Ca\,{\sc ii} H$\&$K at $>$ 6-$\sigma$. The signature of Fe\,{\sc i} was also detected in the CARMENES data-sets at $>$ 6-$\sigma$. Also, we confirmed the previous detection of Fe\,{\sc ii}, Ca\,{\sc ii} IRT and Na\,{\sc i} D in all data-sets that we analysed. We constrained the systemic velocity of KELT-20/MASCARA-2 to $-22.06\pm0.35$\,km s$^{-1}$ and $-22.02\pm0.47$\,km s$^{-1}$ using HARPSN N2 and CARMENES data-sets respectively and found a significant blue-shift in the Fe\,{\sc i} signal only ($>$ 3 km s$^{-1}$ at $>$ 5.3-$\sigma$). Using a new likelihood-mapping method, we were able to show that the absorption features of the detected species extend to different altitudes in the atmosphere. It also shows that our chemical equilibrium model has underestimated the line-contrast of the detected species except for Fe\,{\sc i}, which might indicate an inversion layer in the upper atmosphere. We detected no significant signature of other thermal inversion agents. Through the injection and recovery tests, we showed that our data are sensitive to most of the atomic/molecular species that we considered assuming the line lists are accurate. The non-detection of Ti- and V- bearing species suggests the presence of non-chemical equilibrium mechanisms, e.g. cold-traps, that removes them from the upper atmosphere. With these results, we predict that KELT-20b/MASCARA-2b either has no observable inversion layer or, if one does exist, then it might be caused by non TiO/VO-related mechanisms, most likely by UV and optical wavelength stellar absorption by Fe\,{\sc i} and Fe\,{\sc ii}.

Finally, in our analysis we detected a double-peak structure in the $K_{\mathrm{p}}-\Delta V$ map of Fe\,{\sc i}. In Section \ref{subsec:pecFeI}, we have shown that this structure is unlikely to have originated from either the noise or residuals of the data reduction processes that we know of or the spectrum template that we use. If it is real, this could be a signature of atmospheric dynamics. However, further investigation is needed to confirm the origin and nature of this structure. 

\section*{Acknowledgements}
S.K.N. and C.A.W. would like to acknowledge support from UK Science Technology and Facility Council grant ST/P000312/1. N. P. G. gratefully acknowledges support from Science Foundation Ireland and the Royal Society in the form of a University Research Fellowship. H.K. is supported by a Grant-in-Aid from JSPS (Japan Society for the Promotion of Science), Nos. JP17K14246, JP18H01247, and JP18H04577. This work was also supported by the JSPS Core-to-Core Program "Planet$^{2}$". We are grateful to the anonymous referee for constructive and insightful comments that greatly improved the quality of this paper. sWe would like to thank Simon Grimm for supporting us in adapting {\sc HELIOS-K} to our needs; Daniel Kitzmann for providing the updated database of {\sc FastChem}; and Vivien Parmentier for a fruitful discussion about the possible interpretation of the double-peak structure. This work is based on observations made with the Italian Telescopio Nazionale Galileo (TNG) operated on the island of La Palma by the Fundaci\'on Galileo Galilei of the INAF (Instituto Nazionale di Astrofisica) at the Spanish Observatorio del Roque de los Muchachos of the Instituto de Astrofisica de Canarias and CARMENES which is an instrument for the Centro Astron\'omico Hispano-Alem\'an de Calar Alto (CAHA, Almer\'ia, Spain) funded by the German Max-Planck-Gesellschaft (MPG), the Spanish Consejo Superior de Investigaciones Cient\'ificas (CSIC), the European Union through FEDER/ERF FICTS-2011-02 funds, and the members of the CARMENES Consortium. We are also grateful to the developers of the {\sc Numpy}, {\sc Scipy}, {\sc Matplotlib}, {\sc Jupyter Notebook}, and {\sc Astropy} packages, which were used extensively in this work  \citep{2020SciPy-NMeth,Hunter:2007, Kluyver:2016aa,astropy:2013, astropy:2018}
%%%%%%%%%%%%%%%%%%%%%%%%%%%%%%%%%%%%%%%%%%%%%%%%%%

%%%%%%%%%%%%%%%%%%%% REFERENCES %%%%%%%%%%%%%%%%%%

% The best way to enter references is to use BibTeX:

\bibliographystyle{mnras}
%\bibliography{example} % if your bibtex file is called example.bib
\bibliography{kelt20b_nugroho} % your references Yourfile.bib

% Alternatively you could enter them by hand, like this:
% This method is tedious and prone to error if you have lots of references

%%%%%%%%%%%%%%%%%%%%%%%%%%%%%%%%%%%%%%%%%%%%%%%%%%

%%%%%%%%%%%%%%%%% APPENDICES %%%%%%%%%%%%%%%%%%%%%

% \appendix

% \section{Some extra material}

% If you want to present additional material which would interrupt the flow of the main paper,
% it can be placed in an Appendix which appears after the list of references.

%%%%%%%%%%%%%%%%%%%%%%%%%%%%%%%%%%%%%%%%%%%%%%%%%%

% Don't change these lines
\bsp	% typesetting comment
\label{lastpage}
\end{document}